\documentclass[hyper,12pt,letterpaper]{JHEP3}
\usepackage{amsmath,amssymb,multirow,array}

\newcommand{\half}{\frac{1}{2}}

\numberwithin{equation}{section}

\newcommand{\nn}{\nonumber}

\title{Constraints on Fluid Dynamics from Equilibrium Partition Functions}
\author{Nabamita Banerjee$^{a,d}$, Jyotirmoy Bhattacharya$^c$, Sayantani Bhattacharyya$^b$,
Sachin Jain$^a$, Shiraz Minwalla$^a$ and Tarun Sharma$^a$\\
$^a$Dept. of Theoretical Physics, Tata Institute of Fundamental Research, Homi Bhabha Rd,
Mumbai 400005, India. \\
$^b$Harish-Chandra Research Institute, Chhatnag Road, Jhunsi, Allahabad-211019.\\
$^c$Institute for the Physics and Mathematics of the Universe (IPMU), University of Tokyo,
Kashiwa, Chiba 277-8582, Japan \\
$^d$ Institute for Theoretical Physics, Utrecht University,
Leuvenlaan 4, 3584 CE Utrecht, The Netherlands\\
Email:\ \ {\bf N.Banerjee@uu.nl, jyotirmoy.bhattacharya@ipmu.jp, sayanta@hri.res.in,
minwalla@theory.tifr.res.in, sachin@theory.tifr.res.in, tarun@theory.tifr.res.in
}}

\abstract{We study the thermal partition function of quantum field
theories on arbitrary stationary background spacetime, and with arbitrary
stationary background gauge fields, in the long wavelength expansion.
We demonstrate that the equations of relativistic hydrodynamics are significantly
constrained by the requirement of consistency with any
partition function. In examples at low orders in the derivative expansion we demonstrate that these
constraints coincide precisely with the equalities between hydrodynamical
transport coefficients that follow from the local form of the second law
of thermodynamics. In particular we recover the results of Son and Surowka
on the chiral magnetic and chiral vorticity flows,
starting from a local partition function that manifestly reproduces the
field theory anomaly, without making any reference to an entropy current.
We conjecture that the relations between transport coefficients that
follow from the second law of thermodynamics agree to all orders in the
derivative expansion with the constraints described in this paper.  }

\preprint{TFR/TH/12-05\\IPMU12-0037}
\begin{document}
\maketitle

\section{Introduction and Summary}

In this paper we explore the structural constraints imposed on the equations of
relativistic hydrodynamics by two related physical requirements. First that
these equations admit a stationary solution on an arbitrarily
weakly curved stationary background spacetime.
Second that the conserved currents (e.g. the stress tensor) on the
corresponding solution follow from an equilibrium partition function.

Landau-Lifshitz \cite{Landau:1987gn}, and several subsequent authors, have emphasized
another
source of constraints on the equations of hydrodynamics, namely that
the equations are consistent with a local form of the second law of
thermodynamics. As is well known, this requirement imposes inequalities on
several parameters (like viscosities and conductivities) that appear in the
equations of hydrodynamics. It is perhaps less well appreciated that
the requirement of local entropy increase also yields equalities relating
otherwise distinct fluid dynamical
parameters, and so reduces the number of free parameters that
appear in the equations of fluid dynamics (see e.g. \cite{Landau:1987gn,putterman}, for more recent work inspired by the AdS/CFT correspondence 
see e.g. \cite{Son:2009tf,Bhattacharya:2011tra,
Bhattacharya:2011eea,Herzog:2011ec,Neiman:2011mj, Loganayagam:2011mu, 
Romatschke:2009kr, Bhattacharyya:2012ex,Dubovsky:2011sk}). In three specific examples
we demonstrate below that the equalities obtained from the comparison with
equilibrium (described in the previous paragraph) agree exactly with
the equalities between coefficients obtained from the local second law of
thermodynamics. These results lead us
to conjecture that the constraints obtained from these two a naively distinct
physical requirements infact always coincide. We leave a fuller investigation
of this conjecture to future work.

In this paper we nowhere utilize the AdS/CFT correspondence. However our
work is motivated by the potential utility of our results in an investigation
of the constraints imposed by the second law of
thermodynamics on higher derivative corrections to Einstein's equations 
\cite{st}, via the Fluid - Gravity map of AdS/CFT (\cite{Bhattacharyya:2008jc}, see \cite{Hubeny:2010ry,Rangamani:2009xk} for reviews).

In the rest of this section we summarize our procedure and results in
detail. In subsection \ref{epf} below
we describe the structure of equilibrium partition functions for field theories
on stationary spacetimes in an expansion
in derivatives of the background spacetime metric (and gauge fields). In subsection 
\ref{ec} we then describe the constraints on the equations of relativistic hydrodynamics 
imposed by the structure of the partition
functions described in subsection \ref{epf}. In three
examples we compare these constraints to those obtained from the requirement
of entropy increase and find perfect agreement in each case.

\subsection{Equilibrium partition functions on weakly curved manifolds}\label{epf}

Consider a relativistically invariant quantum field theory on
a manifold with a timelike killing vector. By a suitable
choice of coordinates, any such manifold may be put in the form
\begin{equation} \label{metgf}
ds^2=-e^{2 \sigma(\vec x)} \left( dt+ a_{i}(\vec x) dx^i \right)^2
+g_{ij}({\vec x}) dx^i dx^j
\end{equation}
where $i = 1 \ldots p$. $\partial_t$ is the killing vector on this
manifold, while the coordinates ${\vec x}$ parametrize spatial slices. Here $\sigma, a_{i}, g_{ij}$ are smooth functions
of coordinates ${\vec x}$.

Let $H$ denote the Hamiltonian that generates translations of the
time coordinate $t$. In this subsection we address the following
question. What can we say, on general symmetry grounds, about the
dependence of the the partition function of the system
\begin{equation}\label{pftrd}
Z={\rm Tr} e^{-\frac{H}{T_0}},
\end{equation}
on $\sigma$, $g_{ij}$ and $a_i$? In this paper we focus on the long wavelength limit,
i.e. on manifolds whose curvature length scales are much larger than the
`mean free path' of the thermal fluid \footnote{Equilibrium Partition functions 
special curved manifolds or with particular background gauge fields have been 
studied before in \cite{Bhattacharyya:2007vs,Loganayagam:2012pz,Jensen:2011xb}}.  In this limit the question
formulated above may addressed using the techniques of effective
field theory. In the long wavelength limit the
background manifold may be thought of
as a union of approximately flat patches, in each
of which the system is in a local flat space thermal equilibrium
at the locally red shifted temperature
\begin{equation}\label{Td}
T(x)=e^{-\sigma} T_0 + \ldots
\end{equation}
(where $T_0$ is the equilibrium temperature of the system and
the $\ldots$ represent derivative corrections, see below).
Consequently the partition function of the system is given by
\begin{equation}\label{approxpft}
\ln Z=\int d^p x \sqrt{g_{p}} \frac{1}{T(x)} P(T(x)) + \ldots
\end{equation}
where $P(T)$ is the thermodynamical function that computes the
pressure as a function of temperature in flat space.
Substituting \eqref{Td} into \eqref{approxpft} we find
\begin{equation}\label{approxpf}
\ln Z=\int d^p x \sqrt{g_{p}} \frac{e^{\sigma}}{T_0} P(T_0 e^{-\sigma}) + \ldots
\end{equation}

The $\ldots$ in \eqref{approxpf}
denote corrections to $\ln Z$ in an expansion in derivatives of the background
metric. At any given order in the derivative expansion these correction
are determined, by the requirement of diffeomorphism invariance,
in terms of a finite number of unspecified  functions of $\sigma$. For example, to
second order in the derivative expansion, $p$ dimensional diffeomorphism
invariance and $U(1)$ gauge invariance of the
Kaluza Klein field $a$ constrain the action to take the form
\begin{equation}\label{pf2o}
\begin{split}
\log Z = W 
&= -\frac{1}{2} \Bigg(  \int d^px
\sqrt{g_{p}} \frac{e^{\sigma}}{T_0} P(T_0 e^{-\sigma})\\ &+
\int d^p x \sqrt{g_{p}}
\big(P_1(\sigma) R+T_0 ^2 P_2(\sigma)(\partial_i a_j -\partial_j a_i)^2
+P_3(\sigma)(\nabla \sigma)^2 \big) \Bigg)
\end{split}
\end{equation}
where $P_1(\sigma)$, $P_2(\sigma)$ and $P_3(\sigma)$
are arbitrary functions. It is possible to demonstrate on general
grounds that the temperature dependence of these functions is given by
\begin{equation}\label{tdf}
P_i(\sigma)={\tilde P}_i(T_0 e^{-\sigma})
\end{equation}
so that 
\begin{equation}\label{pf2ot}
\begin{split}
\log Z = W 
&= -\frac{1}{2} \Bigg(  \int d^px
\sqrt{g_{p}} \frac{e^{\sigma}}{T_0} P(T_0 e^{-\sigma})\\ &+
\int d^p x \sqrt{g_{p}}
\big( {\tilde P}_1(T_0 e^{-\sigma}) R+T_0 ^2 {\tilde P}_2(T_0e^{-\sigma})(\partial_i a_j -\partial_j a_i)^2
+ {\tilde P}_3(T_0 e^{-\sigma})(\nabla \sigma)^2 \big) \Bigg)
\end{split}
\end{equation}

The discussion above is easily generalized to the study of a relativistic
fluid which possesses a conserved current $J_\mu$ corresponding to a
global $U(1)$ charge.  We work on the manifold \eqref{metgf} in the presence
of a time
independent background $U(1)$ gauge connection
\begin{equation} \label{gf}
{\cal A}= {\cal A}_0(\vec x) dx^0+ {\cal A}_i (\vec x) dx^i
\end{equation}
and study the partition function
\begin{equation}\label{pftrdn}
Z={\rm Tr} e^{-\frac{H -\mu_0 Q}{T_0}}
\end{equation}
Later in the paper we present a detailed study of the
special case of charged fluid dynamics in $p=3$ and
$p=2$ spatial dimensions, at first order in the derivative expansion,
without imposing the requirement of parity invariance.
Let us first consider the case $p=3$. The requirements of three dimensional
diffeomorphism invariance, Kaluza Klein gauge invariance, and $U(1)$
gauge invariance upto an anomaly\footnote{In this paper we only consider the effect of $U(1)^3$ anomalies 
ignoring the effects of . for instance, mixed gravity-gauge  anomalies. A systematic study of the effect of these 
anomalies in fluid dynamics would require us to extend our analysis of charged fluid dynamics to 2nd order, 
a task we leave for the future (see however
section \ref{count}). It is possible that $C_2$ above will turn out to be determined in terms of such an anomaly coefficient. We thank
R. Loganayagam for pointing this out to us.}
 (see below) force the partition function
to take the form\footnote{Our convention is 
$$ \half \int XdY = \int d^3 x \sqrt{g_3} \epsilon^{ijk}X_{i}\partial_{j}Y_{k}~, 
\quad \half \int dY = \int d^2 x \sqrt{g_2} \epsilon^{ij}\partial_{i}Y_{j}~. $$}
\begin{equation}\label{cfacn} \begin{split}
\ln Z & = W^0+W^1_{inv} +W^1_{anom}\\
W^0 &= \int  \sqrt{g_3} \frac{e^{\sigma}}{T_0} P\left(T_0 e^{-\sigma}, e^{-\sigma}A_0 \right) \\
W^1_{inv}&= \frac{C_0}{2}  \int A d A
+ \frac{T_0 ^2 C_1}{2} \int  a d a + \frac{T_0 C_2}{2} \int A d a   \\
W^1_{anom}&= \frac{C}{2} \left( \int \frac{A_0}{3 T_0} A d A
+ \frac{A_0^2}{6 T_0} A d a \right)
\end{split}
\end{equation}
where $A_i$
\begin{equation}\label{sgf} \begin{split}
A_0&={\cal A}_0+\mu_0\\
A_i&={\cal A}_i -A_0 a_i\\
\end{split}
\end{equation}
\eqref{cfacn} is written in terms of $A_i$ because $A_i$,  unlike ${\cal A}_i$, is
Kaluza Klein gauge invariant\footnote{The background data can be taken as gauge field ${\cal A}=({\cal A}_0, {\cal A}_i)$ 
with constant chemical potential $\mu_0$ and temperature $T_0$. Equivalently we can think of the system to have background 
gauge field $B = ({\cal A}_0 + \mu_0, {\cal A}_i)$ with no chemical potential. These two are equivalent physical statements as 
$\mu_0$ can be absorbed in the constant part of ${\cal A}_0$. }.

$W^0$ in \eqref{cfacn} is zero derivative contribution to the partition
function, and is the patchwise approximation to equilibrium, in the
spirit of $\eqref{approxpf}$. $W^1_{inv}$ is the most general diffeomorphism
and gauge invariant one derivative correction to $W^0$. Note that
$W^1$ is the sum of a Chern Simons term for the connection $A$, a Chern
Simons term for the connection $a$ and a mixed Chern Simons term
in $A$ and $a$. As usual, gauge invariance forces
 the coefficients $C_0$, $C_1$ and $C_2$ of these
Chern Simons terms to be constants.

\eqref{cfacn} is the most general form of the partition function of our system
that satisfies the requirements of 3 dimensional diffeomorphism invariance and
gauge invariance. If we, in addition, impose the requirement of CPT invariance of
the underlying four dimensional field theory then it turns out that
$C_0=C_1=0$ (see subsection \ref{cpt}). In other words, the requirement of CPT invariance allows only the mixed
Chern Simons term, setting the `pure' Chern Simons terms to zero.

$W^1_{anom}$ is the part of the effective action that is not gauge
invariant under $U(1)$ gauge transformations. \footnote{It is striking that 
the effect of the anomaly can be captured by a local term in the $3$ dimensional 
effective action. Note that $W^1$ cannot be written as the dimensional reduction of 
a local contribution to the 4 dimensional action, in agreement with 
general expectations.} Its gauge variation under
$A_\mu \rightarrow A_\mu+\partial_\mu \phi({\vec x})$ is given by
\begin{equation}\label{vpg}
\delta W^1_{anom}=  \frac{C}{24 T_0} \int d^3 x \sqrt{-g_4} *( {\cal F} \wedge {\cal F})~\phi(x)
\end{equation}
As we explain in much more detail below, this is exactly the variation
of the effective action predicted by the anomalous conservation equation
\begin{equation}\label{gie}
\nabla_\mu {\tilde J}^\mu = -\frac{C}{8} *({\cal F} \wedge {\cal F})
\end{equation}
where ${\tilde J}$ is the gauge invariant $U(1)$ charge current, and
$*$ denotes the Hodge dual.

Let us now turn to parity violating charged fluid dynamics in $p=2$ spatial
dimensions. In this case there is no anomaly in the system and the parity odd
sector is qualitatively much different from its $p=3$ spatial
dimension counterpart. For this system we primarily focus on the 
parity odd sector upto the first order in derivative expansion 
and the manifestly gauge invariant partition function in this case takes the form
\begin{equation}
 \ln Z = {\cal W}^0 + {\cal W},
\end{equation}
where
\begin{equation}\label{action3dintro}
\begin{split}
 {\cal W}^0 &=\int \sqrt{g_2} \frac{e^{\sigma}}{T_0} P\left(T_0 e^{-\sigma}, e^{-\sigma}A_0 \right) \\
 {\cal W} &= \frac{1}{2}\int  \left( \alpha(\sigma, A_0) ~dA + T_0 ~\beta(\sigma, A_0)~da \right).
\end{split}
\end{equation}
Where $A_0$ and $A_i$ are defined in \eqref{sgf} and $\alpha$ and $\beta$ are arbitrary functions. 

It is straightforward, if tedious, to generalize the form of the
partition function  presented in special examples
above to higher orders in the derivative
expansion. To any given order
in the derivative expansion, the dependence of $\ln Z$,
on $g_{ij}$, $a_i$, $\sigma$, $A_0$ and $A_i$
is fixed by the requirements of $p$ dimensional diffeomorphism invariance and
gauge invariance in terms of a finite number of unspecified functions of
two variables, $\sigma$ and $A_0$.

We will now define some terminology that will prove useful in the sequel.
Let $s_e^n$ denote the number of independent gauge invariant scalar expressions
that one can construct out of $\sigma$, $a_i$ (and $A_0$ and $A_i$ in
the case that the fluid is charged) at $n^{th}$ order in the
derivative expansion. In a similar manner, $v_e^n$ and $t^n_e$ will
denote the number of $n^{th}$ order independent gauge invariant vectors and
(traceless symmetric two index) tensors formed out of the same quantities.
Finally let $st_{e}^n$ denote the total number of $n^{th}$ order scalar
expressions that happen to be total derivatives (including the 
contribution of a coefficient function) and so integrate to zero
\footnote{For example, the two derivative scalar 
$\nabla_\mu h(\sigma) \nabla^\mu \sigma$ is a total derivative for
arbitrary $h(\sigma)$.}
It is clear that at $n^{th}$ order in the derivative expansion, the
equilibrium action $\ln Z$ depends on $s_e^n-st_e^n$ unknown
functions of two variables.

\subsection{Constraints on Fluid Dynamics from stationary equilibrium}\label{ec}

\subsubsection{Relativistic Hydrodynamics}\label{rh}

In this subsubsection we present a lightening review of the structure of the
equations of
charged relativistic hydrodynamics. The equations of hydrodynamics are simply
the equations of conservation of the stress tensor and the charge current
\begin{equation}\label{EOMNE}
\nabla_{\mu} T^{\mu}_{\nu}= {\cal F}_{\nu \mu}\tilde J^{\mu}, \quad \nabla_{\mu}\tilde J^{\mu}= -\frac{C}{8} *({\cal F} \wedge {\cal F}),
\end{equation}
where ${\cal F}$ is the field strength of the gauge field ${\cal A}$ in \eqref{gf}.
These equations constitute a closed dynamical system when supplemented
with constitutive relations that express $T_{\mu\nu}$ and $J_\mu$ as
a function of the fluid temperature, chemical potential and velocity.
These constitutive relations are presented in an expansion in
derivatives and take the form
 \begin{equation}\label{consrel}
T^{\mu\nu}=(\epsilon+P) u^{\mu}u^{\nu}+ P g^{\mu\nu}+\pi^{\mu\nu}, \quad J^{\mu}=
q u^{\mu}+J^{\mu}_{diss},
\end{equation}
The pressure $P$, proper energy density $\epsilon$ and
proper charge density $q$ are those functions of $T$ and
$\mu$ predicted by flat space equilibrium thermodynamics.
$\pi_{\mu\nu}$ refers to the sum of all corrections to the stress tensor
that are of first or higher order in the derivative expansion
(the derivatives in question could act either on the $T$,  $\mu$, $u^\mu$,
or the background metric and gauge field $g_{\mu \nu}$ and ${\cal A}_\mu$).
Similarly $J^{\mu}_{diss}$ refers to corrections to the perfect fluid
charge current that depend on atleast one spacetime derivative.
 Field redefinitions of the
$T$ $\mu$ and $u^\mu$ may be used to impose $p+2$ constraints
on $\pi_{\mu\nu}$ and $J^\mu_{diss}$; throughout this paper we will work in
the so called Landau Frame in which
\begin{equation}\label{frame}
u^\mu \pi_{\mu\nu}=0, ~~~ u^\mu J_\mu^{diss}=0
\end{equation}
Terms in $\pi_{\mu\nu}$ and $J^\mu_{diss}$ are both graded according to the
number of spacetime derivatives they contain, i.e.
\begin{equation}\label{ceq} \begin{split}
\pi^{\mu\nu}&=\pi_{(1)}^{\mu\nu} +  \pi_{(2)}^{\mu\nu} + \pi_{(3)}^{\mu\nu} + \ldots \\
J^\mu_{diss}&=J^\mu_{diss, (1)}+ J^\mu_{diss, (2)}+ J^\mu_{diss, (3)} + \ldots \\
\end{split}
\end{equation}
where the subscript counts the number of derivatives.

Symmetry considerations immediately constrain the possible expansions for
$\pi_{\mu\nu}$ and $J^\mu_{diss}$ as follows. At any given point in spacetime,
the fluid velocity $u^\mu$ is a particular timelike vector. The
value of the velocity breaks the local $SO(p,1)$ Lorentz symmetry
of the theory down to the rotational subgroup $SO(p)$. In the Landau frame
\eqref{EOMNE}
$\pi_{\mu\nu}$ may be decomposed into an $SO(p)$ tensor and $SO(p)$ scalar.
$J^\mu_{diss}$ is an $SO(p)$ vector.

In order to parameterize freedom in the equations of hydrodynamics, it is
useful to define some terminology. Let $t_f^n$, $v_f^n$ and $s_f^n$
respectively denote the number of onshell
inequivalent tensor, vector and scalar expressions that can be formed
out expressions made up of a total of $n$ derivatives acting on
$T$, $u^\mu$, $\mu$, $g_{\mu\nu}$ and $A_\mu$. It follows immediately
that the most general symmetry allowed expression for $\pi^{\mu\nu}_{(n)}$
is given
in terms of $t_f^n + s_f^n$ unknown functions of the two variables $T$ and
$\mu$. In a similar manner the most general expression for the
$J^\mu_{diss (n)}$, permitted by symmetries,
 is given in terms of $v_f^n$ unknown functions of the same two
variables.

It turns out that the $(t_f^n+s_f^n+v_f^n)$  $n^{th}$ order transport
coefficients are not all independent. The requirement that
the hydrodynamical equations are consistent with the existence of an entropy
current that is of positive divergence in every conceivable fluid flow
imposes several relationships between these coefficients cutting down the number
of parameters in these equations; we refer the
reader to \cite{Landau:1987gn,Son:2009tf,Bhattacharya:2011eea,Bhattacharyya:2012ex},
for example, for a fuller discussion. We now turn to a
description of  a simpler physical principal that appears predict the same relations between
these coefficients.These relations may all be constructively determined by comparison of the equations of hydrodynamics with a
partition function.

\subsubsection{Constraints from stationary equilibrium}
\label{cfd}

As we have explained in the previous subsubsection, it follows from
symmetry considerations that the equations of charged hydrodynamics, at
$n^{th}$ order in the derivative expansion, are parameterized by
$t_f^n+v_f^n+s_f^n$ unknown functions of two variables (or
$t_f^n+s_f^n$ functions of one variable for uncharged hydrodynamics).
We will now argue that these functions are not all independent, but
instead are determined in terms of a smaller number of functions.

 It is easy to
verify that the equations of perfect fluid hydrodynamics (hydrodynamics
at lowest order in the derivative expansion) admit a stationary
`equilibrium' solution in the backgrounds \eqref{metgf} and \eqref{gf} given by
\begin{equation}\label{pfeq}
u_{(0)}^\mu({\vec x})=e^{-\sigma}(1,0, \ldots, 0), ~~~~T_{(0)}({\vec{x})}=T_o e^{-\sigma}, ~~~~
\mu_{(0)}({\vec x})=e^{-\sigma}A_0
\end{equation}
As explained above, this is also the equilibrium solution one expects of the fluid
on intuitive ground.
At higher order in the derivative expansion this solution is corrected;
the corrected solution may be expanded in derivatives
\begin{equation}\label{pfeqg} \begin{split}
u^\mu&=u_{(0)}^\mu +  u_{(1)}^\mu + u_{(2)}^\mu + \ldots \\
T&=T_{(0)}+ T_{(1)}+ T_{(2)} + \ldots \\
\mu&=\mu_{(0)}+ \mu_{(1)}+ \mu_{(2)} + \ldots \\
\end{split}
\end{equation}
where $u^\mu_{(n)}$, $T_{(n)}$ and $\mu_{(n)}$ are expressions of $n^{th}$
order in derivatives acting on $\sigma$, $A_0$, $a_i$, $A_i$ and $g_{ij}$.
What can we say about the form of the corrections
$u^\mu_{(n)}$, $T_{(n)}$ and $\mu_{(n)}$? Adopting the notation defined in the
last paragraph of the previous subsection,  symmetries determine the expression
for $u^\mu_{(n)}$ in terms of $v_e^n$ as yet unknown
functions of $\sigma$ and $A_0$, while $T$ and $\mu$ are each determined
in terms of $s_e^n$ as yet unknown equations of $A_0$ and $\sigma$.

The stress tensor and charge current in equilibrium are given by plugging
\eqref{pfeqg} into \eqref{ceq}. The result is an expression for
$\pi^{\mu\nu}$ and $J^\mu_{diss}$ written
entirely in terms of $\sigma$, $A_0$, $a_i$, $A_i$, $g_{ij}$ and their
derivatives.

This expressions
for the stress tensor and charge current so obtained depend
only on a subset of the transport coefficients that appear
in the expansion of $\pi^{\mu\nu}$ and $J^\mu_{diss}$. For instance,
the expansion of the $n^{th}$ order tensor part of $\pi^{\mu\nu}$
has $t_f^n$ terms in general. When evaluated on \eqref{pfeq}, however,
this expression reduces to a sum over $t_e^n \leq t_f^n$ terms.
The coefficients of these terms define $t_e^n$ subspace of the $t_f^n$
dimensional set of $n^{th}$ order transport coefficients. We refer to
this subspace as the subspace of {\it non dissipative} transport
coefficients.

In this paper we demand that the expressions for the equilibrium
stress tensor and charge current, obtained as described in the
previous paragraph, agree with the corresponding expressions obtained
by differentiating the equilibrium partition function of subsection
\ref{epf} with respect to  the
background gauge field and metric. This
requirement yields a set of $t_e^n+ 2v_e^n + 3s_e^n$ equations\footnote{The counting goes as follows.
The stress tensor decomposes into one $SO(p)$, tensor, one vector
and two scalars. The charge current decomposes into a vector and a scalar.
Equating the hydrodynamical equilibrium stress tensor and charge current to
the expressions obtained by varying the equilibrium yields
$3s_e^n+2v_e^n + t_e^n$ equations.}
that completely determine both the $n^{th}$ order corrections
to the equilibrium solutions $T_n$ $\mu_n$ and $u^\mu_n$
 ($v_e^n+2s_e^n$ coefficients in all) as
well as the $t_e^n+v_e^n+ s_e^n$ non dissipative hydrodynamical transport
coefficients. Note that the number of variables precisely equals the number
of equations. Dissipative hydrodynamical transport coefficients
are completely unconstrained by this procedure.

We emphasize that the shifted equilibrium velocities, temperatures and
chemical potentials obtained from the procedure just described automatically
obey the equations of hydrodynamics. By construction, the shifted
fluid variables, together with the constitutive relations determined above
yield the stress tensor that follows from the functional variation of an equilibrium
partition function, and the stress tensor obtained from the variation of {\it any}
diffeomorphically invariant functional is automatically conserved. Very similar
remarks apply to the charge current.

Let us summarize. In general $\pi^{\mu\nu}$ and
$J^\mu_{diss}$ are expanded in terms of $t_f^n+s_f^n$ and $v_f^n$ transport
coefficients, each of which is a function of temperature and chemical
potential. However $t_f^n-t_e^n+s_f^n-s_e^n$ of these coefficients in
$\pi^{\mu\nu}$ and $v_f^n-v_e^n$ of these coefficients in $J^\mu_{diss}$
evaluate to zero on the `equilibrium' configuration \eqref{pfeq}.
The remaining $t_e^n+v_e^n+s_e^n$  non dissipative
transport coefficients multiply
expressions that do not vanish on \eqref{pfeq}.  Comparison with the
equilibrium partition function algebraically determines all non dissipative
transport coefficients in terms of the $s_e^n-st_e^n$ functions
(and derivatives thereof) that appear as coefficients in
the derivative expansion of the partition function. In other words
the  $t_e^n+v_e^n+s_e^n$ non dissipative transport coefficients
are not all independent; there
exist $t_e^n+v_e^n+st_e^n$ relations between these coefficients.

The procedure described above may also be used to derive constraints 
on the form of the fluid entropy current. The entropy current must obey 
two constraints. First its divergence must vanish on all the equilibrium 
configurations derived above. Second, the integral over the entropy density
(obtained from the entropy current) must equal the thermodynamical 
entropy that follows from the partition function \eqref{pftrdn}. 
These requirements impose constraints on the form of the (non dissipative) part of 
the most general symmetry allowed hydrodynamical entropy current.

We have implemented the procedure described above in detail
in three separate examples which we describe in more detail
immediately below.  In each case we have obtained detailed
expressions for all non dissipative hydrodynamical coefficients
in terms of the parameters that appear in the action.
In each case, the relations obtained between non dissipative
transport coefficients, after eliminating the action parameters,
agree exactly with the relations obtained between the same
quantities by previous investigations based on the study study
of the second law of thermodynamics.

In the case of parity violating first order fluid dynamics in $3+1$ dimensions,
the results for transport coefficients computed from \eqref{cfacn}
match perfectly\footnote{See \cite{Gao:2012ix,Pu:2010as} for an alternate (using quantum kinetic theory)
 derivation  of  hydrodynamic coefficients related to the chiral anomaly without making any reference to an entropy current.} with those of Son and Surowka \cite{Son:2009tf} (generalized in
\cite{Neiman:2010zi},\cite{Bhattacharya:2011tra}) once we impose the additional requirement of CPT invariance.
\footnote{Before imposing the requirement of CPT invariance, we have
an additional one parameter freedom that is not captured by the
the generalized Son-Surowka analysis. The reason for this is that Son and
Surowka (and subsequent authors) assumed that the entropy current was
necessarily gauge invariant. This does not seem to us to be physically
necessary. It seems to us that an entropy current whose divergence is
gauge invariant - and whose integral over a compact manifold in equilibrium
is gauge invariant - is perfectly acceptable. As we explain below,
it is easy to find a one parameter generalization of the Son-Surowka
solution that meets these conditions, and that gives rise to the additional
term $C_0$ in the partition function \eqref{cfacn}. However it turns out that
the requirement of CPT invariance sets $C_0$ (along with $C_1$)
to zero in \eqref{cfacn}, so this possible ambiguity is never realized
in the hydrodynamical description of a quantum field theory.}

In the case of parity preserving fluid dynamics in 3+1 dimensions, the
results obtained from the partition function \eqref{pf2o} agree perfectly
with those of Bhattacharyya  \cite{Bhattacharyya:2012ex}. Finally, in the case of parity non
preserving charged fluid dynamics in 2+1 dimensions, the results from section
\ref{Sec:chargedhydro3d} agree perfectly with those of \cite{Jensen:2011xb}.

In ending this introduction let us note the following. As we have described
at the beginning of the introduction, the physical principles that yield
constraints on the transport relations of fluid dynamics are twofold.
First, that these equations are consistent with the existence of a stationary
solution in every background of the form \eqref{metgf}, \eqref{gf}. Second,
that the stress tensor and charge current evaluated on
this equilibrium configuration obeys the integrability constraints
that follow if these expressions can be obtained by differentiating a
partition function. In the presentation described above we have mixed
these two conditions together (as the partition function is the
starting point of our discussion). However it is also possible to separate
these two conditions. For each of the three examples discussed above, in
Appendix \ref{Sec:eqcal1}, \ref{Sec:eqcal3} and \ref{Sec:eqcal2} we present a detailed study of the constraints on the equations
of fluid dynamics obtained merely from the existence of stationary
solutions in arbitrary backgrounds of the form  \eqref{metgf}, \eqref{gf}.
In each case we find that all of the relations between transport coefficients,
derived in this paper, are implied already by this weaker condition.
In these three examples, once equilibrium exists, the requirement that
it follows from a partition function turns out to be automatic. We do
not expect this always to be the case. In more complicated cases we expect
the existence of a partition function to imply further constraints than those 
implied merely by the existence of equilibrium.
However we leave the study of such effects to future work\footnote{After this paper was completed we were informed of two upcoming works \\ 
(1)``Towards hydrodynamics without an entropy current''
of K. Jensen, M. Kaminski, P. Kovtun, A. Ritz, R. Meyer and A. Yarom \\
(2)``Triangle Anomalies, Thermodynamics, and Hydrodynamics'' of K. Jensen \\
which have some overlap with our current work.}.

\section{Preparatory Material}

In this section we present background material that we will need in the
main part of the paper. In subsection \ref{kkm} we present
some Kaluza Klein reduction formulae for metrics of the form \eqref{metgf}.
In subsection \ref{kkgt} we describe the transformation properties of various
quantities of interest under Kaluza Klein gauge transformations.
In subsection \ref{stuc} we discuss how the stress tensor and charge current
of our system is related to the partition function. We also discuss the thermodynamical
energy, entropy and entropy of our system, and compare these quantities to those obtained
from integrals over local currents. In subsection \ref{cca} we discuss the relation between
consistent currents (those obtained from the variation of an action) and gauge invariant
currents in systems with a $U(1)$ anomaly. In \ref{pfh} we describe how the
equations of perfect fluid hydrodynamics may be `derived' starting from a zero derivative
equilibrium partition function.

\subsection{Kaluza Klein Reduction Formulae}\label{kkm}

As explained in the introduction, in this paper we study theories on
 metric and gauge fields in the Kaluza Klein form
\begin{equation} \label{metgfa}
\begin{split}
ds^2&=-e^{2 \sigma(\vec x)} \left( dt+ a_{i}(\vec x) dx^i \right)^2
+ g_{ij}(\vec x) dx^i dx^j\\
{\cal A}^{\mu}&=( A^0(\vec x), {\cal A}^i (\vec x))\
\end{split}
\end{equation}
The inverse of this metric is given by

\[
g^{\mu\nu}=
\left( {\begin{array}{cc}
(-e^{-2 \sigma}+a^2) & -a^i  \\
 -a^i & g^{ij}  \\
\end{array} } \right)
\]
where the first row and column refer to time and  $g^{ij}$ is the inverse of $g_{ij}$.
Christoffel symbols, ${\tilde \Gamma}$, of the $p+1$ dimensional metric
are given in terms of those of the $p$ dimensional Christoffel symbols $\Gamma$
by
\begin{eqnarray}\label{Gamma}
\tilde \Gamma^{0}_{00}&=& - e^{2 \sigma} (a.\partial) \sigma \nn \\
\tilde \Gamma^{i}_{00}&=&  e^{2 \sigma} g^{im}\partial_m \sigma \nn \\
\tilde \Gamma^{0}_{i0}&=& \partial_i \sigma - e^{2 \sigma}(a.\partial) \sigma  a_i + \frac{e^{2 \sigma} f_{im}a^m}{2} \nn \\
\tilde \Gamma^{i}_{j0}&=& e^{2 \sigma} g^{ik}( - \frac{1}{2} f_{jk}+ \partial_k \sigma a_j) \nn \\
\tilde \Gamma^{0}_{ij}&=& - a_n \Gamma^n_{ij} +\frac{e^{2 \sigma}}{2}\bigg[a_j a^m\partial_i a_m+a_i a^m \partial_j a_m\bigg] \nn \\
&-&\frac{1}{2}a.\partial (e^{2 \sigma} a_ia_j)+ \frac{e^{-2 \sigma}}{2}\bigg[\partial_i(e^{2 \sigma}a_j)
+ \partial_j(e^{2 \sigma}a_i)\bigg] \nn \\
\tilde \Gamma^{k}_{ij}&=& \Gamma^{k}_{ij} -\frac{ e^{2 \sigma}}{2}g^{km}\bigg[a_j \partial_i a_m + a_i \partial_j a_m \bigg]\nn \\
&+& \frac{1}{2}g^{km}\partial_m(e^{2 \sigma}a_ia_j)
\end{eqnarray}
Curvature symbols of the $p+1$ dimensional metric (e.g. the Ricci scalar
${\tilde R}$) are given in terms of $p$ dimensional curvature data (e.g.
the $p$ dimensional Ricci Scalar $R$) by\footnote{The definitions we adopt in this paper are 
\begin{equation}\begin{split}\label{defr}
 R_{\mu\nu\rho}^{~~~~\sigma}&=\partial_{\nu}\Gamma_{\mu\rho}^{\sigma}-\partial_{\mu}\Gamma_{\nu\rho}^{\sigma}+\Gamma_{\mu\rho}^{\alpha}\Gamma_{\alpha\nu}^{\sigma}-\Gamma_{\nu\rho}^{\alpha}\Gamma_{\alpha\mu}^{\sigma},\\ 
R_{\mu\nu}&= R_{\mu\sigma\nu}^{~~~~\sigma}.
\end{split}
\end{equation}
We always use the mostly positive
signature.}

\begin{eqnarray}\label{sodata}
\tilde R&=& R+\frac{1}{4}e^{2\sigma}f^2- 2 (\nabla \sigma)^2-2 \nabla^2 \sigma \nn \\
\tilde R^{ij}&=& R^{ij}-\nabla^i \sigma \nabla^j \sigma-\nabla^i \nabla^j \sigma
+\frac{1}{2} e^{2 \sigma}f^{i m}f^j_{\,\ m} \nn \\
K^{ij}&\equiv &\tilde R_{0 \,\, 0}^{ \,\ i \,\ j} (u^0)^2= \nabla^i \sigma \nabla^j \sigma +\nabla^i \nabla^j \sigma
+\frac{1}{4} e^{2 \sigma}f^{i m}f^j_{\,\ m}, \nn \\
\end{eqnarray}
where $f_{ij}=\partial_i a_j - \partial_j a_i$

Let us define $u^\mu_K$ to be the unit normalized vector in the Killing direction. In components
\begin{equation}\label{ukd}
 u^\mu_K=e^{-\sigma}(1, 0, \ldots , 0)
\end{equation}
Let ${\cal P_K}^{\mu\nu}$ denote the projector orthogonal to $u^\mu_K$
\begin{equation}\label{pd}
 {\cal P_K}^{\mu\nu}=g^{\mu\nu} +u^\mu_K u^\nu_K
\end{equation}
Explicitly in matrix form
\[
({\cal P}_K)_{\mu\nu}=\left( \begin{array}{cc}
           0 & 0 \\
           0 & g_{ij} \\
         \end{array} \right)
\]

Let us also define the shear tensor, vorticity and expansion and acceleration
 of this Killing `velocity' field by
\begin{equation}\label{kds} \begin{split}
&\Theta_K = \nabla. u_K = \text{Expansion},~~
{\mathfrak a}_K^\mu = (u_K.\nabla) u^\mu_K = \text{Acceleration}\\
&\sigma^{\mu\nu}_K =
P^{\mu\alpha} P^{\nu\beta}\left(\frac{\nabla_\alpha (u_K)_\beta + \nabla_\beta (u_K)_\alpha}{2}
 - \frac{\Theta_K}{3}g_{\alpha_\beta}\right) = \text{Shear tensor}\\
&\omega^{\mu\nu}_K =
P^{\mu\alpha} P^{\nu\beta}\left(\frac{\nabla_\alpha (u_K)_\beta
- \nabla_\beta (u_K)_\alpha}{2}\right)=\text{Vorticity}\\
\end{split}
\end{equation}
A straightforward computation yields
\begin{equation}\label{kds} \begin{split}
&\Theta_K =0,~~
({\mathfrak a}_K)_\mu = ({\cal P_K})_{\mu i} \nabla^i \sigma \\
&\sigma^{\mu\nu}_K = 0\\
&(\omega_K)_{\mu\nu} = \frac{e^\sigma}{2} ({\cal P_K})_{\mu i} ({\cal P_K})_{\nu j} f^{ij}
\end{split}
\end{equation}

\subsection{Kaluza Klein gauge transformations}\label{kkgt}

The form of the metric and gauge fields in
\eqref{metgfa} is preserved by $p$ dimensional spatial diffeomorphisms
together with redefinitions of time of the form
\begin{equation}\label{KKgauge}
t'=t+ \phi(\vec x),\quad x'=x.
\end{equation}
Under coordinate changes of the form \eqref{KKgauge} the Kaluza Klein
gauge field $a_i$ transforms like a connection:
$$a_i'=a_i -\partial_i \phi.$$
Let us now examine the transformation of $p+1$ dimensional tensors
under the coordinate transformations \eqref{KKgauge}. It is not difficult
to verify that {\it upper spatial} indices and {\it lower temporal} indices
are gauge invariant. So, for instance, if $A_{\mu\nu}$ is any $p+1$
dimensional two tensor, the $p$ dimensional scalar $A_{00}$, the $p$
dimensional vector $A_0^i$ and the $p$ dimensional tensor $A^{ij}$
are all Kaluza Klein gauge invariant. On the other hand lower spatial
indices and upper temporal indices transform under the Kaluza Klein
gauge transformation \eqref{KKgauge} according to
\begin{equation}
V'_i=
V_i- \partial_i \phi V_0,
\quad (V')^0=
V^0+ \partial_i \phi V^i.
\end{equation}
Note that the $p$ dimensional oneforms
$$g_{ij} V^j = V_i-a_i V_0$$
are gauge invariant. In the sequel we will make heave use of the
$p$ dimensional oneforms
\begin{equation}\label{shift}
A_i={\cal A}_i- a_i A_0
\end{equation}
This oneform is Kaluza Klein gauge invariant
and transform as connections under
$U(1)$ gauge transformations. This is the reason that the partition function
\eqref{cfacn} was written in terms of $A_i$ rather than ${\cal A}_i$.

\subsection{Stress Tensor and $U(1)$ current} \label{stuc}

The $p+1$ dimensional tensors that will be of most interest to us in this
paper are the stress tensor, the charge current and the entropy current.
The stress tensor and charge current are defined in terms of variation
of the action with respect to the higher dimensional metric and gauge field
according to the formulas
\begin{equation}\label{defnstc}
\delta S= \int dx^{p+1}\sqrt{-g_{p+1}}
\left( -\frac{1}{2} T_{\mu\nu} \delta g^{\mu\nu}  + J^\mu \delta {\cal A^\mu}
\right)
\end{equation}
As we have described in the introduction, in this paper we will be
interested in the partition function $\ln Z$ of our system on the
background \eqref{metgfa}. This partition function may be thought of as
the Euclidean action of our system on the metric \eqref{metgf} with
coordinate time $t$ compactified on a circle of length $\frac{1}{T_0}$.
The change of $\ln Z$ under time independent variations of the metric
and gauge field is thus given by
\begin{equation}\label{defnstc} \begin{split}
\delta \ln Z & = \int dx^{p+1}\sqrt{-g_{p+1}}
\left( -\frac{1}{2} T_{\mu\nu} \delta g^{\mu\nu}  + J^\mu \delta {\cal A_\mu}
\right)\\
&= \frac{1}{T_0} \int dx^{p}\sqrt{-g_{p+1}}
\left( -\frac{1}{2} T_{\mu\nu} \delta g^{\mu\nu}  + J^\mu \delta {\cal A_\mu}
\right)\\
\end{split}
\end{equation}
It follows that
\begin{equation} \begin{split} \label{stj}
T_{\mu\nu} & =  -2 T_0\frac{\delta \ln Z}{\delta g^{\mu\nu}} \\
J^\mu &= T_0\frac{\delta \ln Z}{\delta {\cal A}_\mu}
\end{split}
\end{equation}
The formulae \eqref{stj} are not written in the
most useful form for the purposes of this paper.
 As we have described in the introduction, we find it
useful to regard our partition function as a functional of
\begin{equation}\label{partionfunc}
\ln Z = W(e^\sigma,A_0, a_i, A_i, g^{ij}, T_0, \mu_0).
\end{equation}
By application of the chain rule to the formulas \eqref{defnstc} we find
\begin{eqnarray}\label{stcurrent}
T_{00}&=& -\frac{T_0 e^{2 \sigma}}{\sqrt{-g_{(p+1)}} }\frac{\delta W}{\delta \sigma},
\quad T_0^i= \frac{T_0}{\sqrt{-g_{(p+1)}} }\bigg(\frac{\delta W}{\delta a_i}-
A_0 \frac{\delta W}{\delta A_i}\bigg) , \nn \\
T^{ij}&=& -\frac{2 T_0}{\sqrt {-g_{(p+1)}}} g^{il}g^{jm}\frac{\delta W}
{\delta g^{lm}}, \quad
J_0 =-\frac{e^{2\sigma} T_0}{\sqrt{-g_{(p+1)}}}\frac{\delta W}{\delta A_0},
\quad J^i=\frac{T_0}{\sqrt{-g_{(p+1)}}}\frac{\delta W}{\delta A_i}.
\end{eqnarray}
where, for instance, the derivative w.r.t $A_0$ is taken at constant $\sigma$, $a_i$, $A_i$,
$g^{ij}$, $T_0$ and $\mu_0$.

\subsubsection{Dependence of the partition function on $T_0$ and
$\mu_0$} \label{tm}

From the viewpoint of a Euclidean path integral, the parameter $T_0$ in
the partition function \eqref{pftrdn} is the coordinate length of the time
circle. Moreover, every quantum field of charge $q$ is twisted by the phase 
$q \frac{\mu_0}{T_0}$ as it winds the temporal circle in Euclidean space. 
As usual, such a twist is gauge equivalent to a shift in the '
temporal gauge field ${\cal A}_0 \rightarrow {\cal A}_0+\mu_0 =A_0$ \footnote{In this formula
${\cal A}_0$ is refers to the gauge field in Lorentzian space. Note that $\mu_0$ is gauge equivalent
to an imaginary shift of ${\cal A}_0$ in Euclidean space.}  holding ${\cal A}_i$ fixed.
It follows that $\ln Z$ is a function of ${\cal A}_0$, ${\cal A}_i$ and $\mu_0$ only in the combination
$ A_0$ and $ A_i$ . The dependence of $\ln Z$ on $T_0$
may be deduced in a similar fashion. The Euclidean time coordinate $t'=t T_0$ has unit
periodicity. When rewritten in terms of $t'$, the metric and gauge field retain the
form \eqref{metgfa} with
$$e^{\sigma'}=\frac{e^{\sigma}}{T_0}, ~~~a_i'=a_i T_0, ~~~A_0'=\frac{A_0}{T_0}$$
It follows from all these considerations that
\begin{equation}\label{pft}
 W(e^\sigma,{\cal A}_0, a_i, {\cal A}_i, g^{ij}, T_0, \mu_0)
= {\cal W}(\frac{e^\sigma}{T_0},\frac{A_0}{T_0}, T_0 a_i, A_i, g^{ij}).
\end{equation}
We will never use the function ${\cal W}$ below; all our explicit formulae will be written in
terms of the function $W$. Nonetheless \eqref{pft} will allow us to relate thermodynamical derivatives
w.r.t. $T_0$ and $\mu_0$ to functional derivatives of the partition function w.r.t.
background fields.

\subsubsection{Conserved charges and entropy} \label{cce}

In this subsubsection we will compute the $U(1)$ charge and energy of our system
from integrals over the appropriate charge currents, and compare the expressions so
obtained with thermodynamical formulas.

The $U(1)$ charge of our system in equilibrium is given by
\begin{equation}\label{uic}
Q =\int d^{p}x \sqrt{-g_{p+1}} J^0
\end{equation}
where the integral is taken over the $p$ dimensional spatial manifold. Let us now
define the (conserved) energy of our system. Whenever the divergence of the stress
tensor vanishes, the current $-v^\lambda T_\lambda^\mu$ is conserved provided $v^\lambda$
is a killing vector field.  We cannot directly apply this result to the killing vector
field $v^\lambda=(1, \ldots,  0)$, as the stress tensor is our paper is not divergence free
in general (see \eqref{EOMNE}). However it is easily verified that the shifted current
\begin{equation}\label{scc}
J_E^\mu = -T_0^\mu - {\cal A}_0 J^\mu 
\end{equation}
is conserved in equilibrium. As a consequence we define
\begin{equation}\label{uie}
E =\int d^{p}x \sqrt{-g_{p+1}} J_E^0= \int d^{p}x \sqrt{-g_{p+1}}
\left( -T_0^0 -{\cal A}_0 J^0 \right)
\end{equation}
$Q$ and $E$ defined in \eqref{uic} and \eqref{uie} may be shown to be Kaluza Klein
gauge invariant. For instance, the Kaluza Klein gauge variation of the RHS of \eqref{uic} is
given by
\begin{equation}\label{conscur} \begin{split}
&\int d^p x \sqrt{-g_{p+1}} J^i \partial_i \phi \\
&=  \int d^p x \sqrt{-g_{p+1}} J^\mu \partial_\mu \phi   \\
&=- \int d^p x \sqrt{-g_{p+1}}~ \phi  \nabla_\mu J^\mu=0
\end{split}
\end{equation}
(where we have used the fact that the gauge parameter $\phi$ is independent of $t$,
integrated by parts, and used the fact that $J^\mu$ is a conserved current). The gauge invariance
of $E$ follows from an almost identical argument.

We will now demonstrate that the expressions \eqref{uic} and \eqref{uie} agree exactly with
the thermodynamical definitions of the charge and energy that follow from the partition function.
In great generality, the charge of any thermodynamical system may be obtained from its partition function
\eqref{pftrdn} via the thermodynamical formula
$$Q= T_0\frac{ \partial W }{\partial \mu_0}$$
where the partial derivative is taken at constant $T_0, {\cal A}_0, {\cal A}_i, g^{ij}, a_i, \sigma$.
In the current context
\begin{equation}\label{uict} \begin{split}
T_0\frac{ \partial W }{\partial \mu_0} 
&=T_0 \int d^{p}x
\left( \frac{ \delta {\cal W}}{\delta A_0(x)} - a_i \frac{ \delta {\cal W} }{\delta A_i(x)} \right) \\
=& \int d^{p}x \sqrt{-g_{p+1}} J^0=Q\\
\end{split}
\end{equation}
where we have used \eqref{pft}, $J^0=-e^{- 2 \sigma} J_0 -a_i J^i$
(this follows from the fact that $J_0=g_{00} J^0 + g_{0i} J^i$) and explicit expressions
for $J_0$ and $J^i$ listed in \eqref{stcurrent}. Let us note that, in the presence of anomaly \ref{EOMNE}
 current $J^{\mu}$ 
is neither gauge invariant nor conserved\footnote{One can construct a conserved current which is given by
$$\hat J^{\mu} = J^\mu + \frac{C}{12}\epsilon^{\mu\nu\rho\sigma}{\cal A}_\nu F_{\rho\sigma}.$$}.

The thermodynamical energy
$$ T_0^2 \frac{ \partial W }{\partial T_0} +  \mu_0 Q $$ 
(where the partial derivative is taken at constant $\mu_0, {\cal A}_0, {\cal A}_i, g^{ij}, a_i, \sigma$).
may be
processed, in the current context, as
\begin{equation}\label{uiet} \begin{split} 
& T_0^2 \frac{ \partial W }{\partial T_0} +  \mu_0 Q= \\
&= T_0 \int 
\bigg[ - \frac{\delta {\cal W} }{\delta \sigma} + a_i \frac{\delta {\cal W}}{\delta a_i}
-\frac{\delta {\cal W}}{\delta A_0}A_0 
 + \frac{\delta {\cal W}}{\delta A_0}\mu_0 -\mu_{0} a_{i}\frac{\delta {\cal W}}{\delta A_i} \bigg]\\
&=\int \sqrt{-g_{p+1}} \left[ \left( e^{-2 \sigma} T_{00}+ a_i T_0^i \right) - {\cal A}_0 J^0 \right]\\
&=\int \sqrt{-g_{p+1}} \left[- T_0^0- {\cal A}_0 J^0 \right]\\
&=E\\
\end{split}
\end{equation}
where we have used \eqref{pft}, the fact that
$- T_0^0=e^{-2 \sigma} T_{00}+ a_i T_0^i$
and the explicit expressions for $T_{00}$ and $T_0^i$ in \eqref{stcurrent}).
In summary
\begin{equation}\label{euiem} \begin{split}
E&=T_0^2 \frac{ \partial W }{\partial T_0} +  \mu_0  Q\\
Q&=T_0\frac{ \partial W }{\partial \mu_0}\\
\end{split}
\end{equation}
Even in the presence of anomaly one can show that the current $J_E^{\mu}$  in \ref{scc} 
remains conserved, where $J^{\mu}$ is defined as in \ref{stj}. Thus, the thermodynamic formula \ref{uiet}
holds for anomalous system as well.

 We conclude that the conserved charge and energy in our system are given, in terms of the
partition function, by the usual thermodynamical formulae. It follows that the
entropy of our system should also be given by the standard statistical formula
\begin{equation}\label{entexp}
S=\frac{\partial(T_0 W) }{\partial T_0}
\end{equation}
Later in this paper we obtain constraints on the entropy current of our system by equating
\eqref{entexp} with $\int d^{p}x \sqrt{-g_{p+1}} J_S^0$.

\subsection{Consistent and Covariant Anomalies}\footnote{
We would like to thank S. Trivedi and S. Wadia for discussions and on this 
topic and S. Wadia for referring us to \cite{Bardeen:1984pm}.}\label{cca} 
In this section we discuss the relationship between the consistent charge current (the current obtained by 
differentiating the partition function w.r.t. the background gauge field) and the gauge invariant charge 
current in arbitrary $3+1$ dimensional $U(1)$ gauge theories with a $U(1)^3$ anomaly. 
Readers who are familiar with the issue of consistent and covariant anomalies in quantum field theories can 
skip this section. The equations which will be used later are \eqref{tj},\eqref{consanom},\eqref{consid}.

In this paper we will have occasion to study field theories in 4 spacetime dimensions
whose $U(1)$ current obeys the anomalous conservation
\begin{equation} \label{consanom}
\nabla_\mu J^\mu = - \frac{C}{24} *({\cal F} \wedge {\cal F})
\end{equation}
$J^\mu$ in \eqref{consanom} is the so called `consistent' current  defined by
$J^\mu= \frac{\delta W}{\delta {\cal A}_\mu}$.
As all gauge fields in this paper are always time independent
\begin{equation} \label{bbb}
*({\cal F} \wedge {\cal F})=-8 e^{-\sigma} \epsilon^{ijk} \partial_i A_0 {\partial_j} {\cal A}_k
\end{equation}
(here $\epsilon^{123}= \frac{1}{\sqrt{g_3}}$) so that the anomaly equation may be rewritten as
\begin{equation} \label{consanoma}
\nabla_\mu J^\mu = \frac{C}{3} e^{-\sigma} \epsilon^{ijk} \partial_i A_0 {\partial_j} {\cal A}_k
\end{equation}
\footnote{In order to forestall all possible confusion we list our conventions.
${\cal F}_{\mu \nu}= \partial_\mu {\cal A}_{\nu}-\partial_\nu {\cal A}_\mu$~, ~~
$*({\cal F} \wedge {\cal F})= \epsilon^{\mu\nu\alpha \beta} {\cal F}_{\mu \nu} {\cal F}_{\alpha \beta}$ where
$\epsilon^{0123}=\frac{1}{\sqrt{-g_4}}$. The variation of the gauge field
${\cal A}_\mu$ under a gauge transformation is given by
$\delta {\cal A}_\mu=\partial_\mu \phi$.}
It follows that the variation of the action under a gauge transformation is given by
\begin{equation}\label{vagt}
\delta S= \int \sqrt{-g_4} \frac{\delta S}{\delta {\cal A}_\mu} \partial_\mu \phi
= \frac{C}{24} \int d^4 x \sqrt{-g_4} \phi *({\cal F} \wedge {\cal F})
=-\frac{C}{3} \int d^4 x \sqrt{g_3} \phi \epsilon^{ijk} \partial_i A_0 {\partial_j} {\cal A}_k
\end{equation}
We now follow the discussion of Bardeen and Zumino \cite{Bardeen:1984pm} to determine the gauge transformation
property of $J^\mu$. The principle that determines this transformation law is simply that the
result of first performing an arbitrary variation of the gauge field
${\cal A}_\mu \rightarrow \delta{\cal A}_\mu$ and then a
gauge transformation generated by $ \delta \phi$ must be the same as that obtained upon
reversing the order of these operations.  The variation of the action under the first
order of operations, to quadratic order in variations, is given by
$$\int \sqrt{-g_4} \delta J^\mu (\delta {\cal A}_\mu) $$
(where $\delta J^\mu$ denotes the variation of the consistent current $J^\mu$ under the gauge
transformation $\delta \phi$). The reverse order gives
$$ \frac{C}{24} \int \delta \phi  \frac{\delta ({\cal F} \wedge {\cal F}) }{\delta
{\cal A}_\mu}  \delta {\cal A}_\mu
=\frac{C}{24} \int \delta \phi  \frac{\delta ({\cal F} \wedge {\cal F}) }{\delta
{\cal A}_\mu}  \delta {\cal A}_\mu
= \frac{C}{6} \int \sqrt{-g_4}  \delta {\cal  A}_\alpha \epsilon^{\alpha \beta \gamma \delta}
\partial_\beta \phi  {\cal F}_{\gamma \delta}$$
Comparing the two expressions it follows that under a gauge transformation
\begin{equation}\label{jugt}
\delta J^\alpha = \frac{C}{6} \epsilon^{\alpha \beta \gamma \delta} \partial_\beta \phi
{\cal F}_{\gamma \delta}
\end{equation}
It follows that the shifted current
\begin{equation} \label{tj}
{\tilde J}^\mu = J^\mu -\frac{C}{6}
\epsilon^{\mu \nu \gamma \delta} {\cal A}_\nu {\cal F}_{\gamma \delta}
\end{equation}
is gauge invariant. ${\tilde J}^\mu$ is the current that is most familiar to most field
theorists; for instance it is the current whose divergence is computed by the usual
triangle diagram in standard text books. It follows from \eqref{tj} that the divergence of
${\tilde J}^\mu$ is given by
\begin{equation} \label{consanomc}
\nabla_\mu {\tilde J}^\mu = - \frac{C}{8} *({\cal F} \wedge {\cal F})
\end{equation}

Using \eqref{bbb},  the anomaly equations may be rewritten as
\begin{equation} \label{consanomb} \begin{split}
\nabla_\mu J^\mu& = \frac{C}{3} e^{-\sigma} \epsilon^{ijk} \partial_i A_0 {\partial_j} {\cal A}_k\\
\nabla_\mu {\tilde J}^\mu& = C e^{-\sigma} \epsilon^{ijk} \partial_i A_0 {\partial_j} {\cal A}_k\\
\end{split}
\end{equation}

Let us summarize. ${\tilde J}^\mu$ is the gauge invariant current that we will use in the
fluid dynamical analysis in our paper. It obeys the anomalous conservation equation
\eqref{consanomc}. On the other hand the non gauge invariant current
$J^\mu$ is simply related to the action $W$ (it is the functional derivative of
$W$ w.r.t. ${\cal A}_\mu$). These two currents are related by \eqref{tj}.

To end this subsection we will now derive the stress tensor conservation equation
\eqref{EOMNE} in the presence of a potential anomalous background gauge field. We start by
noting that the variation of $W$ under an arbitrary variation of $g^{\mu\nu}$ and ${\cal A}_\mu$ is
given by
\begin{equation}\label{gv}
\delta W= \int \sqrt{-g_4} \left( - \frac{1}{2} \delta g^{\mu\nu} T_{\mu\nu}
+ J^\mu \delta {\cal A}_\mu \right)
\end{equation}
Let us now choose the variations of the metric and gauge fields to be of the form generated
by an infinitesimal coordinate transformation, i.e.
$$ \delta g^{\mu\nu}= \nabla^\mu \epsilon^\nu + \nabla^\nu \epsilon^\mu, ~~~ \delta {\cal A}_\mu
=- \left( \nabla_\mu \epsilon^\nu \cal A_{\nu} + \epsilon^\nu \nabla_\nu {\cal A}_\mu \right)$$
General coordinate invariance (which we assume to be non anomalous)
demands that $\delta W=0$ in this special case. Plugging these variations into \eqref{gv},
setting the LHS to zero and integrating by parts yields
\begin{equation} \label{mi}
\int d^4 x \sqrt{-g_4} \epsilon^\nu \bigg( \nabla^\mu T_{\mu\nu} - J^\mu
\left( \nabla_\nu {\cal A}_\mu- \nabla_\mu {\cal A}_\nu \right) + \nabla_\mu J^\mu  {\cal A}_\nu
\bigg)
\end{equation}
Using \eqref{consanom} together with the identity
\begin{equation}\label{ident}
{\cal A}_\sigma \epsilon^{\mu\nu \alpha \beta} {\cal F}_{\mu \nu} {\cal F}_{\alpha \beta}= -4 \epsilon^{\mu \nu \alpha \beta} {\cal A}_\nu {\cal F}_{\alpha \beta}
{\cal F}_{\sigma \mu}
\end{equation}
we conclude that
\begin{equation} \label{consid}
\nabla_\mu T^{\mu}_{\nu} = {\cal F}_{\nu \mu} (J^\mu -\frac{C}{6}\epsilon^{\mu\sigma\alpha\beta} {\cal A}_\sigma {\cal F}_{\alpha\beta})= {\cal F}_{\nu \mu} {\tilde J}^\mu
\end{equation}
Thus the two equation of motion of charged fluids are given by \eqref{consanom},\eqref{consid}.

\subsection{Perfect fluid hydrodynamics from the zero derivative partition function} \label{pfh}

It is well known (and obvious on physical grounds) that the equations of perfect fluid dynamics
are completely determined by the equation of state of the fluid (i.e, for instance, the
pressure as a function of temperature and velocity).

In this section we will `rederive' the fact that the equations of hydrodynamics, at zero
derivative order, are determined in terms of a single function of two variables, by comparison
with the equilibrium partition function on a general background of the form \eqref{metgf}.
The the results we obtain in this subsection are obvious on physical grounds.
However this subsection illustrates the basic
idea behind the work out in subsequent sections.

At zero order in the derivative expansion, the most general symmetry allowed
constitutive relations of fluid dynamics take
the form
\begin{equation}\label{perfectfluidCR}
T^{\mu\nu}= (\epsilon + {\cal P} ) u^{\mu}u^{\nu}+ {\cal P} g^{\mu \nu}, \quad J^{\mu}= q u^{\mu},
\end{equation}
At this stage $\epsilon$, ${\cal P}$ and $q$ are arbitrary functions of any two thermodynamical
fluid variables. $\epsilon$, ${\cal P}$ and $q$ (which will, of course,  eventually turn out to be the fluid
energy density, pressure and charge density) are as yet independent and arbitrary functions of the temperature and velocity.

We will now show that $\epsilon$, ${\cal P}$ and $q$ cannot be independent functions, but
are all determined in terms of a single `master' function of two variables. In order to do
that we note that the most general $p$ dimensional gauge and
diffeomorphism invariant partition function for our system on \eqref{metgf}
must take the form
\begin{equation}\label{perfectfluidA}
W = \ln Z = \int d^3 x \sqrt{g_3}\,\  \frac{e^{\sigma}}{T_0}
P \left(T_0 e^{-\sigma}, e^{-\sigma} A_0 \right)
\end{equation}
for some function of two variables $P$ (it is convenient to regard $P$ as a function of
$e^{-\sigma}$ and $e^{-\sigma}A_0$ rather than simply $\sigma$ and $A_0$ as we will see below).
The stress tensor and charge current that follows from the partition function \eqref{perfectfluidA}
are easily evaluated using \eqref{stcurrent}. The results are most simply written once we
introduce some notation. Let
$$a= e^{-\sigma} T_0, ~~~b=e^{-\sigma}A_0$$
Let $P_a$ denote the partial derivative of $P$ w.r.t its first argument, and $P_b$ the
partial derivative of $P$ w.r.t. its second argument. Below, unless
otherwise specified, the functions $P$, $P_a$ and $P_b$ will always evaluated at $(a,b)$, and
we will notationally omit the dependence of these functions on their arguments.
In terms of this notation
\begin{eqnarray} \label{stc}
T^{ij}&=& P g^{ij}, \quad T_{00}= e^{2 \sigma} \left(P - a P_a -b P_b \right) , \quad
\quad J^0 = e^{-\sigma} P_b
\\
T_0^i&=& 0, \quad J^i=0,
\end{eqnarray}
Comparing the expression for $J^i$ in  \eqref{perfectfluidCR} with the same quantity in
\eqref{stc} we conclude that
$$u^\mu= e^{-\sigma} (1, 0, \ldots, 0)$$
Comparing the other quantities it follows that
\begin{equation}\label{jj}
{\cal P}=P, ~~~\epsilon= -P +aP_a +b P_b, ~~~q = P_b
\end{equation}
In the special case of flat space the variables $a$ and $b$ reduce to the
temperature and chemical potential. It is clear on physical grounds that
${\cal P}$, $\epsilon$ and $q$ are functions only of local values of thermodynamical variables.
Consistency requires us to identify the local value of the temperature with $a$ and the local
chemical potential with $b$. The function $P$ that appears in the partition is simply
the pressure as a function of $T$ and $\mu$. Standard thermodynamical identities
then allow us to identify $\epsilon$ with the energy density of the fluid and
$q$ with the the charge density of the fluid.

Let us summarize the net upshot of this analysis. Symmetries determine the form of the
perfect fluid constitutive relations upto three undetermined functions
$\epsilon$, ${\cal P}$ and $q$, of the temperature and
chemical potential. On the other hand the equilibrium partition function is given by a
single unknown function, $P$,  of two variables. Comparison of the
partition function with the fluid hydrodynamics allow us to
determine ${\cal P}$, $\epsilon$ and $q$ in terms of $P$; as a bonus we also find
expressions for the temperature and chemical potential in equilibrium on an arbitrary
background of the form \eqref{metgf}, \eqref{gf}.

As the results of this subsection are obvious, and very well known. However a
similar procedure leads non obvious
constraints for higher derivative corrections of the fluid constitutive relations, as
we now explain.

\section{3 + 1 dimensional Charged fluid dynamics at first order in the derivative expansion}\label{Sec:4dcf}

In this section we will derive the constraints imposed on the equations of charged fluid dynamics,
at first order in the derivative expansion, by comparison with the most general equilibrium
partition function.

The final results of this section agree with the slight generalization of
Son and Surowka \cite{Son:2009tf} presented in \cite{Neiman:2010zi},\cite{Bhattacharya:2011tra} as we now explain.

Recall that \cite{Son:2009tf} argued that the hydrodynamic
charge currents in field theories with a $U(1)^3$ anomaly must contain a term
proportional to the vorticity and another term proportional to the background
magnetic field. \cite{Son:2009tf} used the principle of entropy increase to
find a set of differential equations that constrain these coefficients, and
determined one solution to these differential equations. It was later demonstrated
that the most general solution to these differential equations is a two parameter
generalization of the Son Surowka result \cite{Neiman:2010zi},\cite{Bhattacharya:2011tra}. The further requirement
of CPT invariance disallows one of these two additional coefficients.

As we describe in detail below, our method for determining the hydrodynamical expansion
starts with the action \eqref{cfacnu}, and then proceeds to determine the coefficients terms in
the charge current proportional to vorticity and the magnetic field in a purely algebraic manner.
Nowhere in this procedure do we solve a differential equation, so our procedure generates no integration
constants. However the starting point of our procedure, the partition function
\eqref{cfacnu} itself,  depends on the three constants $C_0$, $C_1$ and $C_2$.
As we demonstrate below, $C_1$ and $C_2$ map to the integration constants obtained
from the differential equations of \cite{Son:2009tf}. The third constant $C_0$ is new, and
does not arise from the analysis of \cite{Son:2009tf}. As we explain below, this coefficient
corresponds to the freedom of adding a $U(1)$ gauge non invariant term to the entropy current,
subject to the physical requirement that the contribution to entropy production from this
term is gauge invariant. It turns out, however, that the requirement of CPT invariance forces
$C_0$ to vanish. As a consequence this new term cannot arise in the hydrodynamical expansion
of any system that obeys the CPT theorem.

\subsection{Equilibrium from Hydrodynamics}

\begin{table}
\centering
\begin{tabular}[h]{|c|c|c|}
\hline
Type & Data& Evaluated at equilibrium \\
 & & $T=T_0e^{-\sigma}, ~ \mu=e^{-\sigma} A_0, ~ u^\mu= u^\mu_K$\\
\hline
 Scalars & $\nabla.u$  & 0 \\
\hline
 Vectors & $E_\mu = F_{\mu\nu}u^{\nu}$,  & $e^{-\sigma} \partial_i A_0$ \\
         & ${\cal P}^{\mu\alpha} \partial_\alpha T ,$   & -$ T_{0}e^{-\sigma}\partial^{i}\sigma$ \\
         &$\left(E^\mu -T {\cal P}^{\mu\alpha} \partial_\alpha \nu  \right)$     & $0$\\
\hline
Pseudo-Vectors & $\epsilon_{\rho \lambda \alpha \beta} u^\lambda \nabla^\alpha u^\beta$ & $ \frac{e^{\sigma}}{2}\epsilon_{ijk}f^{jk}$\\
               &$B_{\mu} =\half\epsilon_{\rho \lambda \alpha \beta} u^\lambda F^{\alpha\beta}$ & $B_{i} = \half g_{ij}\epsilon^{jkl}(F_{kl} + A_0 f_{kl}) $\\
\hline
Tensors & ${\cal P}_{\mu\alpha} {\cal P}_{\nu\beta}\big(\frac{\nabla^{\alpha} u^{\beta}+\nabla^{\beta} u^{\alpha}}{2}-\frac{\nabla.u}{3}g^{\alpha\beta}\big)$ & 0 \\
\hline
\end{tabular}
\caption{One derivative fluid data}
\label{odfd}
\end{table}

\begin{table}
\centering
\begin{tabular}[h]{|c|c|}
\hline
 Scalars & None  \\
\hline
 Vectors & $\partial^i A_0$ ,  $\partial^i \sigma$ \\
\hline
Pseudo-Vectors & $\epsilon^{ijk} \partial_j A_k~,~~ \epsilon^{ijk} \partial_j a_k$ \\
\hline
Tensors & None \\
\hline
\end{tabular}
\caption{One derivative background data}
\label{odbd}
\end{table}

In Table (\ref{odfd}) we have listed all scalar, vector and tensor expressions
that one can form out of fluid fields and background metric and gauge fields (not necessarily
in equilibrium) at first order in the derivative expansion. It follows from the listing
of this table that the most general symmetry allowed one derivative expansion of the
constitutive relations is given by
\begin{equation} \label{cr}\begin{split}
\pi^{\mu\nu}&= -\zeta \theta {\cal P}_{\mu\nu} - \eta \sigma_{\mu\nu} \\
J^\mu_{diss}&= \sigma
\left(E_\mu -T {\cal P}_\mu^\alpha \partial_\alpha \nu  \right)
+ \alpha_1  E^\mu + \alpha_2 {\cal P}^{\mu\alpha} \partial_\alpha T
+ \xi_\omega \omega^\mu + \xi_B B^\mu
\end{split}
\end{equation}
where the shear viscosity $\eta$, bulk viscosity $\zeta$, conductivity $\sigma$ and 
the remaining possible transport coefficients  $b_1$, $b_2$, $b_3$ and $b_4$ are arbitrary functions of $\sigma$ and $A_0$.

We are interested in the stationary equilibrium solutions of these equations.
In general, every fluid variable can receive derivative corrections in terms of derivatives of the back ground data. The equilibrium temperature, chemical potential
and velocity of our system to first order is given by,
\begin{eqnarray}
 T = T_{(0)} + \delta T = T_0 e^{-\sigma} + \delta T, && \mu= \mu_{(0)} + \delta \mu = e^{-\sigma}A_0 + \delta \mu, \nn \\
 u^\mu = u^\mu_{(0)} + \delta u^\mu  &=& e^{-\sigma}(1,0,0,0) + \delta u^\mu , \nonumber 
\end{eqnarray}
$\delta u^0$ is determined in terms of $\delta u^i$ (which we would specify in a moment) as follows.
Since both $u^\mu$ and $u^\mu_{(0)}$ is normalized to $(-1)$, we have
\begin{equation}\label{norfixing}
{u_{(0)}}_\mu\delta u^\mu = 0~~\Rightarrow ~~\delta u^0 =-a_i \delta u^i.
\end{equation}
Thus, the nontrivial part of velocity correction $\delta u^\mu$ is encoded in $\delta u^i$.

Solutions in equilibrium are determined entirely by the background fields
$\sigma$, $A_0$, $a_i$, $A_i$ and $g^{ij}$. In Table(\ref{odbd},\ref{odfd}) we have listed
all coordinate and gauge invariant
one derivative scalars, vectors and tensors constructed out of this background
data. As Table (\ref{odbd},\ref{odfd}) lists no one derivative scalars, it follows immediately that
the equilibrium temperature field $T(x)=e^{-\sigma} T_0$ and chemical potential field
$\mu(x)=e^{-\sigma} A_0$ receive no corrections at first order in the derivative
expansion. The velocity field in equilibrium can, however, be corrected. The most
general correction to first order is proportional to the vectors and pseudo vectors
listed in Table (\ref{odbd},\ref{odfd}) and is given by
\begin{equation}\begin{split} \label{vc}
\delta u^i =
-\frac{e^{-\sigma} b_1}{4} \epsilon^{ijk}f_{jk} + b_2 B^i_K+b_{3} \partial^i \sigma + b_4 \partial^i A_0
\end{split}
\end{equation}
where
\begin{equation}\label{not} \begin{split}
f_{jk}& = \partial_j a_k -\partial_k a_j\\
F_{jk}&= \partial_j A_k-\partial_k A_j\\
A_j &={\cal A}_j-a_j A_0\\
B^{i}_K &= \frac{1}{2}\epsilon^{ijk}(F_{jk}+ A_{0} f_{jk})\\
\epsilon^{123}&=\frac{1}{\sqrt{g_3}}\\
\end{split}
\end{equation}
The fluid stress tensor evaluated on this equilibrium configuration evaluates to \eqref{stc}
corrected by an expression of first order in the derivative expansion. The one derivative
corrections have two sources.

The first set of corrections arises from the corrections \eqref{cr} evaluated on the zero order
equilibrium fluid configuration \eqref{pfeq}. \footnote{When
$u^\mu \propto (1, 0 \ldots, 0)$ the Landau frame condition employed in this paper sets
$\pi_{00}=\pi_{0i}=J^{diss}_0=0$. Consequently $T_{00}$, $T_{0i}$ and $J_0$ receive no one
derivative corrections of this sort.} Using Table(\ref{odbd}), we then conclude that
the change
in the stress tensors and charge current due to the modified constitutive relations is given by
\begin{equation}\label{ccm} \begin{split}
\delta T_{00}&=\delta T_{0}^i=\delta J_0=\delta T^{ij}=0\\
\delta {\tilde J}^i &= \alpha_{1} e^{-\sigma} \partial^{i}A_{0}-\alpha_{2} T_{0}e^{-\sigma}\partial^{i}\sigma+ \frac{1}{2}(\xi_{B} A_{0}-\frac{1}{2}\xi_{\omega}e^{\sigma} )\epsilon^{ijk}f_{jk}+\frac{1}{2}\xi_{B} \epsilon^{ijk}F_{jk} \\
\end{split}
\end{equation}

The second source of corrections arises from inserting the velocity correction \eqref{vc}
into the zero order (perfect fluid) constitutive relations. At the order at which we work
these velocity corrections do not modify $T_{00}$, $J_0$ or $T^{ij}$. A short calculation shows
that the modification of the stress tensor and charge corrections due to these corrections
takes the form
\begin{equation}\label{ccvm} \begin{split}
\delta T_{00}&=\delta J_0=\delta T^{ij}=0\\
\delta T_{0}^i &=-e^{\sigma}(\epsilon+P)\Big[\frac{1}{2}(b_{2} A_{0}-\frac{1}{2}b_{1}e^{\sigma} )\epsilon^{ijk}f_{jk}+\frac{1}{2}b_{2} \epsilon^{ijk}F_{jk}-b_{3}T_{0}e^{-\sigma}\partial^{i}\sigma+b_{4} \partial^{i}A_{0}\Big] \\
\delta {\tilde J}^i &= \Big[\frac{1}{2}\Big( q b_{2} A_{0}-\frac{1}{2}q b_{1}e^{\sigma}\Big )\epsilon^{ijk}f_{jk}+\frac{1}{2}q b_{2} \epsilon^{ijk}F_{jk}\\
&-q b_{3}T_{0}e^{-\sigma}\partial^{i}\sigma+q b_{4} \partial^{i}A_{0}\Big] \\
\end{split}
\end{equation}
The net change in $T_0^i$ and $J^i$ is given by summing \eqref{ccvm} and \eqref{ccm} and is given by
\begin{equation}\label{ccvmf} \begin{split}
\delta T_{0}^i &=-e^{\sigma}(\epsilon+P)\Big[\frac{1}{2}(b_{2} A_{0}-\frac{1}{2}b_{1}e^{\sigma} )\epsilon^{ijk}f_{jk}+\frac{1}{2}b_{2} \epsilon^{ijk}F_{jk}-b_{3}T_{0}e^{-\sigma}\partial^{i}\sigma+b_{4} \partial^{i}A_{0}\Big] \\
\delta {\tilde J}^i &= \Big[\frac{1}{2}\Big((\xi_{B}+ q b_{2}) A_{0}-\frac{1}{2}(\xi_{\omega}+q b_{1})e^{\sigma}\Big )\epsilon^{ijk}f_{jk}+\frac{1}{2}(\xi_{B}+q b_{2}) \epsilon^{ijk}F_{jk}\\
&-(q b_{3}+\alpha_{2})T_{0}e^{-\sigma}\partial^{i}\sigma+(q b_{4}+\alpha_1) \partial^{i}A_{0}\Big] .\\
\end{split}
\end{equation}

\subsection{Equilibrium from the Partition Function}

We now turn to the study of the first correction to the perfect fluid
equilibrium partition function \eqref{perfectfluidA} at first order in the
derivative expansion. From the fact that Table (\ref{odbd},\ref{odfd}) lists no gauge invariant
scalars, one might be tempted to conclude that the equilibrium partition function can
have no gauge invariant one derivative corrections. We have already explained in the
introduction that this is not the case; the three (constant) parameter set of Chern Simons
terms listed in the third line of \eqref{cfacn} yield perfectly local and gauge invariant
contributions to the partition function, even though they cannot be written
as integrals of local gauge invariant expressions. In addition to these gauge invariant
pieces we need a term in the action that results in its anomalous gauge transformation property
\eqref{vagt}. This requirement is precisely met by the term in the last line of \eqref{cfacn}.
\footnote{In order to see this we first note that the last line of \eqref{cfacn} may be rewritten
as
$$\frac{C}{3} \int d^3x \sqrt{g_3} A_0 \epsilon^{ijk} {\cal A}_i \partial_j{\cal A}_{k} $$
The variation of this term under a gauge transformation is given by
\begin{equation}\label{vtgt}
-\frac{C}{3} \int d^3x \sqrt{g_3} \epsilon^{ijk}\phi\partial_i A_0  \partial_j{A}_{k}
\end{equation}
in perfect agreement with \eqref{vagt}. }

With the action \eqref{cfacn} in hand it is straightforward to use
\eqref{stcurrent} to obtain the stress tensor and current corresponding to this equilibrium
solution. We find
\begin{eqnarray}\label{stcuchaction}
T_{00}&=&0, \quad T^{ij}=0, \nn\\
T_0^i&=& e^{-\sigma}\epsilon^{ijk}\bigg[(-\half C A_{0}^{2}+ 2 C_{0} A_{0}+ C_{2})\nabla_j A_k+(2 C_1-\frac{C}{6}A_{0}^{3}  - C_{2} A_{0})\nabla_j a_k\big]\nn\\
\quad J_0&=&-e^{\sigma}\epsilon^{ijk}\bigg[ \frac{C}{3} A_i \nabla_j A_k + \frac{C}{3} A_{0} A_i \nabla_j a_k\bigg] \nn \\
J^i&=&e^{-\sigma}\epsilon^{ijk}\bigg[2 \left(\frac{C}{3} A_{0}+C_{0}\right) \nabla_j A_k + \left(\frac{C}{6}A_{0}^{2}+C_{2}\right) \nabla_j a_k + \frac{C}{3}  A_k\nabla_j A_0 \bigg], \nn \\
\end{eqnarray}
Using \eqref{tj} it follows that
\begin{equation}\begin{split}\label{gicurrent}
\tilde J_0&=0,\\ \quad \tilde J^i&= e^{-\sigma} \epsilon^{ijk}\big[ ( C A_0 + 2 C_0)\nabla_j A_k + (\half C A_0^2+  C_2)\nabla_j a_k\big],
\end{split}\end{equation}

\subsection{Constraints on Hydrodynamics}
Equating the coefficients of independent terms in the two expressions for $T_0^i$
\eqref{ccvmf},\eqref{stcuchaction} determines the one derivative corrections of the
velocity field in equilibrium. We find.
\begin{eqnarray}\label{SSsolv}
b_1&=&\frac{T^3}{\epsilon+P}\big(\frac{2}{3} \nu^3 C+ 4 \nu^2  C_0 - 4 \nu  C_2 + 4 C_1 \big), \nn \\
b_2&=& \frac{T^2}{\epsilon+P}\big(\frac{1}{2}\nu^2 C+ 2 \nu  C_0- C_2\big), \nn \\
b_3&=&b_4=0.
\end{eqnarray}
where $\nu = \frac{\mu}{T}=\frac{A_{0}}{T_{0}}$.

Equating coefficients of independent terms in $J^{i}$ in equations \ref{ccvmf} and
\ref{gicurrent} and using \eqref{SSsolv} gives
\begin{eqnarray}\label{SSsolc}
\xi_{\omega}&=&C \nu^2 T^2 \big(1 - \frac{2 q}{3(\epsilon+P)}\nu T\big)+ T^2 \big[(4 \nu  C_0 - 2  C_2)- \frac{q T}{\epsilon+P}(4 \nu^2  C_0 - 4 \nu  C_2+ 4 C_1)\big], \nn \\
\xi_{B} &=& C \nu T\big(1-\frac{ q}{2(\epsilon+P)}\nu T\big)+ T \big(2  C_0 -\frac{q T}{\epsilon+P}(2 \nu  C_0 -  C_2)\big), \nn \\
\alpha_1&=&\alpha_2=0
\end{eqnarray}

Let us summarize. We have found that the hydrodynamical charge current and stress tensor are
given by
\begin{equation} \label{crn}\begin{split}
\pi^{\mu\nu}&= -\zeta \theta {\cal P}_{\mu\nu} - \eta \sigma_{\mu\nu} \\
J^\mu_{diss}&= \sigma
\left(E_\mu -T {\cal P}_\mu^\alpha \partial_\alpha \nu  \right)
+ \xi_\omega \omega^\mu + \xi_B B^\mu
\end{split}
\end{equation}
In \eqref{crn} the viscosities $\zeta$ and $\eta$ together with the conductivity $\sigma$
are all dissipative parameters. These parameters multiply expressions that vanish
in equilibrium and are completely unconstrained by the analysis of this subsection.
On the other hand $\zeta_\omega$ and $\zeta_B$ - together with $\alpha_1$ and $\alpha_2$
in \eqref{cr} -  are non dissipative parameters. They multiply
expressions that do not vanish in equilibrium. The analysis of this subsection has
demonstrated that $\alpha_1$ and $\alpha_2$ vanish and that $\zeta_\omega$ and
$\zeta_B$ are given by \eqref{crn}. The expressions \eqref{crn} agree exactly with
the results of Son and Surowka - based on the requirement of positivity of the
entropy current - upon setting $C_0=C_1=C_2=0$. Upon setting
$C_0=0$ they agree with the generalized results of \cite{Neiman:2010zi} (see also \cite{Bhattacharya:2011tra},\cite{Loganayagam:2011mu}).
We will return to the role of the additional parameter $C_0$ later in this section.

\subsection{The Entropy Current}

The entropy of our system is given by
\begin{equation}\label{fent}
 \begin{split}
  S&=\frac{\partial}{\partial T_0}(T_0 \log Z) \\
&=\int d^3x \sqrt{g_3}\epsilon^{ijk}\big[ C_0 A_i\nabla_j A_k+ 3 C_1 T_0^2 a_i \nabla_j a_k + 2  C_2 T_0 A_i \nabla_j a_k\big].
 \end{split}
\end{equation}

In this subsection we determine the constraints on the entropy current $J^\mu_S$ of our system
from the requirement that \eqref{fent} agree with the local integral
\begin{equation}\label{entn}
 S=\int d^3 x \sqrt{-g_4} J^0_S
\end{equation}

Notice that the first term in \eqref{fent} (the term proportional
to $C_0$) cannot be written as the integral of a $U(1)$ gauge invariant entropy density.
It follows immediately that \eqref{entn} and \eqref{fent} cannot agree unless $J^\mu_S$ has
a non gauge invariant term proportional to $C_0$. Is it permissible for the entropy current
of a system to be non gauge invariant (and therefore ambiguous)? Entropy in equilibrium is
physical and should be well defined. Moreover, if we start a system in equilibrium, kick the
system (by turning on time dependent background metric and gauge fields) and let it settle back
into equilibrium, then the difference between the entropy of the initial and final state, is
also unambiguous. It follows that the entropy production (i.e. divergence of the entropy current)
as well as \eqref{entn} are necessarily gauge invariant. However these requirements leaves room
for the entropy current itself to be gauge dependent.

Over the next few paragraphs we find it useful to dualize the
entropy current to a 3 form.  The addition of an exact form to the entropy three form
contributes neither to entropy production nor to the total integrated value of the
entropy in equilibrium. For this reason we regard any two entropy 3-forms that differ by
an exact three form as equivalent. With this understanding, the unique non gauge
invariant entropy 3 form whose exterior derivative (the Hodge dual of entropy production) is gauge
invariant is given by $${\cal A}\wedge d {\cal A}$$
The requirement that the exterior derivative of this
3 form to be gauge invariant forces its coefficient to be constant.\footnote{Naively, another 
candidate for a non gauge invariant contribution to the entropy three form is given by
\begin{equation}\label{ttt}
{\cal A}  \wedge d \left( h(T, \mu) U \right)
\end{equation}
where $U = u_\mu dx^\mu$ and $h$ is an arbitrary function of temperature and chemical potential.
But this term  can be rewritten as follows.
\begin{equation*}
\begin{split}
{\cal A}  \wedge d \left( h(T, \mu)~ U \right)= d\left(h(T, \mu)~U\wedge{\cal A}  \right)- h(T, \mu) ~U\wedge d{\cal A}
\end{split}
\end{equation*}
 It follows that this addition is actually equivalent to a gauge invariant addition to the
entropy 3 form. }

The most general physically allowed form for the entropy current, at one derivative order,
may then be read off from Table \ref{odbd}
\begin{equation} \label{mjecn}
\begin{split}
J_S^\mu=&~s u^\mu -\nu J^\mu_{diss} + D_\theta \Theta u^\mu
 + D_c\left(E^\mu -T {\cal P}^{\mu\alpha} \partial_\alpha \nu  \right)
+ D_E E^\mu+ D_a {\mathfrak a}^\mu\\
& + D_\omega \omega^\mu + D_B B^\mu + h \epsilon^{\mu\nu\lambda\sigma}{\cal A}_\nu\partial_{\lambda}{\cal A}_\sigma\\
&\text{where $h$ is a constant}
\end{split}
\end{equation}
How is the entropy current \eqref{mjecn} constrained by the requirement that its integral
agrees with \eqref{fent}? The one derivative entropy, as computed from the formula
$\int d^3x \sqrt{g_4}J_S^0$ has two sources. First, the perfect fluid entropy current
$su^0$ has a first derivative piece that comes from the one derivative correction of the
equilibrium fluid velocity (see above). Second, from the one derivative correction to the
entropy current (evaluated on the leading order equilibrium fluid configuration). 
The terms with coefficients $D_\theta$ and $D_c$ vanish on the leading order 
equilibrium fluid configuration. Therefore these two coefficients can not be determined by
 comparing with
the total entropy as derived from action.
All the other correction
terms computed from this procedure are parity odd, except those multiplying
$D_a$ and $D_E$. It is possible to verify that the integrals of the terms multiplying
$D_a$ and $D_E$ are nonvanishing and linearly independent.  As all
first derivative entropy corrections in \eqref{fent} are parity odd, it follows immediately
that
$$D_a=D_E=0.$$
Therefore the zero component of the entropy current
at first derivative order is given by the following expression.
\begin{equation}\label{zerocompent}
 \begin{split}
  J_S^0|_{correction} &= s \delta u^0 + \left( -\nu \xi_B + D_B\right) B^0 +
\left( -\nu \xi_\omega + D_\omega\right) \omega^0
+h\epsilon^{0\nu\lambda\sigma}{\cal A}_\nu\partial_{\lambda}{\cal A}_\sigma
 \end{split}
\end{equation}

Using
\begin{equation}\label{sampark}
\begin{split}
\nu & =\frac{ A_0}{T_0}\\
&B^0  = -\epsilon^{ijk}a_i \partial_j\left(A_k +T_0\nu a_k\right)\\
&\omega^0 = \frac{e^\sigma}{2}\epsilon^{ijk} a_i \partial_j a_k\\
&\epsilon^{0\nu\lambda\sigma}{\cal A}_\nu\partial_{\lambda}{\cal A}_\sigma =
e^{-\sigma}\epsilon^{ijk}
\left[A_i\partial_j A_k + 2T_0\nu a_i \partial_j A_k + T_0^2\nu^2 a_i \partial_j a_k
+\partial_i\left(T_0\nu a_j A_k\right)\right] \\
&\delta u^0 = -a_i \delta u^i =
b_1 \left[\frac{e^\sigma}{2} \epsilon^{ijk} a_i \partial_j a_k\right]
- b_2 \left[ \epsilon^{ijk} a_i \partial_j\left(A_k +T_0\nu a_k\right)\right]
\end{split}
\end{equation}
and the expressions for $\xi_B$, $\xi_\omega$, $b_1$ and $b_2$
as computed in the previous subsection (see \eqref{SSsolv} and \eqref{SSsolc}), we find
\begin{equation}\label{totalent}
\begin{split}
&\int d^3 x \sqrt{-g_4} J^0_s|_{correction}\\
=~&\int d^3 x \sqrt{g_3}\epsilon^{ijk}\bigg[T_0^2\left(3 C_1 + h\nu^2
+\frac{d_\omega}{2} - \nu d_B\right) a_i\partial_j a_k\\
&~~~~~~~~~~~~~~~~~~~~~~~~~~~~~+ T_0 (2C_2 + 2 h \nu - d_B) a_i\partial_j A_k + h A_i\partial_j A_k\bigg]
\end{split}
\end{equation}

where
$$d_B  = \frac{D_B}{T} - \left(\frac{C\nu^2}{2} - C_2\right),~~~
 d_\omega = \frac{D_\omega}{T^2} - \left(\frac{C\nu^3}{3} - 2 C_2 \nu + 2 C_1\right)$$

Comparing this expression with \eqref{fent} we find
\begin{equation}\label{naammil}
\begin{split}
&h = C_0,~~~d_B = 2 C_0 \nu ,~~~d_\omega = 2 C_0 \nu^2
\end{split}
\end{equation}
This result agrees precisely with that of Son and Surowka as
generalized in \cite{Bhattacharya:2011eea}

\subsection{Entropy current with non-negative divergence}
In the previous subsection we have determined the entropy current by comparing with the total
entropy derived from the equilibrium partition function and we have allowed for terms which
 are not gauge invariant provided their divergence is gauge-invariant.

Now we shall try to constrain the most general entropy current (as given in \eqref{mjecn})
by demanding that its divergence is always non-negative for every possible fluid flow,
consistent with the equations of motion. The analysis will be a small modification of
\cite{Son:2009tf} because of the new gauge non-invariant term with constant coefficient
 $C_0$ added.
The steps are as follows.
\begin{itemize}
 \item First we have to compute the divergence of the current given in \eqref{mjecn}.
The new term in the entropy current contributes to the divergence in the following way.
$$\nabla_\mu \left[C_0\epsilon^{\mu\nu\alpha\beta} {\cal A}_\nu\nabla_\alpha{\cal A}_\beta\right]
 =
\frac{C_0}{4}\epsilon^{\mu\nu\alpha\beta}F_{\mu\nu} F_{\alpha\beta} = -2 C_0 E_\mu B^\mu$$
The full divergence of the entropy current is given by
\begin{equation}\label{divng1}
\begin{split}
\nabla_\mu J^\mu_S
=&~\eta \sigma_{\mu\nu}\sigma^{\mu\nu} +\zeta \Theta^2 + \sigma T Q_\mu Q^\mu 
-\xi_E (Q_\mu E^\mu) -\xi_a (Q_\mu {\mathfrak a}^\mu)\\
&+ \Theta (u.\nabla)D_\theta + ({\mathfrak a}.\nabla)D_a + (Q.\nabla)D_c + (E.\nabla)D_E\\
&+D_\theta (u.\nabla)\Theta + D_a (\nabla.{\mathfrak a}) + D_c (\nabla.Q) + D_E (\nabla.E)\\
&+\left[\frac{\partial D_B}{\partial T} -\frac{D_B}{T}\right] (B^\mu \partial_\mu T)
+ \left[\frac{\partial D_\omega}{\partial T} -2\frac{D_\omega}{T}\right]
 (\omega^\mu \partial_\mu T)\\
&+\left[\frac{\partial D_B}{\partial \nu} -C T\nu - 2 C_0 T\right] (B^\mu \partial_\mu \nu)
+ \left[\frac{\partial D_\omega}{\partial \nu} -2D_B\right]
 (\omega^\mu \partial_\mu \nu)\\
&+\left[-\xi_\omega -\frac{2 qT}{\epsilon + P} D_\omega + 2 D_B T \right] (\omega_\mu Q^\mu)\\
&+ \left[-\xi_B -\frac{qT}{\epsilon + P} D_B +C T \nu + 2C_0 T \right] (B_\mu Q^\mu)
\end{split}
\end{equation}

where
$$ Q_\mu = \partial_\mu \nu - \frac{E_\mu}{T}$$
and
$$J^\mu_{diss} =- \sigma T Q^\mu +\xi_E E^\mu + \xi_a {\mathfrak a}^\mu+ \xi_\omega \omega^\mu 
+ \xi_B B^\mu,~~\text{and}
~~ \pi^{\mu\nu} = -\eta \sigma^{\mu\nu} - \zeta \Theta P^{\mu\nu}$$

\item As explained in \cite{Bhattacharya:2011tra}, the divergence computed in \eqref{divng1}
can be non-negative if $D_\theta,~D_E,~D_c,~\xi_E$ and $\xi_a$ are set to zero 
in the parity even sector.
\item Since there is no $B^2$ or $\omega^2$ term present in \eqref{divng1}, for positivity,
 in the parity odd sector we need
all the terms that are linear in $B_\mu$ and $\omega_\mu$ to vanish. This condition imposes
the following 6 constraints.

\begin{equation}\label{6const}
 \begin{split}
  &\frac{\partial D_B}{\partial T} -\frac{D_B}{T}=0,
~~~~~\frac{\partial D_\omega}{\partial T} -\frac{2D_\omega}{T}=0\\
&\frac{\partial D_B}{\partial \nu} -C T\nu - 2 C_0 T=0,
~~~~~~\frac{\partial D_\omega}{\partial \nu} -2D_B =0\\
&-\xi_\omega -\frac{2 qT}{\epsilon + P} D_\omega + 2 D_B T=0\\
&-\xi_B -\frac{qT}{\epsilon + P} D_B +C T \nu + 2C_0 T=0
 \end{split}
\end{equation}

\item We can determine $\xi_\omega$, $\xi_B$, $D_B$ and $D_\omega$ by solving these equations.
The solution
 is identical to the solution determined from the partition function
(as given in \eqref{SSsolc} and \eqref{naammil}).
\end{itemize}

\subsection{CPT Invariance} \label{cpt}
In this subsection we explore the constraints imposed on the partition function \eqref{cfacn} by the
requirement of 4 dimensional CPT invariance. In Table \ref{cpt} we list the action of CPT on
various fields appearing in the partition function.
\begin{table}
\centering
\begin{tabular}[h]{|c|c|c|c|c|}
\hline
Field & C & P & T & CPT \\
\hline
$\sigma$ & $+$ & $+$ & $+$ & $+$ \\
\hline
$a_i$ & $+$ & $-$ & $-$ & $+$ \\
\hline
$g_{ij}$ & $+$ & $+$ & $+$ & $+$ \\
\hline
$A_0$ & $-$ & $+$ & $+$ & $-$ \\
\hline
$A_i$ & $-$ & $-$ & $-$ & $-$ \\
\hline
\end{tabular}
\caption{Action of CPT}
\label{cpt}
\end{table}
Using this table one can easily see that the terms with coefficient $C_1$ and $C_0$ change
sign under CPT transformation while the terms with coefficient $C_2$ and $C$ remains invariant.
Thus the requirement of CPT invariance of the partition function forces $C_1 = 0$ and $C_0 = 0$.
Further it also tell us that the function $P$ appearing in the perfect fluid partition function,
$W^0$, must be an even function of $A_0$ (i.e. that equilibrium does not distinguish between 
positive and negative charges).


\section{Parity odd first order charged fluid dynamics in 2+1 dimensions}\label{Sec:chargedhydro3d}

In this section we will derive the constraints imposed on the equations of 2+1 dimensional charged fluid dynamics,
at first order in the derivative expansion, by comparison with the most general equilibrium
partition function. The parity even constraints are identical to the ones found in 3+1 dimensions
(which has been extensively discussed in \S \ref{Sec:4dcf}). Therefore in this section we shall primarily focus
on the parity odd constraints which are qualitatively much different from their 3+1 dimensional counterpart.
These constraints have been obtained using a local form of the second law of thermodynamics in \cite{Jensen:2011xb}, which
we shall reproduce starting from the most general equilibrium partition function.

\subsection{Equilibrium from Hydrodynamics}

Partially borrowing some notations from equation (1.2) in \cite{Jensen:2011xb}, the the most
general symmetry allowed one derivative expansion of the constitutive relations is given by
\footnote{Note that in this constitutive relation the parity even constraint, namely the Einstein relation, have
already been taken into account.}
\begin{subequations}
\label{constrel3d}
\begin{eqnarray}
  &&   T^{\mu\nu} = \epsilon  u^\mu u^\nu
  			    +\left(P - \zeta \nabla_\alpha u^\alpha - \tilde{\chi}_B B - \tilde{\chi}_\Omega \Omega\right) P^{\mu\nu}
			    -\eta \sigma^{\mu\nu} - \tilde{\eta} \tilde{\sigma}^{\mu\nu} \,,  \\
  &&  J^\mu = \rho u^\mu + \sigma V^{\mu} + \tilde{\sigma} \tilde{V}^{\mu}
+ \tilde{\chi}_E \tilde{E}^{\mu} + \tilde{\chi}_T \tilde{T}^{\mu}  \,.
\end{eqnarray}
\end{subequations}
The various quantities appearing in the constitutive relations \eqref{constrel3d} are defined as
\begin{subequations}
\label{defs3d}
\begin{align}
\label{E:OandB}
	 & \Omega = -\epsilon^{\mu\nu\rho}u_{\mu} \nabla_{\nu} u_{\rho}, &
	& B = -\frac{1}{2} \epsilon^{\mu\nu\rho}u_{\mu} F_{\nu\rho}, \\
	& E^{\mu}  =  F^{\mu\nu}u_{\nu}, &
	& V^{\mu}  = E^{\mu} - T P^{\mu\nu}\nabla_{\nu} \frac{\mu}{T}, \\
\label{E:sigmaDef}
	& P^{\mu\nu} = u^{\mu}u^{\nu} + g^{\mu\nu}, &
	& \sigma^{\mu\nu}  = P^{\mu\alpha} P^{\nu\beta} \left(\nabla_{\alpha}u_{\beta}
+ \nabla_{\beta} u_{\alpha} - g_{\alpha\beta} \nabla_{\lambda} u^{\lambda} \right) \,,
\intertext{and}
	&\tilde{E}^{\mu}  = \epsilon^{\mu\nu\rho}u_{\nu}E_{\rho}\,,&
	&\tilde{V}^{\mu}  = \epsilon^{\mu\nu\rho}u_{\nu} V_{\rho}\,, \\
	&\tilde{\sigma}^{\mu\nu} = \frac{1}{2} \left( \epsilon^{\mu\alpha\rho} u_{\alpha} \sigma_{\rho}^{\phantom{\rho}\nu}
+  \epsilon^{\nu\alpha\rho} u_{\alpha} \sigma_{\rho}^{\phantom{\rho}\mu} \right)\,, && \tilde{T}^{\mu} = \epsilon^{\mu\nu\rho}u_{\nu} \nabla_{\rho} T .
\end{align}
\end{subequations}
The thermodynamic quantities $P$, $\epsilon$ and $\rho$ are the values of the pressure,
energy density and charge density respectively in equilibrium.
The transport coefficients $\tilde\chi_B,\ \tilde \chi_\Omega,\ \tilde\chi_E$ and
$\tilde\chi_T$ are arbitrary functions of $\sigma$ and $A_0$.
The only non-zero quantities in
equilibrium are $B,\ \omega, \tilde E^{\mu}$ and $\tilde T^{\mu}$. The rest of the first order
quantities appearing on the RHS of \eqref{constrel3d} vanish on our equilibrium configuration.
In Table \ref{T:odbd3d} we list all the parity odd diffeomorphism invariant background field data.
In Table \ref{T:odfd3d} we list the first order quantities occurring in the constitutive relations
that are non-zero in equilibrium and express them in terms of the background metric and
gauge fields\footnote{In the following, we shall use $\epsilon^{12} = \frac{1}{\sqrt{g_{2}}}$}.

\begin{table}
\centering
\begin{tabular}[h]{|c|c|}
\hline
 Pseudo-scalars & $\epsilon^{ij} \partial_i A_j$ ,  $\epsilon^{ij} \partial_i a_j$ \\
\hline
 Pseudo-vectors & $\epsilon^{ij} \partial_i A_0$ ,  $\epsilon^{ij} \partial^i \sigma$ \\
\hline
Pseudo-tensors & None \\
\hline
\end{tabular}
\caption{One derivative parity odd diffeomorphism and gauge invariant background data. 
Here $\epsilon^{ij}$ is defined so that $\epsilon^{12} = \frac{1}{\sqrt{g_{2}}}$.}
\label{T:odbd3d}
\end{table}

\begin{table}
\centering
\begin{tabular}[h]{|c|c|c|}
\hline
Type & Data & Evaluated at equilibrium\\
\hline
 Pseudo-Scalars & $B$ & $\epsilon^{ij} \partial_i A_j + A_0 \epsilon^{ij} \partial_i a_j$ \\
\hline
& $\Omega$ &  -$e^{ \sigma} \epsilon^{ij}\partial_i a_j$ \\

\hline
 Pseudo-Vectors & $\tilde{E}^{\mu}$ & $\tilde{E}_{0} = 0$, $\tilde{E}^{i} = e^{ -\sigma} \epsilon^{ij} \partial_j A_0 $ \\
                & $\tilde{T}^{\mu}$ & $\tilde{T}_{0} = 0$, $\tilde{T}^{i} = -e^{ -\sigma} \epsilon^{ij} \partial_j \sigma $ \\
\hline
 Pseudo-Tensors & none & \\
\hline
\end{tabular}
\caption{One derivative fluid data which are non-zero in equilibrium.}
\label{T:odfd3d}
\end{table}

We are interested in the stationary equilibrium solutions of the fluid equations arising from
constitutive relations \eqref{constrel3d}. Solutions in equilibrium are determined entirely by the background fields
$\sigma$, $A_0$, $a_i$, $A_i$ and $g^{ij}$. Just like in 3+1 dimensions the zeroth order solution of the
fluid fields are given by
\begin{equation}\label{pfeq3d}
 u^{\mu}_{{0}} = \{e^{-\sigma}, 0, 0 \}; ~~ T_{(0)} = T_0 e^{-\sigma};~~ \mu_{(0)} = e^{-\sigma} A_0 .
\end{equation}

The unit normalized vector in the killing direction is, $$ u^{\mu}_K= e^{-\sigma}(1,0,0)$$
In Table(\ref{T:odbd3d},\ref{T:odfd3d}) we have listed
all coordinate and gauge invariant
one derivative parity odd scalars, vectors and tensors constructed out of this background
data. Since there are 2 pseudo-scalars and 2 pseudo-vectors we can have the following
most general parity odd corrections to the fluid fields at first order
\begin{equation}\label{ffcor}
\begin{split}
 u_{\mu} &= u^{\mu}_{(0)} + \xi_E ~\tilde E^{\mu}_K + \xi_T ~\tilde T^{\mu}_K, \\
 T &= T_{(0)} + \tau_B ~B_K + \tau_\Omega ~\Omega_K, \\
 \mu &= \mu_{(0)} + m_B ~B_K + m_\Omega ~\Omega_K,
\end{split}
\end{equation}
where $\xi_E, \xi_T, \tau_B, \tau_\Omega, m_B$ and $m_\Omega$ are taken to be arbitrary
functions of $\sigma$ and $A_0$ to be determined by matching with the equilibrium
partitions function in \S \ref{sSec:match3d}. $E^{\mu}_K,T^{\mu}_K,B_K,\omega_K$ are the vectors and scalars, 
defined in equations\ref{E:OandB} and \ref{E:sigmaDef},  
velocity $u$ replaced by $u_K$.

Just like in 3+1 dimensions the fluid stress tensor evaluated on this
equilibrium configuration evaluates to \eqref{stc}
corrected by an expression of first order in the derivative expansion.
The one derivative corrections again have two sources.

The first set of corrections arises from the corrections \eqref{constrel3d} evaluated on the zero order
equilibrium fluid configuration \eqref{pfeq3d}. \footnote{When
$u^\mu \propto (1, 0 \ldots, 0)$ the Landau frame condition employed in this paper sets
$\pi_{00}=\pi_{0i}=J^{diss}_0=0$. Consequently $T_{00}$, $T_{0i}$ and $J_0$ receive no one
derivative corrections of this sort.}
The second source of corrections arises from inserting the fluid field corrections in \eqref{ffcor}
into the zero order (perfect fluid) constitutive relations. The net change in the stress tensor
and the charge current at first order is obtained by summing these two contributions
and is given by
\begin{equation}\label{TJcor3d}
 \begin{split}
  \delta T^{\mu \nu} &= \left( \frac{\partial P}{ \partial T} \tau_B + \frac{\partial P}{\partial \mu} m_B - \tilde \chi_B \right) B_{K} P_{(0)}^{\mu \nu} +
\left( \frac{\partial P}{ \partial T} \tau_\Omega + \frac{\partial P}{\partial \mu} m_\Omega - \tilde \chi_\Omega \right) \Omega_{K} P_{(0)}^{\mu \nu} \\
& + \left( \frac{\partial \epsilon}{ \partial T} \tau_B + \frac{\partial \epsilon}{\partial \mu} m_B \right) B_{K}  u_{(0)}^{\mu}u_{(0)}^{\nu} +
\left( \frac{\partial \epsilon}{ \partial T} \tau_\Omega + \frac{\partial \epsilon}{\partial \mu} m_\Omega \right) \Omega_{K} u_{(0)}^{\mu}u_{(0)}^{\nu} \\
& + (\epsilon + P) \xi_E (u_{(0)}^{\mu} \tilde E^{\nu}_{K} + u_{(0)}^{\nu} \tilde E^{\mu}_{K} ) + (\epsilon + P) \xi_T (u_{(0)}^{\mu} \tilde T^{\nu}_{K}
+ u_{(0)}^{\nu} \tilde T^{\mu}_{K}).  \\
 \delta J ^{\mu} &= \left( \frac{\partial \rho}{ \partial T} \tau_B + \frac{\partial \rho}{\partial \mu} m_B \right) B_{K}  u_{(0)}^{\mu}+
\left( \frac{\partial \rho}{ \partial T} \tau_\Omega + \frac{\partial \rho}{\partial \mu} m_\Omega \right) \Omega_{K} u_{(0)}^{\mu} \\
& + (\tilde \chi_E + \rho \xi_E) \tilde E^{\mu}_{K} + (\tilde \chi_T + \rho \xi_T) \tilde T^{\mu}_{K}
 \end{split}
\end{equation}
For future reference it will be convenient to to write down some of the components of the stress tensor and current in
 \eqref{TJcor3d} purely in terms of the background fields using the expressions listed in the third column of
Table \ref{T:odfd3d}.
\begin{equation}\label{tjcorcompo}
 \begin{split}
\delta T^{ij} &= \left( \frac{\partial P}{ \partial T} \tau_B + \frac{\partial P}{\partial \mu} m_B - \tilde \chi_B \right) B_{K} g^{ij} +
\left( \frac{\partial P}{ \partial T} \tau_\Omega + \frac{\partial P}{\partial \mu} m_\Omega - \tilde \chi_\Omega \right) \Omega_{K} g^{ij}\\
  \delta T_{00} &= e^{2 \sigma}
\left( \left( \frac{\partial \epsilon}{ \partial T} \tau_B + \frac{\partial \epsilon}{\partial \mu} m_B \right) \left( \epsilon^{ij} \partial_i A_j
+ A_0 \epsilon^{ij} \partial_i a_j \right)
- e^\sigma \left( \frac{\partial \epsilon}{ \partial T} \tau_\Omega
+ \frac{\partial \epsilon}{\partial \mu} m_\Omega \right)  \epsilon^{ij} \partial_i a_j \right), \\
  \delta T_0^i &= \left(  - (\epsilon + P ) \xi_E \epsilon^{ij} \partial_j A_0 +
  (\epsilon + P )  \xi_T  \epsilon^{ij} \partial_j \sigma \right) \\
\delta J_0 &= e^{\sigma}
\left(- \left( \frac{\partial \rho}{ \partial T} \tau_B + \frac{\partial \rho}{\partial \mu} m_B \right) \left( \epsilon^{ij} \partial_i A_j
+ A_0 \epsilon^{ij} \partial_i a_j \right)
+ e^\sigma \left( \frac{\partial \rho}{ \partial T} \tau_\Omega
+ \frac{\partial \rho}{\partial \mu} m_\Omega \right)  \epsilon^{ij} \partial_i a_j \right), \\
\delta J^i &= e^{-\sigma}\left( \left( \tilde \chi_E + \rho \xi_E \right) \epsilon^{ij} \partial_j A_0
- (\tilde \chi_T + \rho \xi_T) \epsilon^{ij} \partial_j \sigma \right).
 \end{split}
\end{equation}

\subsection{Equilibrium from the Partition Function} \label{sSec:eqlpartfn3d}

We now turn to the study of the first correction to the perfect fluid
equilibrium partition function \eqref{perfectfluidA} at first order in the
derivative expansion. From the fact that Table (\ref{T:odbd3d},\ref{T:odfd3d}) lists two gauge invariant
Hence the most general parity odd equilibrium partition function is given by
\begin{equation}\label{action3d}
 {\cal W} = \frac{1}{2}\int  \left( \alpha(\sigma, A_0) dA + T_0 \beta(\sigma, A_0)da \right),
\end{equation}
where $\alpha$ and $\beta$ are two arbitrary functions in terms of which all the 4 transport coefficients and the 6 
first order corrections to the velocity, temperature and chemical potential are to be determined.

With the action \eqref{action3d} in hand it is straightforward to use
\eqref{stcurrent} to obtain the stress tensor and current corresponding to this equilibrium
solution. We find
\begin{equation}\label{TJfrmaction3d}
 \begin{split}
  T^{ij} &= 0, \\
  T_{00} &= -T_0 e^{ \sigma}    \left( \frac{\partial \alpha}{\partial \sigma} \epsilon^{ij} \partial_i A_j
+ T_0 \frac{\partial \beta}{\partial \sigma} \epsilon^{ij} \partial_i a_j\right), \\
  T^i_0 &= T_0 e^{ -\sigma}  \left( \left( T_0 \frac{\partial \beta}{\partial \sigma}- A_0 \frac{\partial \alpha}{\partial \sigma}\right)
\epsilon^{ij}\partial_j \sigma + \left( T_0 \frac{\partial \beta}{\partial A_0}- A_0 \frac{\partial \alpha}{\partial A_0}\right)
\epsilon^{ij} \partial_j A_0 \right), \\
  J_0 &= -T_0 e^{ \sigma}  \left( \frac{\partial \alpha}{\partial A_0} \epsilon^{ij} \partial_i A_j + 
 T_0 \frac{\partial \beta}{\partial A_0} \epsilon^{ij} \partial_i a_j\right),\\
  J^i &= T_0 e^{ -\sigma}  \left( \frac{\partial \alpha}{\partial \sigma} \epsilon^{ij} \partial_j \sigma
 + \frac{\partial \alpha}{\partial A_0} \epsilon^{ij} \partial_j A_0 \right).
 \end{split}
\end{equation}

\subsection{Constraints on Hydrodynamics} \label{sSec:match3d}

In this subsection we shall equate the coefficients of independent terms in \eqref{TJcor3d} (or \eqref{tjcorcompo}) with those in
\eqref{TJfrmaction3d}, to determine the first order transport coefficients and fluid corrections in terms of the
two arbitrary functions in the action \eqref{action3d}.

The fact that $T^{ij}$ as evaluated from the action \eqref{action3d} vanishes immediately implies from \eqref{tjcorcompo}
\begin{equation}\label{Tij3d}
 \begin{split}
  \tilde \chi_B &= \frac{\partial P}{\partial T} \tau_B + \frac{\partial P}{\partial \mu} m_B, \\
  \tilde \chi_\Omega &= \frac{\partial P}{\partial T} \tau_\Omega + \frac{\partial P}{\partial \mu} m_\Omega, \\
 \end{split}
\end{equation}

Comparing $T_{00}$ from \eqref{tjcorcompo} and \eqref{TJfrmaction3d} we have
\begin{equation}\label{T003d}
 \begin{split}
  \frac{\partial \epsilon}{\partial T} \tau_B + \frac{\partial \epsilon}{\partial \mu} m_B &= - T_0 e^{-\sigma} \frac{\partial \alpha}{\partial \sigma},\\
\frac{\partial \epsilon}{\partial T} \tau_\Omega + \frac{\partial \epsilon}{\partial \mu} m_\Omega
&=  T_0 e^{-2\sigma} \left( T_0 \frac{\partial \beta}{\partial \sigma} - A_0 \frac{\partial \alpha}{\partial \sigma} \right),\\
 \end{split}
\end{equation}

Comparing $T_{0}^i$ from \eqref{tjcorcompo} and \eqref{TJfrmaction3d} we have
\begin{equation}\label{Tio3d}
 \begin{split}
  \xi_E &= - \frac{T_0 e^{-\sigma}}{(\epsilon + P)} \left( T_0 \frac{\partial \beta}{\partial A_0} - A_0 \frac{\partial \alpha}{\partial A_0} \right),\\
  \xi_T &=  \frac{ T_0 e^{-\sigma}}{(\epsilon + P)} \left( T_0 \frac{\partial \beta}{\partial \sigma} - A_0 \frac{\partial \alpha}{\partial \sigma} \right).\\
 \end{split}
\end{equation}

Comparing $J_{0}$ from \eqref{tjcorcompo} and \eqref{TJfrmaction3d} we have
\begin{equation}\label{J03d}
\begin{split}
  \frac{\partial \rho}{\partial T} \tau_B + \frac{\partial \rho}{\partial \mu} m_B &= T_0 \frac{\partial \alpha}{\partial A_0},\\
  \frac{\partial \rho}{\partial T} \tau_\Omega + \frac{\partial \rho}{\partial \mu} m_\Omega &= -T_0 e^{-\sigma}
\left( T_0 \frac{\partial \beta}{\partial A_0} - A_0 \frac{\partial \alpha}{\partial A_0} \right),
\end{split}
\end{equation}

Finally, comparing $J^i$ from \eqref{tjcorcompo} and \eqref{TJfrmaction3d} we have
\begin{equation}\label{Ji3d}
 \begin{split}
  \tilde \chi_E + \rho \xi_E &= T_0 \frac{\partial \alpha}{\partial A_0}, \\
  \tilde \chi_T + \rho \xi_T &= - \frac{\partial \alpha}{\partial \sigma}.
 \end{split}
\end{equation}

In order to compare the constraints obtained in this section with that in
\cite{Jensen:2011xb} we find the following thermodynamical identities useful
\begin{equation}\label{thermid}
 \begin{split}
  \frac{\partial P}{\partial \epsilon}
&= \bigg(\frac{\partial P}{\partial T}\frac{\partial \rho}{\partial \mu}- \frac{\partial P}{\partial \mu} \frac{\partial \rho}{\partial T}\bigg)
\bigg/\bigg(\frac{\partial \rho}{\partial \mu}\frac{\partial \epsilon}{\partial T}- \frac{\partial \rho}{\partial T}
\frac{\partial \epsilon}{\partial \mu}\bigg), \\
\frac{\partial P}{\partial \rho}
&= \bigg(-\frac{\partial P}{\partial T}\frac{\partial \epsilon}{\partial \mu}+ \frac{\partial P}{\partial \mu} \frac{\partial \epsilon}{\partial T}\bigg)
\bigg/\bigg(\frac{\partial \rho}{\partial \mu}\frac{\partial \epsilon}{\partial T}- \frac{\partial \rho}{\partial T}
\frac{\partial \epsilon}{\partial \mu}\bigg), \\
 \end{split}
\end{equation}

Now solving for $\tau_B, \tau_\Omega, m_B$ and $m_\Omega$ from \eqref{T003d} and \eqref{J03d}, plugging the answer in to
\eqref{Tij3d} and using the thermodynamical identities \eqref{thermid}, we have
\begin{equation}\label{chiBO}
 \begin{split}
  \tilde \chi_B &= \frac{\partial P}{\partial \epsilon} \bigg( - T_0 e^{-\sigma} \frac{\partial \alpha}{\partial \sigma}\bigg)
+ \frac{\partial P}{\partial \rho} \bigg( T_0 \frac{\partial \alpha}{\partial A_0}\bigg), \\
  \tilde \chi_\Omega &= \frac{\partial P}{\partial \epsilon} \bigg( T_0 e^{-2 \sigma} \left( T_0 \frac{\partial \beta}{\partial \sigma}
- A_0 \frac{\partial \alpha}{\partial \sigma}\right) \bigg)
+ \frac{\partial P}{\partial \rho} \bigg( - T_0 e^{-\sigma} \left( T_0 \frac{\partial \beta}{\partial A_0}
- A_0 \frac{\partial \alpha}{\partial A_0}\right)\bigg), \\
 \end{split}
\end{equation}
Finally plugging in the values of $\xi_E$ and $\xi_T$ from \eqref{Tio3d} into \eqref{Ji3d} we have
\begin{equation}\label{chiET}
 \begin{split}
  \tilde \chi_E &= \bigg( T_0 \frac{\partial \alpha}{\partial A_0}\bigg) - \frac{\rho}{\epsilon + P}
\bigg( - T_0 e^{-\sigma} \left( T_0 \frac{\partial \beta}{\partial A_0}
- A_0 \frac{\partial \alpha}{\partial A_0}\right)\bigg)   \\
  T \tilde \chi_T &= \bigg( - T_0 e^{-\sigma} \frac{\partial \alpha}{\partial \sigma}\bigg) - \frac{\rho}{\epsilon + P}
\bigg( T_0 e^{-2 \sigma} \left( T_0 \frac{\partial \beta}{\partial \sigma}
- A_0 \frac{\partial \alpha}{\partial \sigma}\right) \bigg)
 \end{split}
\end{equation}

Thus through \eqref{chiBO} and \eqref{chiET} we are able to express the 4 transport coefficients in terms
two arbitrary functions in the action. The two dimensional manifold of allowed transport coefficients is
identical to that in equation (1.8) in \cite{Jensen:2011xb}\footnote{Note that the additional
function $f_\Omega(T)$ may be reabsorbed into a redefinition of $M_\Omega (\mu, T)$ in equation
(1.8) in \cite{Jensen:2011xb}.}. In particular it easy to eliminate $\alpha$ and $\beta$ from \eqref{chiBO} and \eqref{chiET}
so as to obtain the following relation between the transport coefficients
\begin{equation}
\tilde \chi_B - \frac{\rho}{\epsilon + P} ~\tilde \chi_\Omega =
 \frac{\partial P}{\partial \rho} \tilde \chi_E + \frac{\partial P}{\partial \epsilon} ~T ~\tilde \chi_T.
\end{equation}
Note that this relation is identical to equation (4.29) in  \cite {Jensen:2011xb}.

\subsection{The Entropy Current}\label{sSec:entcur3d}

In this system $\ln Z$ is simply given by the action
\begin{equation}
 \ln Z = \half \int \left(  \alpha(\sigma, A_0) dA + T_0 \beta(\sigma, A_0) da \right)
\end{equation}
The entropy that follows from this partition function is 
\begin{equation}
\begin{split}\label{entfrmAction}
 S &= \frac{\partial}{\partial T_0} \left( T_0 \ln Z \right) \\
    &= \half \int \left( \alpha - \frac{\partial \alpha}{\partial \sigma} -A_0 \frac{\partial \alpha}{\partial A_0}\right) dA 
+T_0 \left( 2\beta - \frac{\partial \beta}{\partial \sigma} - A_0 \frac{\partial \beta}{\partial A_0}\right) da
\end{split}
\end{equation}
We will now utilize \eqref{entfrmAction} to constrain the hydrodynamical entropy current of the system. 
The entropy current must take the `canonical' form $s u^\mu -\nu J^\mu_{diss}$ corrected by first derivative terms. 
As in the rest of this section we keep track only of parity odd terms. It follows from Table \ref{T:odfd3d} that the most 
general one derivative entropy current is given by \footnote{The map between the corrections to the entropy current in our paper to that in 
\cite{Jensen:2011xb}, considering first order terms which are non-zero in the equilibrium, is given by
\begin{equation}
 n_T = \tilde \nu_1 + \frac{\tilde \nu_5}{T};~~ n_E = \tilde \nu_2 + \tilde \nu_4 + \frac{\tilde \nu_3}{T};~~
 n_B = \tilde \nu_4;~~ n_\Omega = \tilde \nu_5.
\end{equation}
}
\begin{equation}
 J_{(s)}^{\mu} = s u^{\mu} - \frac{\mu}{T} \left( \tilde \chi_E \tilde E^{\mu} + \tilde \chi_T \tilde T^{\mu}\right)
+\left( n_E \tilde E^{\mu} + n_T \tilde T^{\mu}\right) + \left( n_B B + n_\Omega \Omega \right) u^{\mu},
\end{equation}
$n_E, n_T, n_B$ and $n_\Omega$ are functions of temperature and the chemical potential. On substituting 
the equilibrium values of temperature and chemical potential they turn into functions of $\sigma$ and $A_0$.
We find it convenient to define the quantities 
\begin{equation}
\begin{split}
 \tilde n_E &= n_E - \frac{\mu}{T} \tilde \chi_E+ s \xi_E, \\
 \tilde n_T &= n_T - \frac{\mu}{T} \tilde \chi_T+ s \xi_T. 
\end{split}
\end{equation}
in terms of which the first order part of the entropy current is given by
\begin{equation}
\begin{split}
 \delta J_{(s)}^{\mu} =& \left( \left( \frac{\partial s}{ \partial T} \tau_B +\frac{\partial s}{ \partial \mu} m_B + n_B\right) B
+ \left( \frac{\partial s}{ \partial T} \tau_\Omega +\frac{\partial s}{ \partial \mu} m_\Omega + n_\Omega \right) \Omega \right) u_{(0)}^\mu
\\ &+ \tilde n_E \tilde E^{\mu} + \tilde n_T \tilde T^{\mu}
\end{split}
\end{equation}

As we have explained above, the entropy current is necessarily divergence free in equilibrium. This condition yields 
one condition  
\begin{equation}\label{cond}
 \frac{\partial \tilde n_E}{\partial \sigma}= - T_0 \frac{\partial \tilde n_T}{\partial A_0}.
\end{equation}
\eqref{cond} is solved by the ansatz 
\begin{equation}\label{nET}
 \tilde n_E = T_0 \frac{\partial n}{\partial A_0}; ~~~ \tilde n_T = - \frac{\partial n}{\partial \sigma},
\end{equation}
where $n$ is a arbitrary function of $\sigma$ and $A_0$. Plugging in this solution, we now have a 3 parameter 
set of entropy currents parameterized by $n_B$, $n_\omega$ and $n$. The entropy \eqref{entfrmAction} is an integral 
over the two parity odd scalars of the system. Equating \eqref{entfrmAction} with $\int d^3 x\sqrt{-g_3}J_S^0$, and equating 
the coefficients of these two scalars, yields two equations for $n_B$, $n_\omega$ and $n$. We now explain how this works in 
more detail

Using the fact 
\begin{equation}
 \tilde E^0 = - e^{-\sigma} \epsilon^{ij} a_i \partial_j A_0 ; ~~ \tilde T^0 = e^{-\sigma} \epsilon^{ij} a_i \partial_j \sigma,
\end{equation}
and the expressions of $B$ and $\Omega$ in terms of the background field (from Table \ref{T:odfd3d}), the entropy can 
be evaluated from the entropy current in a manifestly Kaluza-Klein gauge invariant way 
\begin{equation}\label{entfrmcur}
\begin{split}
 S &= \int d^2 x \sqrt{-g_3} J^{0}_{(s)} \\
   &= \half \int \Bigg( \left( \frac{\partial s}{\partial T} \tau_B + \frac{\partial s}{\partial \mu} m_B + n_B \right) 
\left( dA + A_0 da \right) \\
&- \left( \frac{\partial s}{\partial T} \tau_{\Omega} + \frac{\partial s}{\partial \mu} m_{\Omega} + n_{\Omega} \right) e^{\sigma} da
- T_0 n da \Bigg).
\end{split}
\end{equation}
Comparing \eqref{entfrmcur} with \eqref{entfrmAction} and using the 
thermodynamic identities 
\begin{equation}
 \frac{\partial s}{\partial \epsilon} = \frac{1}{T} , ~~ \frac{\partial s}{\partial \rho} = -\frac{\mu}{T},
\end{equation}

we get the following simple expressions
\begin{equation}\label{nBO}
 \begin{split}
  n_B &= \alpha \\
  n_\Omega &= T_0 e^{-\sigma} \left( A_0 \alpha - 2 \beta - n \right)
 \end{split}
\end{equation}
In other words, we have managed to evaluate $n_B$, and one linear combination of $n_\Omega$ and  $n$ in terms 
of the functions, $\alpha$ and $\beta$, that appear in the partition function of our system. Note that we have not 
been able to completely determine the non dissipative part of the entropy current using our method (the method 
based on positivity of the entropy current achieves this determination). However, it straightforward to verify that 
the constraints (equations 3.11, 3.17, 3.18, and 3.20) in  \cite{Jensen:2011xb} on the corrections to the entropy current 
from the second law of thermodynamics, are consistent with the relations \eqref{nBO} and \eqref{nET}.

\subsection{Comparison with Jensen et.al. }
In this subsection we shall give a  precise connection between  partition function coefficients $\alpha,~\beta$ in equation \eqref{action3d} and $M_{\Omega},~M$ that appears in
\cite{Jensen:2011xb}.
Comparing equations \eqref{chiBO},\eqref{chiET} with equation $1.8$ of \cite{Jensen:2011xb}, we get the following differential equations
\begin{equation}\begin{split}\label{mathingeq}
\frac{\partial M}{\partial \mu} &= T_{0} \frac{\partial \alpha}{\partial A_{0}},\\
T    \frac{\partial M}{\partial T}+\mu  \frac{\partial M}{\partial \mu}- M &= -T  \frac{\partial \alpha}{\partial \sigma},\\ 
 \frac{\partial M_{\Omega}}{\partial \mu} -M&=- T \left(T_{0}\frac{ \partial\beta}{\partial A_{0}} - A_{0 }\frac{\partial \alpha}{\partial A_{0}}\right) ,\\
 T    \frac{\partial M_{\Omega}}{\partial T}+\mu  \frac{\partial M_{\Omega}}{\partial \mu}+ f_{\Omega}- 2 M_{\Omega} &= 
T e^{-\sigma}\left(T_{0} \frac{ \partial\beta}{\partial \sigma} - A_{0 }\frac{\partial \alpha}{\partial \sigma}\right).
\end{split}
\end{equation}
By solving first two equations in \ref{mathingeq}, we get
\begin{equation}\label{alpha1}
 \alpha = \frac{M}{T_{0}e^{-\sigma}} + c
\end{equation}where  $c$  is some constant. Infact the entropy current presented in this paper matches 
that of \cite{Jensen:2011xb} only if we set $c=0$ (see equation (3.22) of \cite{Jensen:2011xb}).
Solving last two equations in  \ref{mathingeq}, we get
\begin{equation}\label{beta1}
 \beta = -\frac{e^{2\sigma}}{T_{0}^2}M_{\Omega} + \frac{1}{T_{0}^2}\int f_{\Omega}(T_0 e^{-\sigma}) e^{2\sigma} d\sigma +\frac{A_{0}}{T_{0}} 
\left(\frac{M}{T_{0} e^{-\sigma}}\right) +c_{1},
\end{equation}
 where $c_{1}$ is some other constant. \\

Also comparing \eqref{chiBO} and \eqref{chiET} with equations (3.17) and (3.18)
in \cite {Jensen:2011xb} one can express the entropy current corrections 
in terms of $\alpha$ and $\beta$ in the following way

\begin{equation}
\begin{split}
   T^2 \frac{\partial \tilde \nu_4}{\partial T} &=   - T_0 e^{-\sigma} \frac{\partial \alpha}{\partial \sigma}, \\
 \frac{\partial \tilde \nu_4}{\partial (\frac{\mu}{T})} &=   T_0 \frac{\partial \alpha}{\partial A_0}, \\
  T^2 \left( \frac{\partial \tilde \nu_5}{\partial T} + \tilde \nu_1 \right) &=  T_0 e^{-2 \sigma} \left( T_0 \frac{\partial 
\beta}{\partial \sigma}
- A_0 \frac{\partial \alpha}{\partial \sigma}\right) , \\
  \frac{\partial \tilde \nu_5}{\partial (\frac{\mu}{T})} + \tilde \nu_3 &=  - T_0 e^{-\sigma} \left( T_0 \frac{\partial \beta}{\partial A_0}
- A_0 \frac{\partial \alpha}{\partial A_0}\right).
\end{split}
\end{equation}

Note that with this identification, the equation (3.20) in  \cite {Jensen:2011xb}
automatically follows.

\subsection{Constraints from CPT invariance}\label{CPT3d}

Imposing CPT invariance of the partition function \ref{action3d} constrains  the form of the otherwise arbitrary functions $\alpha,~\beta.$ 
(and hence all transport coefficients determined in terms of $\alpha$ and $\beta$). 
Note that we define parity 
in 2+1 dimensions as $x_1 \rightarrow - x_1$ and $x_2 \rightarrow x_2$. 
In Table \ref{cpt1t} we list the action of CPT on
various fields appearing in the partition function \ref{action3d}.
\begin{table}
\centering
\begin{tabular}[h]{|c|c|c|c|c|}
\hline
Field & C & P & T & CPT \\
\hline
$\sigma$ & $+$ & $+$ & $+$ & $+$ \\
\hline
$a_1$ & $+$ & $-$ & $-$ & $+$ \\
\hline
$a_2$ & $+$ & $+$ & $-$ & $-$ \\
\hline
$A_0$ & $-$ & $+$ & $+$ & $-$ \\
\hline
$A_1$ & $-$ & $-$ & $-$ & $-$ \\
\hline
$A_2$ & $-$ & $+$ & $-$ & $+$ \\
\hline
\end{tabular}
\caption{Action of CPT}
\label{cpt1t}
\end{table}
Based on the Table \ref{cpt1t} we see, in \ref{action3d} $``dA"$ changes sign where as $``da"$ does not,
 which implies $\alpha$ is odd under CPT and $\beta$ is even under CPT.


\section{$3+1$ dimensional uncharged fluid dynamics at second order in the derivative expansion} \label{ucf}

In this section we will derive the constraints imposed on the equations of uncharged fluid dynamics,
at second order in the derivative expansion, by comparison with the most general equilibrium
partition function. We do not assume that our system enjoys invariance under parity transformations.
We do, however, assume that the fluid enjoys invariance under CPT transformations.

Before getting into the details let us summarize our results. Symmetry considerations determine
the expansion of the hydrodynamical stress tensor upto 15 parity even and 5 parity odd transport coefficients.
It turns out the 7 of the parity even and 2 of the odd terms vanish in equilibrium. In other words, on
symmetry grounds our  system has 7 parity odd and 2 parity even dissipative coefficients. In addition
we have 8 parity even and 3 parity odd non dissipative coefficients. The most general second order fluid dynamical
partition function, on the other hand, is given in terms of three functions of $\sigma$ \footnote{The only possible 
first order contribution to the partition function is the term proportional to $C_1$ in \eqref{cfacn}. As explained in 
subsection \ref{cpt} the requirement of CPT invariance forces $C_1$ to vanish. It follows that there are no first order 
contributions to the partition function for an uncharged system. We thank S. Dutta for discussions on this topic.}. It turns out that
this partition function is automatically even under parity transformations. As a consequence,
implementing the procedure spelt out in the introduction, we are able to show that the three nondissipative
parity odd coefficients all vanish. In addition the 8 nondissipative parity even coefficients are
all determined in term of three functions. In other words we are able to derive
5 relations between these 8 parity even coefficients.

The problem of constraining fluid dynamics at second order in the derivative expansion, using the
principle of entropy increase, was studied by one of the authors of this paper  in
\cite{Bhattacharyya:2012ex}. In that work the fluid was assumed to enjoy invariance under
parity transformations. It was demonstrated that the principle of entropy increase indeed
implies 5 relations between the 8 non dissipative transport coefficients. It turns out that
the five relations determined in this paper agree exactly with those of \cite{Bhattacharyya:2012ex}.

Even from a practical point of view the method used in this paper appears to have some advantages
over the more traditional entropy method utilized in \cite{Bhattacharyya:2012ex}. To start with
the algebra required for the analysis in this paper is considerably less formidable than that
employed in \cite{Bhattacharyya:2012ex}. As a consequence we are able, rather effortlessly, to
generalize our results to allow for the possibility of parity violation. Such a generalization
would involve considerable extra effort using the method of \cite{Bhattacharyya:2012ex}, and has not yet been
done.

\subsection{Equilibrium from Hydrodynamics}

In Tables $1,~2,~3,~7$ of \cite{Bhattacharyya:2012ex},  all scalar, vector and tensor expressions
that one can form out of fluid fields and background metric  (not necessarily
in equilibrium) at second order in the derivative expansion are listed. It follows from the listing
of these tables that the most general symmetry allowed two derivative expansion of the
constitutive relations is given by
\begin{equation} \label{cru}\begin{split}
\Pi_{\mu\nu} =~&-\eta\sigma_{\mu\nu} - \zeta P_{\mu\nu} \Theta\\
~&+T\bigg[ \tau ~(u.\nabla)\sigma_{\langle\mu\nu\rangle} + \kappa_1 \tilde{R}_{\langle \mu\nu\rangle} + \kappa_2 K_{\langle \mu\nu\rangle} +\lambda_0~ \Theta\sigma_{\mu\nu}\\
&+ \lambda_1~ {\sigma_{\langle \mu}}^a\sigma_{a\nu\rangle}+ \lambda_2~ {\sigma_{\langle \mu}}^a\omega_{a\nu\rangle}+ \lambda_3~ {\omega_{\langle \mu}}^a\omega_{a\nu\rangle} + \lambda_4~{\mathfrak a}_{\langle\mu}{\mathfrak a}_{\nu\rangle}\bigg]\\
&+TP_{\mu\nu}\bigg[\zeta_1(u.\nabla)\Theta + \zeta_2 \tilde{R} + \zeta_3 \tilde{R}_{00}
 + \xi_1 \Theta^2 + \xi_2 \sigma^2+ \xi_3 \omega^2
+\xi_4 {\mathfrak a}^2 \bigg] \\
&+ T \bigg[ \sum_{i=1}^{4} \delta_{i} t^{(i)}_{\mu\nu} + \delta_{5} P_{\mu\nu} {\mathfrak a}_{\alpha}l^{\alpha} \bigg]
\end{split}
\end{equation}
where
\begin{equation}\label{notation}
\begin{split}
&u^\mu =\text{The normalised four velocity of the fluid}\\
&P^{\mu\nu} = g^{\mu\nu} + u^\mu u^\nu =\text{Projector perpendicular to $u^\mu$}\\
&\Theta = \nabla. u = \text{Expansion},~~
{\mathfrak a}_\mu = (u.\nabla) u_\mu = \text{Acceleration}\\
&\sigma^{\mu\nu} =
P^{\mu\alpha} P^{\nu\beta}\left(\frac{\nabla_\alpha u_\beta + \nabla_\beta u_\alpha}{2}
 - \frac{\Theta}{3}g_{\alpha_\beta}\right) = \text{Shear tensor}\\
&\omega^{\mu\nu} =
P^{\mu\alpha} P^{\nu\beta}\left(\frac{\nabla_\alpha u_\beta
- \nabla_\beta u_\alpha}{2}\right)=\text{Vorticity}\\
&K^{\mu\nu} =\tilde{ R}^{\mu a \nu b}u_a u_b,~~\tilde{R}^{\mu\nu}
= \tilde{R}^{a\mu b\nu}g_{ab}~~(\tilde{R}^{abcd} = \text{Riemann tensor})\\
&\sigma^2 = \sigma_{\mu\nu}\sigma^{\mu\nu},~~~~\omega^2 = \omega_{\mu\nu}\omega^{\nu\mu}
\end{split}
\end{equation}
and
$$
A_{\langle\mu\nu\rangle} \equiv P_\mu^\alpha P_\nu^\beta\left(\frac{A_{\alpha\beta} + A_{\beta\alpha}}{2} -
\left[\frac{A_{ab}P^{ab}}{3}\right]g_{\alpha\beta}\right)~~\text{For any tensor $A_{\mu\nu}$}
$$
The parity odd terms in the last bracket in \eqref{cru} are defined in Table \ref{tdpvfd}.

\begin{table}
\centering
\begin{tabular}[h]{|c|c|c|}
\hline
Type & Data & Evaluated on equilibrium \\
\hline
Pseudo-Scalars & $l^{\mu} \mathfrak{a}_\mu$ & $\half e^{\sigma} \epsilon^{ijk} \partial_i \sigma f_{jk}$  \\
\hline
Pseudo-Vectors & $(\nabla.u)l_\mu$,  & 0 \\
               & $\sigma_{\mu \nu} l^\nu$,  & 0 \\
               & $u.\nabla l_\mu$ & $\half e^{2\sigma}$ ($\epsilon^{ijk} \partial_i \sigma f_{jk}$~,~ $f_{ij} \epsilon^{jkl} f_{kl}$) \\
\hline
Pseudo-Tensors & $t^{(1)}_{\mu\nu} = l_{< \mu} \mathfrak{a}_{\nu >}$, & $\half e^{\sigma} \partial_{\langle i} \sigma \epsilon_{j\rangle kl} f^{kl}$ \\
               & $t^{(2)}_{\mu\nu} = \epsilon^{\lambda \rho \alpha \beta} u_{\lambda} \mathfrak{a}_{\rho} \sigma_{\alpha <\mu} g_{\nu> \beta}$, & 0 \\
               & $t^{(3)}_{\mu\nu} = \epsilon^{\lambda \rho \alpha \beta} u_{\lambda} \nabla_{\rho} \sigma_{\alpha <\mu} g_{\nu> \beta}$, & 0 \\
               & $t^{(4)}_{\mu\nu} = u_b \tilde{R}_{<\mu}^{bcd}\epsilon_{\nu>cdq}u^q$ & $\half d_1 e^\sigma \partial_{\langle i} \sigma \epsilon_{j\rangle kl} f^{kl}$ + $\half d_2 e^\sigma \nabla_{\langle i} \epsilon_{j\rangle kl} f^{kl}$\\
\hline
\end{tabular}
\caption{Two derivative parity violating fluid data(Here $d_{1,2}$ are function of $\sigma$ determined by
evaluating $t^4_{\mu\nu}$ on equilibrium, but we will not need there explicit expression.)}
\label{tdpvfd}
\end{table}
The expansion \eqref{cru} is given in terms of 15 undetermined parity even and five undetermined
parity odd transport coefficients, each of which is, as yet, an arbitrary function of
temperature).

We are interested in the stationary equilibrium solutions of these equations.
Solutions in equilibrium are determined entirely by the background fields
$\sigma$,  $a_i$ and $g^{ij}$. In Table(\ref{odbd},\ref{odfd}) we have seen that the
$\Theta$ and $\sigma_{\mu\nu}$ evaluates to zero in equilibrium.
This sets seven of the fifteen parity even terms in equation \ref{cru} to zero.
Two of the five parity odd terms two terms ($t^{(2,3)}_{\mu\nu}$ in table \ref{tdpvfd})
evaluate to zero in equilibrium. The remaining 8 parity even and 3 parity odd coefficients
are non dissipative; the non dissipative part of $\Pi_{\mu\nu}$ is given by
\begin{eqnarray}\label{uncharge2nd}
\frac{\Pi_{\mu \nu}}{T} &=& \kappa_1 \tilde R_{\langle\mu \nu\rangle}+\kappa_2 K_{\langle\mu \nu\rangle}+\lambda_3
\omega_{\langle\mu}^{\,\,\,\ \alpha} \omega_{\alpha \nu\rangle}+\lambda_4 \mathfrak{a}_{\langle \mu}\mathfrak{a}_{\nu \rangle} \nn \\
&+& P_{\mu \nu}(\zeta_2 \tilde R+\zeta_3 \tilde R_{00}(u^0)^2 +\xi_3 \omega^2+\xi_4 \mathfrak{a}^2 ) \nn\\
&+& \delta_{1} t^{1}_{\mu\nu} +  \delta_{4} t^{4}_{\mu\nu} + \delta_{5} P_{\mu\nu} \mathfrak{a}_{\alpha}l^{\alpha}
\end{eqnarray}

In order to proceed further, we list all coordinate invariant
two derivative scalars, vectors and tensors constructed out of background
data are listed in table (\ref{odbdu}). The temperature and velocity in equilibrium
receives correction at second order. The most general symmetry allowed
form of corrected temperature and velocity is
\begin{equation}\begin{split} \label{vcuf}
u^{\mu}&=b_0 u_0^{\mu} + \left( \sum_{m=1}^{2} v_m V_{(m)}^{i} \right) + \tilde{v} \tilde{V}^{i} ,   \nn \\
T&= T_0 e^{-\sigma} + \left( \sum_{m=1}^{4} t_m S_m \right) + \tilde{t} \tilde{S}
\end{split}
\end{equation}
where, $V_m$($\tilde{V}$) and $S_i$($\tilde{S}$)are Vectors(pseudo) and scalars(pseudo)
respectively that are listed in table \ref{odbdu}. Also $b_0$ can be fixed following equation \ref{norfixing} as,
\begin{equation}
b_0= 1- e^{\sigma}a.\left( \sum_{m=1}^{2} v_m V_{(m)}  + \tilde{v} \tilde{V} \right)
\end{equation}

As in previous sections, the stress tensor in equilibrium received corrections at
second order in the derivative expansion. The two derivative corrections have two sources.
The first set of corrections arises from the corrections \eqref{cru} evaluated on the zero order
equilibrium fluid configuration. Using
\begin{eqnarray}
\tilde R_{00}(u^0)^2&=&\tilde R_{\mu \nu}u^{\mu}u^{\nu}=\frac{1}{4}e^{2\sigma}f^2+ (\nabla \sigma)^2
+ \nabla^2 \sigma \nn \\
\omega^{ij}&=& - \frac{e^{\sigma}}{2} f^{ij}, \quad \mathfrak {a}^i=  g^{im}\partial_m \sigma,
\end{eqnarray}

\begin{eqnarray}\label{basischange}
\tilde R_{<ij>}&=&R_{ij}-\nabla_i \sigma \nabla_j \sigma-\nabla_i\nabla_j \sigma+\frac{1}{2}f_i^{\,\ k}f_{jk}e^{2\sigma} \nn \\
&-&\frac{1}{3}\bigg(R-(\nabla \sigma)^2-\nabla^2 \sigma+\frac{1}{2}f^2e^{2\sigma}\bigg)g_{ij}\nn \\
K_{<ij>}&=&\nabla_i \sigma \nabla_j \sigma+\nabla_i\nabla_j \sigma+\frac{1}{4}f_i^{\,\ k}f_{jk}e^{2\sigma} \nn \\
&-&\frac{1}{3}\bigg((\nabla \sigma)^2+\nabla^2 \sigma+\frac{1}{4}f^2e^{2\sigma}\bigg)g_{ij} \nn \\
\omega_{<i}^a \omega_{a j>}&=&e^{2 \sigma}\bigg(f_i^{\,\ a}f_{ja}-\frac{1}{3}f^2 g_{ij}\bigg) \nn \\
\mathfrak{a}_{<i}\mathfrak{a}_{j>}&=&\nabla_i \sigma \nabla_j \sigma-\frac{1}{3}(\nabla \sigma)^2g_{ij}
\end{eqnarray}
we find that these corrections are given by
\begin{eqnarray}\label{gipi}
\Pi^{eq}_{ij}&=&a_1\bigg(R_{ij}-\frac{R}{2}g_{ij}\bigg)+a_2\bigg(\nabla_i\nabla_j \sigma-\nabla^2 \sigma g_{ij}\bigg)
+a_3\bigg(\nabla_i \sigma \nabla_j \sigma-\frac{(\nabla \sigma)^2}{2}g_{ij}\bigg)\nn\\
&+&a_4 \bigg(f_i^kf_{k j}+\frac{f^2}{4}g_{ij}\bigg)e^{2\sigma}
+g_{ij}\bigg(b_1 R+b_2 \nabla^2 \sigma + b_3 (\nabla \sigma)^2+b_4 f^2e^{2\sigma}\bigg) \\
&+& \half T (\delta_1 + \delta_4 d_1) e^{\sigma} \partial_{\langle i}\sigma \epsilon_{j\rangle kl}f^{kl} + \half T d_{2} \delta_4 e^{\sigma} \epsilon_{kl\langle i}\nabla_{j\rangle}f^{kl}
+ \half T \delta_{5} e^{\sigma} g_{ij} \epsilon_{mlk}\partial^{m}\sigma f^{lk}
~~\mbox{where}, \nn \\
\frac{b_1}{T}&=&\zeta_2+\frac{1}{6}\kappa_1, \quad \frac{b_2}{T}=\frac{2}{3}(\kappa_2- \kappa_1)-
2 \zeta_2+\zeta_3 \nn \\
\frac{b_3}{T}&=&\frac{1}{6}(\kappa_2- \kappa_1+\lambda_4)-2\zeta_2+ \zeta_3+\xi_4 \nn \\
\frac{b_4}{T}&=&\frac{1}{48}(\lambda_3 -\kappa_2 -2\kappa_1)+\frac{1}{4}(\zeta_2+\zeta_3-\xi_3) ,
\quad \frac{a_1}{T}=\kappa_1 \nn \\
\frac{a_2}{T}&=&\kappa_2-\kappa_1, \quad \frac{a_3}{T}=\kappa_2-\kappa_1+\lambda_4, \quad \frac{a_4}{T}=\frac{1}{4}(\lambda_3 -2\kappa_1- \kappa_2).
\end{eqnarray}
Here, the indicies are contracted with the lower dimensional metric $g_{ij}$ and its inverse.
The coefficients  are determined by evaluating the
$t^{(1,4)}_{\mu\nu}$ in equilibrium, but we will not need the explicit expressions.

\begin{table}
\centering
\begin{tabular}[h]{|c|c|}
\hline
 Scalars & $ S_1=R, \,\, S_2=\nabla^2 \sigma, \,\, S_3= (\nabla \sigma)^2, \,\, S_4= f^2 e^{2 \sigma}$  \\
\hline
Pseudo-Scalars & $\tilde{S} = \epsilon^{ijk}\partial_i \sigma f_{jk}$  \\
\hline
 Vectors & $\quad V_1=e^{\sigma} \nabla_i \sigma f^{ij}, \quad V_2=e^{\sigma} \nabla_if^{ij}, $ \\
\hline
Pseudo-Vectors & $\tilde{V}_{i} = f_{ij}f_{kl} \epsilon^{jkl}$ \\
\hline
Tensors & $R_{ij}, f_i^{\,\ k} f_{kj}, \nabla_i\nabla_j \sigma, \nabla_i \sigma \nabla_j \sigma$ \\
\hline
Pseudo-Tensors & $\partial_{\langle i}\sigma \epsilon_{j\rangle kl}f^{kl}$, $\nabla_{\langle i} \epsilon_{j\rangle kl} f^{kl}$ \\
\hline
\end{tabular}
\caption{Two derivative background data}
\label{odbdu}
\end{table}

The second source of corrections arises from inserting the velocity correction \eqref{vc}
into the zero order (perfect fluid) constitutive relations. We find that the
modification of the stress tensor due to these corrections
is given by
\begin{equation}\label{ccvmu} \begin{split}
T^{ij}&= P_T g^{ij} \left( \sum t_m S_m + \tilde{t}\tilde{S} \right) \\
T_{00}&=T_0^2 \frac{P_{TT}}{T} \left( \sum t_m S_m + \tilde{t}\tilde{S} \right)  \\
T_0^i&=-(\epsilon+P)e^{\sigma} \left( \sum v_m V_m^i + \tilde{v}\tilde{V^i} \right)
\end{split}
\end{equation}
The net change in $T_0^i$ and $J^i$ is given by summing\eqref{gipi} and  \eqref{ccvmu} and is given by
\begin{equation}\label{ccvmfu} \begin{split}
T^{ij}&= P_T g^{ij} \left( \sum t_m S_m + \tilde{t}\tilde{S} \right) + \Pi_{eq}^{ij}  \\
T_{00}&=T_0^2 \frac{P_{TT}}{T} \left( \sum t_m S_m + \tilde{t}\tilde{S} \right)  \\
T_0^i&=-(\epsilon+P)e^{\sigma} \left( \sum v_m V_m^i + \tilde{v}\tilde{V^i} \right)
\end{split}
\end{equation}
where $\Pi^{ij}_{eq}$ was listed in \eqref{gipi}.

\subsection{Equilibrium from the Partition Function}

We now turn to the study of the first correction to the perfect fluid
equilibrium partition function \eqref{perfectfluidA} at second order in the
derivative expansion. We observe that the Table (\ref{odbdu}) lists four scalars
and one pseudo-scalar. The most generic partition function for this system at two
derivative order is,

\begin{equation}\label{cfacnu}
\begin{split}
W = \log Z &= -\frac{1}{2}\int d^3x~\sqrt{g_3} \left[ \tilde P_1(T_0 e^{-\sigma}) R + T_0^2 \tilde P_2(T_0 e^{-\sigma}) f_{ij}f^{ij} 
+\tilde P_3(T_0 e^{-\sigma})( \partial \sigma)^2\right] \\
\text{where} &~~\tilde P_i(T_0 e^{-\sigma}) = P_i(\sigma)~~\text{and}~~ P_i' \equiv \frac{d P_i(\sigma)}{d\sigma}~~~(i = 1,2,3)
\end{split}
\end{equation}

where $P_1,P_2,P_3$ are three arbitrary function of $\sigma$ and from now on we will remove
the explicit dependence. In partition function, the fourth scalar $\nabla^2 \sigma$ and the
pseudo-scalar $\epsilon_{ijk}\partial^{i}\sigma f^{jk}$ do not appear as they are total
derivatives.

With the action \eqref{cfacnu} in hand it is straightforward to use analog of
\eqref{stcurrent} for uncharged case\footnote{
The stress tensor can be evaluated as
\begin{eqnarray}
T_{00}&=& -\frac{T_0 e^{2 \sigma}}{\sqrt{-g_{(p+1)}} }\frac{\delta W}{\delta \sigma},
\quad T_0^i= \frac{T_0}{\sqrt{-g_{(p+1)}} }\frac{\delta W}{\delta a_i}, \nn \\
T^{ij}&=& -\frac{2 T_0}{\sqrt {-g_{(p+1)}}} g^{il}g^{jm}\frac{\delta W}
{\delta g^{lm}}.
\end{eqnarray} } to obtain the equilibrium stress tensor. We find
\begin{eqnarray}\label{stcuchactionu}
T^{ij}&=& T P_1 \big(R^{ij}-\frac{1}{2}R g^{ij}\big)+ 2 T_0^2 T P_2 \big(f^{ik}f_{jk}-\frac{1}{4}f^2 g^{ij}\big) +T
(P_3-P_1'')\big(\nabla^i \sigma\nabla^j\sigma \nn \\
&-&\frac{1}{2}(\nabla \sigma)^2g^{ij}\big)-T P_1' \big(\nabla^i\nabla^j \sigma-g^{ij}\nabla^2\sigma \big) + \frac{1}{2}T P_1''(\nabla \sigma)^2
g^{ij} \nn \\
T_{00}&=& \frac{T_0^2} {2T}\big(P_1' R + T_0^2 P_2' f^2 - P_3' (\nabla \sigma)^2- 2 P_3\nabla^2 \sigma)\big) \nn \\
T_0^i&=&2 T_{0}^2 T\big(P_2'\nabla_j\sigma f^{ji}+ P_2 \nabla_j f^{ji}\big),
\end{eqnarray}where $'$ denotes derivative with respect to $\sigma$.

\subsection{Constraints on Hydrodynamics}
Comparing non trivial components of the stress tensor $T_0^i,$ $T_{00}$  in equations
\ref{ccvmfu},\ref{stcuchactionu} and   equating coefficients of independent sources one
obtains the velocity and temperature corrections in terms of the coefficients $P$ appearing
in \ref{cfacnu}. We find
\begin{equation}\begin{split}
v_1 &= -\frac{2T^2}{P_T}P'_2 ,~~  v_2 = -\frac{2T^2}{P_T}P_2,~~    \tilde{v}=0,\\
t_{1}&=\frac{1}{2 P_{TT}}P'_1,~~t_{2} = -\frac{1}{ P_{TT}}P_3,~~t_{3}=-\frac{1}{2 P_{TT}}P'_3,~~t_{4}= \frac{T^2}{2 P_{TT}}P'_2,~~\tilde{t}=0.
  \end{split}
\label{vtcor}
\end{equation}

Now comparing $T_{ij}$  in equations  \ref{ccvmfu},\ref{stcuchactionu},
and using expressions for temperature corrections, one can express the
transport coefficients in terms of the  three coefficients $P$ appearing in \ref{cfacnu}.
We find
\begin{equation}\begin{split}
  a_1&= T P_1,~~ a_{2} = -T P'_1 \quad a_4=- 2 T^3 P_2, \quad a_3= T (P_3- P_1''),\\
 b_{1}&=-\frac{P_{T}}{2 P_{TT}}P'_1 ,~~b_{2} = \frac{P_{T}}{ P_{TT}}P_{3},~~b_{4} = - \frac{P_{T} T^2}{2 P_{TT}}P'_2,\\
b_{3} &= \half T P''_{1}+\frac{P_{T}}{2 P_{TT}}P'_{3} , \,\,\ \delta_1=\delta_4=\delta_5=0   .
\end{split}
\label{Intrel1}
 \end{equation}
One can eliminate the coefficients $P's$ from above set of relations which gives five relations among transport coefficients,
\begin{eqnarray}\label{Intrel}
&&a_1+a_2-T \partial_T a_1 =0, \quad \frac{T P_{TT}}{P_T} b_1 +\frac{1}{2}(a_1 -T \partial_T a_1)  =0 \nn \\
&&\frac{T P_{TT}}{P_T} b_2 +(a_2 -T \partial_T a_2) -a_3 =0, \quad 4 \frac{T P_{TT}}{P_T}b_4-(3 a_4 - T \partial_T a_4)=0, \nn \\
&&2 \frac{T P_{TT}}{P_T} b_3+\big(\frac{T P_{TT}}{P_T} +1\big)(a_2 -T \partial_T a_2)
-T \partial_T(a_2 -T \partial_T a_2)-(a_3 -T \partial_T a_3)=0 .\nn \\
\end{eqnarray}

Note that parity odd contributions, both to the equilibrium value of the temperature
and velocity, as well as to the constitutive relations, are forced to vanish.
The simple reason for this is that the most general two derivative correction to the
partition function \ref{cfacnu} is parity even. Note also that the
eight parity even non dissipative transport coefficients are all determined
in terms of the three functions that parameterize the two derivative partition function.
This leaves us five relations among the transport coefficients; these relations may be
obtained by substituting the definitions of the $a$ and $b$ coefficients in
 \eqref{gipi} into \eqref{Intrel}; we find
\begin{equation}\label{relationsintro}
  \begin{split}
  \kappa_2 =&~ \kappa_1 + T\frac{d\kappa_1}{dT}\\
  \zeta_2 =&~ \frac{1}{2}\left[s\frac{d\kappa_1}{ds} - \frac{\kappa_1}{3}\right]\\
  \zeta_3 = &\left(s\frac{d\kappa_1}{ds} + \frac{\kappa_1}{3}\right) + \left(s\frac{d\kappa_2}{ds} - \frac{2\kappa_2}{3}\right)+\frac{s}{T}\left(\frac{dT}{ds}\right)\lambda_4\\
  \\
  \xi_3=&~\frac{3}{4}\left(\frac{s}{T}\right)\left(\frac{dT}{ds}\right)\left(T\frac{d\kappa_2}{dT} + 2\kappa_2\right) -\frac{3\kappa_2}{4} +\left(\frac{s}{T}\right)\left(\frac{dT}{ds}\right)\lambda_4 \\
  &+\frac{1}{4}\left[s\frac{d\lambda_3}{ds} + \frac{\lambda_3}{3} -2 \left(\frac{s}{T}\right)\left(\frac{dT}{ds}\right)\lambda_3\right]\\
  \xi_4 =&~-\frac{\lambda_4}{6} - \frac{s}{T}\left(\frac{dT}{ds}\right)\left(\lambda_4 + \frac{T}{2}\frac{d\lambda_4}{dT}\right)
  -T\left(\frac{d\kappa_2}{dT}\right)\left(\frac{3s}{2T}\frac{dT}{ds} - \frac{1}{2}\right) \\
  &- \frac{Ts}{2} \left(\frac{dT}{ds}\right)\left(\frac{d^2\kappa_2}{dT^2}\right)
  \end{split}
  \end{equation}
This is in perfect agreement with the relations obtained in \cite{Bhattacharyya:2012ex} using the second
law of thermodynamics.

\subsection{The Entropy Current}

The entropy of our system is given by
\begin{eqnarray}\label{fentu}
S&=&\frac{\partial}{\partial T_0}(T_0 \log Z) \nn \\
\end{eqnarray}
The partition function of our system is given by
\begin{equation}\label{partotion2}
\begin{split}
\log Z = -\frac{1}{2}\int d^3x~\sqrt{g_3}\left[ \tilde P_1(T_0 e^{-\sigma}) R + T_0^2 \tilde P_2(T_0 e^{-\sigma}) f_{ij}f^{ij} +\tilde P_3(T_0 e^{-\sigma})( \partial \sigma)^2\right]
\end{split}
\end{equation}
(we are careful to explicitly keep track of the temperature dependence in the partition function,
see the equation \eqref{cfacnu} for a definition of the functions ${\tilde P}$). The total entropy as
evaluated from this partition function is
\begin{equation}\label{fentu2}
\begin{split}
S&=\frac{\partial}{\partial T_0}(T_0 \log Z) \\
&=\frac{1}{2}\int \sqrt{g} \left[ (P_1' - P_1) R + T_0^2(P_2' -3P_2) f_{ij}f^{ij} + (P_3' - P_3)(\partial \sigma)^2\right]\\
\end{split}
\end{equation}

To second order in the derivative expansion, the most general symmetry allowed  entropy current is given
by \cite{Bhattacharyya:2012ex}
\begin{equation} \label{mjecu2}
 \begin{split}
 J^\mu_S =& s u^\mu + \tilde J^\mu_S\\
 \text{where}&\\
 \tilde J^\mu_S =& \nabla_\nu\left[A_1(u^\mu\nabla^\nu T - u^\nu \nabla^\mu T)\right] + \nabla_\nu \left( A_2 T \omega^{\mu\nu}\right)\\
 & + A_3 \left(\tilde R^{\mu\nu} - \frac{1}{2}g^{\mu\nu} \tilde R\right) u_\nu
+\left[ A_4 (u.\nabla)\Theta  + A_5 \tilde R + A_6 (\tilde R_{\alpha\beta}u^\alpha u^\beta)\right] u^\mu\\ &+(B_1\omega^2 + B_2\Theta^2 + B_3 \sigma^2)u^\mu + B_4\left[(\nabla s)^2
u^\mu + 2 s \Theta \nabla^\mu s\right]\\
&+\left[\Theta \nabla^\mu B_5 - P^{ab}(\nabla_b u^\mu)( \nabla_a B_5 )\right]+ B_6 \Theta {\mathfrak a^\mu} + B_7 {\mathfrak a}_\nu \sigma^{\mu\nu}
 \end{split}
\end{equation}

The terms above with $A_1$ and $A_2$ as coefficients are total derivative and do contribute
to the total entropy. It follows that $A_1$ and $A_2$ are unconstrained by comparison with
equilibrium (even though these terms do not pointwise vanish in equilibrium). Terms with
coefficients $A_4,~B_2,~B_3,~B_6$ and $B_7$ vanish on the equilibrium solution. Consequently
these coefficients are also unconstrained by the considerations of this section.
The entropy current coefficients that can be are constrained by comparison with
\eqref{fentu2} are
$A_3,~A_5,~A_6,~B_1,~B_4$ and $B_5$

As above, there are two sources for the second order correction to the entropy of our system.
The $su^\mu$ part in $J^\mu_S$ contributes to the total entropy at second order in derivative expansion because of the  second order corrections $\delta u^\mu$ to the equilibrium velocity $u^\mu$
and $\delta T$ to the equilibrium temperature.More precisely, if the equilibrium temperature and velocity of our system to second order is given by
$$T = T_{(0)} + \delta T = T_0 e^{-\sigma} + \delta T ~~\text{and}~~u^\mu = u^\mu_{(0)} + \delta u^\mu  = e^{-\sigma }(1,0,0,0) + \delta u^\mu $$ then clearly 
$$ s u^0|_\text{2nd order} = e^{-\sigma}\left(\frac{ds}{dT} \right)\delta T + s \delta u^0$$.

Using \eqref{vtcor} and \eqref{vcuf} we find
\begin{equation}\label{repeat}
\begin{split}
\left(\frac{ds}{dT} \right)\delta T &= \frac{1}{2}\left[P_1'~ R + P_2'~ T_0^2 f^2 - P_3'~ (\partial \sigma)^2 -2P_3 ~\nabla^2\sigma\right]\\
&=\frac{1}{2}\left[P_1'~ R + P_2'~ T_0^2 f^2 +P_3'~ (\partial \sigma)^2 -2 \nabla_i \left(P_3 \nabla^i \sigma\right)\right]\\
\\
s  \delta u^i &= -2e^{-\sigma}T_0^2\left[P_2'~ \nabla_j \sigma f^{ji} + P_2~ \nabla_j f^{ji}\right] = -2 T_0^2 e^{-\sigma}\nabla_j  \left(P_2 ~ f^{ji}\right)
\end{split}
\end{equation}
Therefore using \ref{norfixing}, the second order correction to $J^0_S$, from the perfect fluid piece
$s u^0$,  evaluates to
\begin{equation}\label{pratham}
\begin{split}
s u^0|_\text{2nd order} =\frac{ e^{-\sigma}}{2}\left[P_1'~ R +( P_2'-2 P_2)~ T_0^2 f^2 +P_3'~ (\partial \sigma)^2 \right] + e^{-\sigma} \nabla_j\left[2 T_0^2 P_2~f^{ji}a_i - P_3\nabla^j\sigma\right]
\end{split}
\end{equation}

The second source of two derivative corrections to the entropy current come from the
explicit two derivative corrections to the entropy current \eqref{mjecu2} evaluated on
the perfect fluid equilibrium configurations. Using
\begin{equation}\label{byabohar}
\begin{split}
&f^2\equiv f_{ij}f^{ij}\\
&P^{\mu a}\nabla_a u^\nu = \sigma^{\mu\nu} + \omega^{\mu\nu} + P^{\mu\nu}\frac{\Theta}{3}\\
&\tilde R = R - 2 (\partial \sigma)^2 - 2 \nabla^2\sigma + \frac{e^{2\sigma}}{4} f^2\\
&\tilde R_{\alpha\beta} u^{\alpha} u^\beta =(\partial \sigma)^2 + \nabla^2\sigma + \frac{e^{2\sigma}}{4} f^2\\
&\tilde R^i_0 = \frac{e^{2\sigma}}{2}\left[\nabla_j f^{ji} +3( \nabla_j\sigma) f^{ji}\right]\\
&\tilde R^0_0 = -\left(e^{-2\sigma}\tilde R_{00} + a_i\tilde R^i_0\right)\\
&e^\sigma \omega ^{0i}\partial_i T = -\frac{T e^{2\sigma}}{2}(\partial _i \sigma)f^{ji}a_j
\end{split}
\end{equation}
we find that the zero component of $\tilde J^\mu_S$ evaluates on equilibrium to
\begin{equation}\label{mjcueq}
\begin{split}
 \tilde J_S^0
=&e^{-\sigma}\bigg[ A_3 \left(\tilde R^0_0 - \frac{\tilde R}{2}\right)
+  A_5 \tilde R + A_6 (\tilde R_{00}e^{-2\sigma})
+B_1\omega^2 + B_4(\partial s)^2
  +e^{\sigma} \left(\frac{d B_5}{dT} \right)\omega^{0 i}(\partial_i T)\bigg]\\
  \\
  =& e^{-\sigma}\bigg[\left(A_5-\frac{A_3}{2}\right)R +\left(\frac{2A_5 + 2A_6 -3A_3 - 2B_1}{8} \right) e^{2\sigma}f^2+T^2\left(\frac{ds}{dT}\right)^2B_4 (\partial\sigma)^2\\
  &+(A_6- 2 A_5)\left[\nabla^2\sigma+(\partial\sigma)^2\right]
  - \frac{A_3 e^{2\sigma}}{2}a_i\nabla_jf^{ji} - \frac{\left(3 A_3-T\frac{dB_5}{dT} \right)e^{2\sigma}}{2}a_if^{ji}\partial_j\sigma\bigg]\\
  \\
 =& e^{-\sigma}\bigg[\left(A_5-\frac{A_3}{2}\right)R + \left(\frac{2A_5 + 2A_6 -A_3 - 2B_1}{8} \right) e^{2\sigma}f^2\\
 &~~ +\left[\left(T\frac{ds}{dT}\right)^2B_4+T\frac{d}{dT}(A_6 - 2 A_5)\right](\partial\sigma)^2 + \frac{T}{2}\left(\frac{dB_5}{dT} -\frac{A_3}{T} - \frac{dA_3}{dT}\right)a_if^{ji}\partial_j\sigma\\
 &~~~~-\frac{1}{2}\nabla_j \left(A_3 e^{2\sigma}a_i f^{ji}\right) + \nabla_i\left[(A_6 - 2 A_5) \nabla^i\sigma\right]\bigg]
\end{split}
\end{equation}

Summing \eqref{pratham} and \eqref{mjcueq} and ignoring total derivatives, we find our
final result for the two derivative correction to the total entropy.
\begin{equation}\label{finent}
\begin{split}
&\text{Total Entropy}\\
=& \int d^3x \sqrt{g_3} \bigg[ \left(A_5-\frac{A_3}{2} +\frac{P_1'}{2}\right)R + \left[\frac{2A_5 + 2A_6 -A_3 - 2B_1 +T_0^2 (4 P_2' - 8 P_2)e^{-2\sigma}}{8} \right] e^{2\sigma}f^2\\
 & +\left[\left(T\frac{ds}{dT}\right)^2B_4+T\frac{d}{dT}(A_6 - 2 A_5) +\frac{P_3'}{2}\right]
(\partial\sigma)^2 + \frac{T}{2}\left(\frac{dB_5}{dT} -\frac{A_3}{T} - \frac{dA_3}{dT}\right)a_i
f^{ji}\partial_j\sigma \bigg]\\
\end{split}
\end{equation}

While the first three terms in \eqref{finent} are Kaluza Klein gauge invariant, the last
term is not. Let us pause, for a moment to explain this. In subsubsection \ref{cce} we
have demonstrated that the integral $\int \sqrt{-g_4} J_S^0$ is Kaluza Klein gauge invariant
{\it provided} that $\partial_\mu J_S^\mu =0$. Now it must certainly be true that the correct
entropy current is divergence free in equilibrium. However the most general entropy current
\eqref{mjecu2} is not divergence free in equilibrium. The non gauge invariant term
ion \eqref{finent} results from such terms. The coefficients of these terms must immediately
be set to zero (even without comparison with a particular form of the entropy). The coefficients
of the remaining three terms in \eqref{finent} must be equated with the coefficients of
the corresponding terms in  \eqref{fentu2}. In net we have four equations which allow us to solve
for four of the entropy current coefficients, $B_5,~~A_3,~~B_1$ and $B_4$ in terms of the
other two ($A_5$ and $A_6$)  and $P_i$ (the coefficients that
appear in the partition function ie.the  $P_i$ ).

\begin{equation}\label{compans}
\begin{split}
&\frac{dB_5}{dT} =\frac{A_3}{T}+\frac{dA_3}{dT}\\
&A_3 = P_1 + A_5\\
&B_1 = -\frac{P_1}{2} + 2 T_0^2 e^{-2\sigma} P_2 + A_5 + A_6\\
&\left( T\frac{ds}{dT}\right)^2B_4 = -\frac{P_3}{2} -T\frac{d}{dT}(A_6 - 2 A_5)
\end{split}
\end{equation}

\subsubsection{Entropy current with non-negative divergence}

Above we have discussed the constraints on the entropy current from
comparison with the total entropy of our system. In this subsubsection
we will discuss the relationship between these constraints and those obtained
by imposing the requirement of positivity of the entropy current.

In the study of the positivity of the divergence of the entropy current, it turns
out that some coefficients in the entropy current are determined in terms of
transport coefficients, while others are left free (more precisely these coefficients
are constrained by inequalities involving transport coefficients). The
determined coefficients turn out to be precisely those that multiply terms
that are nonvanishing in equilibrium, namely
$A_3,~A_5,~A_6,~B_1,~B_4$ and $B_5$. The six equations that determine these six parameters
are
\begin{equation}\label{purano}
\begin{split}
A_5&=0\\
A_6&=0\\
\frac{dB_5}{dT} &=\frac{A_3}{T}+\frac{dA_3}{dT}\\
A_3 &= \kappa_1\\
B_1 &= \frac{1}{4}\left[-\lambda_3+ T\frac{d\kappa_1}{dT} + \kappa_1\right]\\
\left(T\frac{ds}{dT}\right)^2 B_4 &= -\frac{1}{2}\left[\lambda_4 + 2 T\frac{d\kappa_1}{dT} + T^2 \frac{d^2\kappa_1}{dT^2}\right]
\end{split}
\end{equation}
The results \eqref{purano} satisfy the constraints \eqref{compans}. In order to verify this one
plugs in explicit results
\begin{equation}\label{transco}
\begin{split}
\xi_3 &= -\frac{P_1}{2}  + \frac{2}{3}(P_1' - T_0^2  e^{-2\sigma} P_2) +\left(\frac{s}{T} \frac{dT}{ds}\right)\left(2 T_0^2  e^{-2\sigma} P_2' + P_3 -\frac{3}{2}P_1'\right)\\
\xi_4 &=\frac{2}{3}\left(P_1'' - P_1'-\frac{P_3}{4}\right) +\left(\frac{s}{T} \frac{dT}{ds}\right)\left(\frac{P_3'}{2} -P_3\right)\\
\zeta_2 &= -\frac{P_1}{6} -\left(\frac{s}{T} \frac{dT}{ds}\right)\frac{P_1'}{2}\\
\zeta_3 &=\frac{2 P_1' - P_1}{3} +\left(\frac{s}{T} \frac{dT}{ds}\right)\left(P_3 - P_1'\right)\\
\lambda_3 &= 3 P_1 - 8 T_0^2  e^{-2\sigma} P_2 - P_1'\\
\lambda_4 &= P_3 + P_1' - P_1''\\
\kappa_2 &= P_1 - P_1'\\
\kappa_1 & = P_1
\end{split}
\end{equation}
for the transport coefficients in terms of action parameters
into \eqref{purano} and checks that the results are consistent with \eqref{compans}

Our results \eqref{compans} are compatible with but weaker than \eqref{purano}.
\eqref{purano} is equivalent to \eqref{compans} together with $A_5=A_6=0$. As $A_5$ and
$A_6$ multiply terms that are nonvanishing in equilibrium, we find it surprising that '
our equilibrium study has not been powerful enough to demonstrate that $A_5$ and $A_5$
must actually vanish. It is possible that we have overlooked a simple principle that
forces these coefficients to vanish without invoking the principle of entropy increase.

\subsection{The conformal limit}
\footnote{This subsection has been worked out in
collaboration with R. Loganayagam.}Let us consider Weyl transformation of the full four dimensional metric 
$$\bar g_{\mu\nu} = g_{\mu\nu} e^{2\phi(x)}.$$
In this subsection first we would like to write an partition function
 which is invariant under this transformation. In order to have conformal invariance this partition
function will have fewer coefficients than the partition function given in \eqref{cfacnu}. Then we shall 
analyze how it will constrain the stress tensor for a conformal fluid.

Under this transformation several three dimensional quantities transform as follows.
\begin{equation}\label{threetran}
 \begin{split}
  \bar \sigma &= \sigma +\phi ,~~
\bar a_i = a_i,~~
\bar g_{ij} = e^{2\phi} g_{ij}\\
(\nabla \bar\sigma)^2 &= e^{-2\phi}\left[(\nabla\sigma)^2 + 2 (\nabla\sigma).(\nabla\phi)
+ (\nabla\phi)^2\right]\\
\bar R &= e^{-2\phi}\left[R - 4 \nabla^2\phi - 2 (\nabla\phi)^2\right]\\
\bar f_{ij}\bar f^{ij} &= e^{-4\phi} f_{ij}f^{ij}\\
\sqrt{\bar g_3} &= e^{3\phi}\sqrt{g_3}\\
 \end{split}
\end{equation}

Using \eqref{threetran} we can see that under this transformation
the partition function (as given in\eqref{cfacnu}) will be invariant
 (assuming that the total derivative terms will integrate to zero)
only if the coefficients $P_i$'s satisfy the following constraints.

\begin{equation}\label{conconstr}
 \begin{split}
  P_1(\sigma) &= e_1 T_0 e^{-\sigma},~~~ P_2(\sigma) = \frac{e_2}{T_0 e^{-\sigma}}~~\text{and}~~
P_3(\sigma) = 2 P_1(\sigma)
 \end{split}
\end{equation}
where $e_1$ and $e_2$ are two dimensionless constants.

Substituting \eqref{conconstr} in \eqref{transco} we find
\begin{equation}\label{finconcon}
 \begin{split}
  &\xi_3 =\xi_4 =\zeta_2 =\zeta_3 = \lambda_4 =0\\
&\kappa_2 = 2\kappa_1=2 e_1 T_0 e^{-\sigma}\\
&\lambda_3 = 4 T_0 e^{-\sigma}(e_1 - 2 e_2)
 \end{split}
\end{equation}

These relations precisely match with our expectation for the independent transport
coefficients of a conformally covariant stress tensor.
Since for a conformally covariant stress tensor only two terms
($\omega_{\langle\mu a}{\omega^a}_{\nu\rangle}$ with coefficient $\lambda_3$ and
$\left[R_{\langle\mu\nu\rangle} + K_{\langle\mu\nu\rangle}\right]$ with coefficient $\kappa_1$)
can be non zero in equilibrium
and a conformally invariant action also has only two free parameters, it follows that the existence
of a partition function does not constrain the stress tensor of a conformal fluid.

\section{Counting for second order charged fluids in 3+1 dimensions}\label{count}

In this section we will use the methods developed in this paper to answer the following
question: how many transport coefficients are needed to specify the fluid dynamics of
a relativistic charged fluid that may not preserve parity, at second order in the derivative
expansion? We do not attempt to derive the detailed form of the equations so obtained; our
presentation is merely at the level of counting. If the conjecture at the heart of this paper
is correct, then an analysis of entropy positivity would yield the same number of transport
coefficients; however that analysis is much more difficult to perform (even at the level
of counting), and we do not attempt it here.

\subsection{Parity Invariant case}
Let us first consider the parity invariant case. Table \ref{soped} list the number of
all the the parity preserving fluid plus background onshell independent data at second order.
\begin{table}
\centering
\begin{tabular}[h]{|c|c|c|}
\hline
Type & fluid+background data & In equilibrium \\
\hline
scalars &  16  &  9 \\
\hline
vectors &  17  &  6 \\
\hline
tensors &  18  &  9 \\
\hline
\end{tabular}
\label{soped}
\caption{parity even data for charged fluids at second order }
\end{table}
From this table this
we see that the total number of symmetry allowed transport coefficients in stress-energy
tensor and charge current in landau frame is
\begin{equation}\label{sopic}
{\rm tensors}(16) + {\rm scalars}(18) + {\rm vectors}(17) = 51 .
\end{equation}
Now let us consider the equilibrium of this system. The third column of table \ref{soped}
also list the number of scalars, vectors and tensors that can be constructed out of
$\sigma$, $A_0$, $a_i$, $A_i$ and $g^{ij}$. The coefficient of these terms that are
survive in equilibrium we refer to as `non dissipative' coefficients while the remaining
we refer to as `dissipative' coefficients. In this case we have a total of 24 non dissipative
coefficients. Now there are 9 scalars than can be constructed in equilibrium. We list them below
\begin{equation}
R_i^i,~~ \nabla^i \sigma \nabla_i \sigma,~~ f_{ij}f^{ij},~~ F_{ij}F^{ij},~~ F_{ij}f^{ij},~~ \nabla^i \sigma \nabla_i A_0,~~
\nabla^i A_0 \nabla_i A_0,~~ \nabla^i\nabla_i \sigma,~~ \nabla^i\nabla_i A_0
\end{equation}
The last two scalars are total derivatives and hence do not appear in the partition function.
This tell us that the 24 non dissipative coefficients are determined in term of 7 independent
coefficients that appear in the partition function which means that there will be 17 relation
among the 24 non dissipative coefficients.

In summary the methods developed in this paper predict that parity invariant charged
fluid dynamics is characterized by 7 non dissipative transport coefficients, together
with 28 dissipative coefficients (7 scalars, 12 vectors and 9 tensors).  Each of these
35 transport coefficients is an unspecified function of $T$ and $\mu$.

\subsection{Parity Violating case}
Let us now consider the parity non invariant charged fluids at second order. Table \ref{sopod}
lists all the parity odd data at second order.
\begin{table}
\centering
\begin{tabular}[h]{|c|c|c|}
\hline
Type & fluid+background & In equilibrium\\
\hline
pseudo scalars &  6  &  4 \\
\hline
pseudo vectors &  9  &  2 \\
\hline
pseudo tensors &  12  &  6\\
\hline
\end{tabular}
\label{sopod}
\caption{parity odd data for charged fluid at second order}
\end{table}
From this table we see that number of transport coefficients in the parity odd sector is
\begin{equation}
{\rm pseudo~tensors}(12) +{\rm pseudo~scalars}(6) + {\rm pseudo~vectors}(9) = 27 .
\end{equation}
The third column of table \ref{sopod} that out of the 28 parity odd transport coefficients 
12 are non dissipative. 

Now we can construct four new scalars (pseudo) out of the sources.
These are listed below
\begin{equation}\label{ps2nd}
\epsilon^{ijk}\partial_i\sigma f_{jk}~,~~ \epsilon^{ijk}\partial_i A_0 f_{jk}~,~~
\epsilon^{ijk}\partial_i\sigma F_{jk}~,~~ \epsilon^{ijk}\partial_i A_0 F_{jk}
\end{equation}
As such all of
these scalars listed in \eqref{ps2nd} are total derivatives by themselves 
but only two of them can actually be 
written as 
total derivatives
in the partition function since the coefficients that they will appear 
 with are arbitrary functions of $\sigma$ and $A_0.$

Any linear combination of the first two scalars can be rearranged as a total derivative term
 and another term 
that can not be written as total derivative.

\begin{equation*}
 \begin{split}
 &K_1(A_0,\sigma) \epsilon^{ijk}\partial_i\sigma f_{jk}
+ K_2(A_0,\sigma)\epsilon^{ijk}\partial_i A_0 f_{jk}\\ 
=&~ \epsilon^{ijk}(\partial_i K )f_{jk} 
+\left(K_2(A_0,\sigma) - \frac{\partial K}{\partial A_0}\right)\epsilon^{ijk}\partial_i A_0 f_{jk}\\
=&~ \nabla_i\left[\epsilon^{ijk} K f_{jk} \right]
+\left(K_2(A_0,\sigma) - \frac{\partial K}{\partial A_0}\right)\epsilon^{ijk}\partial_i A_0 f_{jk}\\
&\text{where}\\
&K = \int d\sigma K_1(A_0, \sigma)
 \end{split}
\end{equation*}
Similar manipulation can be done for the last two scalars listed in \eqref{ps2nd}.

Thus we see that the 12 parity odd non dissipative coefficients are determined in terms
of $2$ parity odd coefficients in the partition function which means that their would
be $10$ relations in parity odd sector.

In summary we predict that, at second order, we have 4 parity odd nondissipative transport
coefficients, together with 2 pseudo scalar, 7 pseudo vector and 6 pseudo tensor dissipative
coefficients, and total of 20 new coefficients.

\section{Open Questions}

The results reported in this paper suggest several natural follow up questions. In this
section we list and discuss some of these questions, leaving an attempt to answer them
to future work.

The main result of our paper is that two apparently different physical requirements,
namely the requirement of existence of equilibrium in appropriate circumstances
and the requirement of the existence
of a point wise positive divergence entropy current, give the same constraints \footnote{
We ignore the inequalities that follow from the principle of entropy increase in this
statement.} on the equations of hydrodynamics in three specific contexts. Two questions
immediately suggest themselves. Do the results of our paper extend to arbitrary order in the
derivative expansion, as we have conjectured in this paper? If so, why is this the case?
Definitive answers to these questions would be very interesting. A proof that 
the 
existence of equilibrium plus certain
inequalities imply the existence of a positive divergence entropy current
could demystify arguments based on the existence of an entropy current,
and lead  towards a fuller understanding of the second law of thermodynamics.

In the main text of this paper we have derived constraints on the 
constitutive relations of hydrodynamics starting from the assumption of 
the existence of a partition function. In the appendices to this paper 
we have, however, demonstrated that all the constraints derived in this paper
may also be derived from the weaker assumption that fluid admit stationary 
equilibrium configurations in stationary backgrounds. The integrability 
conditions from the demand that the currents and stress tensors 
in equilibrium follow from an action turned out to be automatic in the 
three examples studied in this paper. Is this always the case (we find this
unlikely). In appropriate situations, do the Onsager relations follow from 
the demand that equilibrium is generated from a partition function?

Hydrodynamical equations that do not obey the constraints described in this paper are unphysical. 
Nonetheless these equations are well posed as partial differential equations. As 
a mathematical curiosity one could study such equations in their own right. What is the 
long time behaviour of a generic fluid flow with such equations? As equilibrium does not 
generically exist, flows must, presumably, continue to slosh around in a roughly oscillatory 
manner in the long time limit. It seems clear that such indefinite oscillatory behaviour 
is inconsistent with the second law of thermodynamics, in agreement with the results 
and conjectures of this paper\footnote{We thank T. Takayanagi for discussions on this issue. }.

In another direction, the analysis of this paper has led to the
consideration of partition functions dual to equilibrium hydrodynamics as a function of
background metrics and gauge fields. Given a partition
function as a function of sources, it is standard in quantum field theory to Legendre
transform this object in order to obtain an offshell 1PI effective action for the theory.
It may be possible implement this procedure on our partition function
to obtain an offshell action for fluid dynamics (albeit one applies only to equilibrium
configurations) (see \cite{Dubovsky:2011sj} for related work). If so, what is the interpretation
of this action in the context of the fluid gravity map of the AdS/CFT 
correspondence?

We find it interesting that we have been able to encapsulate the effect of 
the chiral anomaly in a $3+1$ dimensional fluid in a {\it local} contribution
to the partition function. It should be possible to generalize this term 
to capture the effects of anomalies in $2n$ dimensions for an arbitrary 
integer $n$ and thereby reproduce the results of \cite{Loganayagam:2011mu}. 

It would be very interesting to generalize the work presented in this paper away from equilibrium.
Time dependent partition functions are not in general local functionals of their sources.
These partition functions are, however usually generated by coupling local field theory
dynamics to sources. Can time dependent correlators (perhaps in a Schwinger - Keldysh set up)
be generated by minimally coupling a local `action' for hydrodynamics to the background
metric or gauge field? How does this tie in with the fluid gravity map
of the AdS/CFT correspondence?

Apart from the traditional requirement of positivity of the entropy current, and the
requirement of the existence of equilibrium, emphasized in this paper, one may also attempt
to constrain the equations of fluid dynamics by demanding that correlation functions computed
from these equations obey all the symmetry properties that follow from the existence of an
underlying action (see e.g. \cite{Jensen:2011xb}).
Any system that posseses a well defined partition function, as studied
in this paper, automatically obeys all these symmetry properties for time independent
correlators. Do the constraints on hydrodynamics that follow from the existence of an
equilibrium partition function automatically also guarantee that the symmetry requirements
on time dependent correlators are also met?

It would be interesting to investigate whether the methods and results of this paper carry over
to the study of relativistic superfluid hydrodynamics, and in particular whether they can
be used to rederive the results of \cite{Herzog:2011ec,Bhattacharya:2011eea,Lin:2011aa,Bhattacharya:2011tra,Neiman:2011mj} for the most
general allowed form of the equations of superfluid hydrodynamics at first order.

Finally, it would be very interesting to investigate the interplay of the principal
constraint described in this paper (namely the existence of equilibrium for an arbitrary
static metric) with the AdS/CFT correspondence. Is this constraint merely from the
structure of AdS/CFT, for an arbitrary bulk Lagrangian, or does it impose constraints on
possible $\alpha'$ corrections to the equations of Einstein gravity? Within gravity can
one prove directly that the existence of equilibrium implies the existence of a
Wald entropy increase theorem (and so the existence 
\cite{Bhattacharyya:2008xc} of a positive divergence
entropy current)(see \cite{Chapman:2012my} for related discussion)?

We feel that several of these questions touch on very interesting, and sometimes deep issues
in the study of both dissipative systems as well as gravity. We hope to report on progress
on some of these issues in the future.

\acknowledgments

We would like to acknowledge useful discussions and correspondences with
S.M. Bhattacharjee, K. Damle, D. Dhar, S. Dutta, G. Mandal, S. Mukherji, Y. Nakayama, A. Sen, T. Takayanagi, S. Trivedi,
S. Wadia and A. Yarom. We would also like to thank R. Loganayagam and T. Takayanagi for helpful
comments on a preliminary draft of this paper. 
S.M., S.B., N.B., S.J. and T.S. would like to acknowledge the hospitality
National Strings Meeting 2012.
The work of S.M. was supported in part by a Swarnajayanti Fellowship.
S.B., S.M., S.J. and T.S. would also like to acknowledge our debt to the
people of India for their generous and steady support to research in the basic sciences.
The work of J.B. is supported by World Premier International 
Research Center Initiative (WPI Initiative), MEXT, Japan. The work of N.B. is supported by
NWO Veni Grant.

\appendix

\section{First order charged fluid dynamics from equilibrium in 3+1 dimensions}\label{Sec:eqcal1}

In this appendix we shall rederive the results obtained in section \ref{Sec:4dcf} making fewer assumptions
than in that section. In this Appendix we make no reference to the equilibrium partition
function, and nowhere assume its existence. The only demand that we make on our system is
that it admit an equilibrium solution in an arbitrary background of the form \eqref{metgf},
\eqref{gf}. We also assume that the zeroth order equilibrium configuration is given by
equation \ref{pfeq}.

As discussed in section \ref{Sec:4dcf} for the parity violating first order charged fluid, one can not
construct any scalar or pseudo scalar at equilibrium and hence
temperature or chemical potential does not get corrected to this order.
To the first order, the velocity corrections can be written as
\begin{equation}\begin{split} \label{vce}
\delta u^i =
-\frac{e^{-\sigma} b_1}{4} \epsilon^{ijk}f_{jk} + b_2 B^i_K+b_{3} \partial^i \sigma + b_4 \partial^i A_0
\end{split}
\end{equation}

The dissipative part of the stress tensor and the current are written in
equation \ref{cr}. Since, $\partial_\alpha \frac{\mu}{T} - \frac{E_\alpha}{T}$,
$\theta,~\sigma_{\mu\nu}$ evaluate to zero on equilibrium, we are left with

\begin{equation} \label{creq}\begin{split}
\pi^{\mu\nu}&= 0 \\
J^\mu_{diss}&= \alpha_1  E^\mu + \alpha_2 {\cal P}^{\mu\alpha} \partial_\alpha T
+ \xi_\omega \omega^\mu + \xi_B B^\mu
\end{split}
\end{equation}

We shall now impose that the equations \ref{creq},\ref{vce} obeys the conservation laws
\begin{equation}\begin{split}\label{consevationlaws}
 \nabla_{\mu}T^{\mu\nu} &= {\cal F}^{\nu\lambda} \tilde{J}_{\lambda}\\
\nabla_{\mu} \tilde{J}^{\mu}&=C E.B
                \end{split}
\end{equation} where
\begin{equation}\begin{split}
 E^{\mu} &= {\cal F}^{\mu\nu}u_{\nu},~~~B^{\mu} = \frac{1}{2}\epsilon^{\mu\nu\rho\sigma}u_{\nu}{\cal F}_{\rho\sigma}  \\
\omega^{\mu}&= \frac{1}{2}\epsilon^{\mu\nu\rho\sigma}u_{\nu}\nabla_{\rho}u_{\sigma}.
                \end{split}
 \end{equation}
For computational simplicity, we shall take thermodynamic variables temperature $T$ and $\nu = \frac{\mu}{ T}$ as the independent ones.
Some useful formulas that are used in computation are
\begin{equation}
\begin{split}\label{usefulformulas}
\nabla_{\mu} \nu &= \frac{E_{\mu}}{T} ,~~  \nabla_{\mu} P =q E_{\mu}+\frac{\epsilon+P}{T} \nabla_{\mu} T \\
\frac{\partial P}{\partial T}\Big|_{\nu }  &= \frac{\epsilon+P}{T} ,~~    \frac{\partial P}{\partial \nu}\Big|_{T} =q T \\
\nabla_{\mu}\omega^{\mu} &= -\frac{2}{T}\omega_{\mu}\nabla^{\mu}T,~~\nabla_{\mu}B^{\mu} =-2\omega_{\mu}E^{\mu}-\frac{1}{T}B_{\mu}\nabla^{\mu}T.
                   \end{split}
\end{equation}
Now using formulas in equation \ref{usefulformulas}, it is straight forward to evaluate the  scalar equations namely
\begin{equation}\begin{split}\label{scalareq}
 u_{\nu}\nabla_{\mu}T^{\mu\nu} &=u_{\nu} {\cal F}^{\nu\lambda} \tilde{J}_{\lambda}\\
\nabla_{\mu}\tilde{J}^{\mu}&=C E.B .
\end{split}
\end{equation}
On setting coefficients of independent data in \ref{scalareq}, one obtains
\begin{equation}\begin{split}\label{constraint}
T\partial_{T}(\xi_\omega+q b_{1})&=2 (\xi_\omega+q b_{1}),~~T\partial_{T}(\xi_B+q b_{2})= (\xi_B+q b_{2})\\
\partial_{\nu}(\xi_\omega+q b_{1})&= 2 T (\xi_B+q b_{2}),~~\partial_{\nu}(\xi_B+q b_{2})= C T\\
\partial_{\nu}[(\epsilon+P)b_{2}] &= T (\xi_B+q b_{2}),~~\partial_{T}[(\epsilon+P)b_{2}]=\frac{2}{T}(\epsilon+P)b_{2}\\
\partial_{\nu}[(\epsilon+P)b_{1}] &= T (\xi_\omega+q b_{1})+ 2 T (\epsilon+P)b_{2},~~\partial_{T}[(\epsilon+P)b_{1}]=\frac{3}{T}(\epsilon+P)b_{1}\\
\alpha_{1}&=\alpha_{2} = b_{3}=b_{4}=0.
\end{split}
\end{equation}
The vector equation ${\cal P}_{\mu\sigma}\nabla_{\mu}T^{\mu\nu} ={\cal P}_{\mu\sigma} {\cal F}^{\nu\lambda} \tilde{J}_{\lambda}$ gives only one new
constraint, which is given by
\begin{equation}\label{constrainttrans}
 2 b_{2} = \frac{(\xi_\omega+q b_{1})}{\epsilon+P}.
\end{equation}

On solving \ref{constraint}, one obtains solution for $b's$ and $\xi's$ but with four
arbitrary constants.
 Now using \ref{constrainttrans} one can eliminate one
of the constants in terms of other one. Finally we obtain
\begin{eqnarray}\label{SSsolveq}
b_1&=&\frac{T^3}{\epsilon+P}\big(\frac{2}{3} \nu^3 C+ 2 \nu^2  z_{0} + 4 \nu  z_2 +  z_1 \big), \nn \\
b_2&=& \frac{T^2}{\epsilon+P}\big(\frac{1}{2}\nu^2 C+  \nu  z_0+ z_2\big)
\end{eqnarray} and
\begin{eqnarray}\label{SSsolceq}
a_1&=&C \nu^2 T^2 \big(1 - \frac{2 q}{3(\epsilon+P)}\nu T\big)+ T^2 \big[(2 \nu  z_0 + 2  z_2)- \frac{q T}{\epsilon+P}(2 \nu^2  z_0 + 4 \nu  z_2+ z_1)\big], \nn \\
a_2 &=& C \nu T\big(1-\frac{ q}{2(\epsilon+P)}\nu T\big)+ T \big(  z_0 -\frac{q T}{\epsilon+P}( \nu  z_0 +  z_2)\big).
\end{eqnarray}Now identifying $z_{0} = 2 C_{0},~z_{2}=C_{2}$ and $z_{1} = 4 C_{1}$ we see that equations \ref{SSsolveq}, \ref{SSsolceq} are
exactly same as equations \ref{SSsolv}, \ref{SSsolc}.

\section{First order parity odd charged fluid dynamics from equilibrium in 2+1 dimension}\label{Sec:eqcal3}

In this section we shall derive the constraints on parity odd charged fluid dynamics in 2+1 dimension at first order using just the assumption that there
exists a equilibrium solution. As discussed in \S \ref{Sec:chargedhydro3d}, in this case there are 4 transport coefficients and there
are 6 corrections to the fluid fields. In \S \ref{Sec:chargedhydro3d}, we were able to express all these 10 functions in terms of 2 arbitrary
functions in the action \eqref{action3d}. This implies that among these 10 functions
only 2 are independent which in turn implies there should exist 8 relations among these 10 functions.
In this appendix we shall present these 8 relations which follows just by demanding that there exists a equilibrium solution.

We consider the corrections to the fluid fields as in \eqref{ffcor} and write down the first order corrections to the stress tensor and the charge
current. The equation of motion of fluid dynamics are given by
\begin{equation}
\begin{split}\label{consevationlaws3d}
\nabla_{\mu}T^{\mu\nu} &= {\cal F}^{\nu\lambda}J_{\lambda}\\
\nabla_{\mu}J^{\mu} &= 0
\end{split}
\end{equation}
Note in particular that the charge current is conserved even in the presence of a
background gauge field due to the absence of any anomaly in 2+1 dimension.

Now the scalar equations $\nabla_{\mu}J^{\mu} = 0$ and $u_{\nu}^{(0)} \left( \nabla_{\mu}T^{\mu\nu} - {\cal F}^{\nu\lambda}J_{\lambda} \right) = 0$
yields the following constraints respectively
\begin{equation}\label{scaleqlconst3d}
 \begin{split}
  & \quad T \frac{\partial}{\partial T} \left(\tilde \chi_E + \rho \xi_E \right) - T \frac{\partial}{\partial \mu} \left(\tilde \chi_T + \rho \xi_T \right)
+ \mu \frac{\partial}{\partial \mu} \left(\tilde \chi_E + \rho \xi_E \right) = 0,\\
T \frac{\partial}{\partial T} &\left((\epsilon + P) \xi_E \right) - T \frac{\partial}{\partial \mu} \left((\epsilon + P) \xi_T \right)
+ \mu \frac{\partial}{\partial \mu} \left((\epsilon + P) \xi_E \right) + T (\tilde \chi_T + \rho \xi_T) = 0.
 \end{split}
\end{equation}
The vector fluid equations $ P_{\rho \nu}^{(0)} \left( \nabla_{\mu}T^{\mu\nu} - {\cal F}^{\nu\lambda}J_{\lambda} \right) = 0$, yields the
rest of the 6 constraints
\begin{equation}\label{veceqlconst3d}
 \begin{split}
  \tilde \chi_B &= \frac{\partial P}{\partial T} \tau_B + \frac{\partial P}{\partial \mu} m_B, \\
  \tilde \chi_\Omega &= \frac{\partial P}{\partial T} \tau_\Omega + \frac{\partial P}{\partial \mu} m_\Omega, \\
  (\epsilon + P) ~\xi_E &=  \frac{\partial \rho}{\partial T} \tau_\Omega + \frac{\partial \rho}{\partial \mu} m_\Omega, \\
  (\epsilon + P) ~T~ \xi_T &=  \frac{\partial \epsilon}{\partial T} \tau_\Omega + \frac{\partial \epsilon}{\partial \mu} m_\Omega, \\
  \tilde \chi_E + \rho \xi_E &= \frac{\partial \rho}{\partial T} \tau_B + \frac{\partial \rho}{\partial \mu} m_B, \\
  T~ \left( \tilde \chi_T + \rho \xi_T \right) &= \frac{\partial \epsilon}{\partial T} \tau_B + \frac{\partial \epsilon}{\partial \mu} m_B. \\
 \end{split}
\end{equation}
Note that the first two constraints in \eqref{veceqlconst3d} are identical to the constraints \eqref{Tij3d} obtained from comparison with
most general equilibrium action in \S \ref{sSec:match3d}. It is straightforward to show that the rest of the constraints in \eqref{scaleqlconst3d}
and \eqref{veceqlconst3d} are solved by the \eqref{T003d}, \eqref{Tio3d}, \eqref{J03d} and \eqref{Ji3d}.

\section{Second order uncharged fluid dynamics from equilibrium in 3+1 dimensions}\label{Sec:eqcal2}

In this appendix we will do a similar computation as done in last two appendices for 3+1 
dimensional uncharged fluids at second order. The non-trivial second order (stress tensor 
conservation equation orthogonal to fluid velocity) equation is,
\begin{equation}\label{equisol}
\nabla_i(T_2 e^{\sigma})+ e^{\sigma} \left(\sum_{n=1}^{2} v_n V^j_n+ \tilde v \tilde V\right) f_{ij}+\frac {e^{\sigma}}{P_T}\nabla_{\mu}\Pi^{\mu}_{i}=0
\end{equation}

Since, temperature correction is a scalar, we can assume the most generalized temperature correction
 to be of the following form,
\begin{equation}
T_2 e^{\sigma}=\sum_{m=1}^{4} t_m S_m + \tilde t \tilde S
\end{equation}
Four dimensional divergence can be expressed as,
\begin{eqnarray}\label{4d3dderi}
\nabla_{\mu}\Pi^{\mu}_{\nu}&=& \frac{1}{\sqrt -g_4}\partial_{\mu}
\bigg(\sqrt -g_4 \tilde g^{\mu \alpha}\Pi_{\alpha\nu}\bigg)-\frac{1}{2}\partial_{\nu}(g_{\alpha\beta})\Pi^{\alpha\beta} \nn \\
&=&\nabla^i\Pi_{ik}+\nabla^i \sigma \Pi_{ik} \nn \\
&=&(\alpha_A - T \partial_T \alpha_A)\nabla^m \sigma \Pi^A_{mi}
+ \alpha_A \nabla^m \Pi^A_{mi},
\end{eqnarray}
where, we have expressed the two derivative correction to equilibrium stress tensor $\Pi$ of \ref{gipi} in a compact form as, $(\Pi_{ij}= \alpha_A \Pi^A_{ij}, \quad A=1,11)$. Using following simple derivative formulae
\begin{eqnarray}
&&\nabla^i\big(R_{ik}-\frac{R}{2}g_{ik}\big)=0,  \quad \nabla^i(\nabla_i\nabla_k-g_{ik} \nabla^2)\sigma= \nabla^i R_{ik},\nn \\
&& \nabla^i(\nabla_i \sigma\nabla_k \sigma-g_{ik} (\nabla \sigma)^2)= \nabla^2 \sigma \nabla_k \sigma, \nn \\
&& \nabla^i\big((f_i^{\,\,\, j}f_{jk}+\frac{f^2}{4}g_{ik})e^{2 \sigma}\big)= e^{2 \sigma}\big( (\nabla^i f_{ij})f^j_{\,\,\ k}+ 2 \nabla^i \sigma (f_i^{\,\,\, j}f_{jk}+\frac{f^2}{4}g_{ik}) \big),
\end{eqnarray}
we solve for complete equilibrium solution. In the equation \ref{equisol}, we get following three different kinds of terms in parity even sector
\begin{equation}
\nabla^i (\mbox{Tensor})_{ik}, \quad \nabla_k (\mbox{Scalar}), \quad \nabla_k \sigma (\mbox{Scalar}),
\end{equation}
and following four kinds of terms in the parity odd sector,
\begin{eqnarray}
&& \epsilon^{mkl}f_{kl}f_{ji}f^j_m, \,\, \epsilon_{imn} f^{mn}\nabla^2 \sigma, \,\, \epsilon_{mni}\nabla^2 f^{mn} \nonumber \\
&& \epsilon^{mnl}\nabla_i(\nabla_m \sigma f_{nl}), \epsilon^{mnl}\nabla_i \sigma \nabla_m \sigma f_{nl}
\end{eqnarray}
Setting the coefficients of $\nabla_k (\mbox{Scalar})$ to zero, we get the temperature correction as 
\footnote{we have used $\nabla^i \sigma \nabla_i\nabla_k\sigma= \frac{1}{2}\nabla_k (\nabla \sigma)^2$.},
\begin{equation}\label{tempcorr}
t_1= -\frac{b_1}{P_T}, \,\ t_2=-\frac{b_2}{P_T}, \,\ t_4=-\frac{b_4}{P_T}, \,\ t_3=-\frac{b_3+\frac{1}{2}(a_2-T \partial_T a_2)}{P_T}.
\end{equation}
Setting the coefficients of the other terms to zero and using \ref{tempcorr}, we get, the velocity corrections as
\begin{equation}
v_1=\frac{3 a_4-T \partial_T a_4}{T^2 P_T}, \quad v_2= \frac{a_4}{T^2 P_T}
\end{equation}
and the relations among the transport coefficients as given in \ref{Intrel}.
Similarly, setting the coefficients of the independent terms in the parity odd sector to zero, we get all parity odd coefficients zero, that is $$ \tilde t=\tilde v= \delta_1= \delta_4= \delta_5 =0. $$

\bibliographystyle{JHEP}
\bibliography{fdpf6july}

\providecommand{\href}[2]{#2}\begingroup\raggedright\begin{thebibliography}{10}

\bibitem{Landau:1987gn}
L.~Landau and E.~Lifshitz, {\it {Textbook on Theoretical Physics. VOL. 6: Fluid
  Mechanics}}, .

\bibitem{putterman}
S.~Putterman, {\it {Superfluid hydrodynamics}}, .

\bibitem{Son:2009tf}
D.~T. Son and P.~Surowka, {\it {Hydrodynamics with Triangle Anomalies}},  {\em
  Phys.Rev.Lett.} {\bf 103} (2009) 191601
  [\href{http://arXiv.org/abs/0906.5044}{{\tt 0906.5044}}].

\bibitem{Bhattacharya:2011tra}
J.~Bhattacharya, S.~Bhattacharyya, S.~Minwalla and A.~Yarom, {\it {A Theory of
  first order dissipative superfluid dynamics}},
  \href{http://arXiv.org/abs/1105.3733}{{\tt 1105.3733}}.

\bibitem{Bhattacharya:2011eea}
J.~Bhattacharya, S.~Bhattacharyya and S.~Minwalla, {\it {Dissipative Superfluid
  dynamics from gravity}},  {\em JHEP} {\bf 1104} (2011) 125
  [\href{http://arXiv.org/abs/1101.3332}{{\tt 1101.3332}}].

\bibitem{Herzog:2011ec}
C.~P. Herzog, N.~Lisker, P.~Surowka and A.~Yarom, {\it {Transport in
  holographic superfluids}},  {\em JHEP} {\bf 1108} (2011) 052
  [\href{http://arXiv.org/abs/1101.3330}{{\tt 1101.3330}}].

\bibitem{Neiman:2011mj}
Y.~Neiman and Y.~Oz, {\it {Anomalies in Superfluids and a Chiral Electric
  Effect}},  {\em JHEP} {\bf 1109} (2011) 011
  [\href{http://arXiv.org/abs/1106.3576}{{\tt 1106.3576}}].

\bibitem{Loganayagam:2011mu}
R.~Loganayagam, {\it {Anomaly Induced Transport in Arbitrary Dimensions}},
  \href{http://arXiv.org/abs/1106.0277}{{\tt 1106.0277}}.

\bibitem{Romatschke:2009kr}
P.~Romatschke, {\it {Relativistic Viscous Fluid Dynamics and Non-Equilibrium
  Entropy}},  {\em Class. Quant. Grav.} {\bf 27} (2010) 025006
  [\href{http://arXiv.org/abs/0906.4787}{{\tt 0906.4787}}].

\bibitem{Bhattacharyya:2012ex}
S.~Bhattacharyya, {\it {Constraints on the second order transport coefficients
  of an uncharged fluid}},  \href{http://arXiv.org/abs/1201.4654}{{\tt
  1201.4654}}.

\bibitem{Dubovsky:2011sk}
S.~Dubovsky, L.~Hui and A.~Nicolis, {\it {Effective field theory for
  hydrodynamics: Wess-Zumino term and anomalies in two spacetime dimensions}},
  \href{http://arXiv.org/abs/1107.0732}{{\tt 1107.0732}}. 9 pages.

\bibitem{st}
S.~Minwalla, {\it {A Framework for Superfluid Hydrodynamics}},  {\em Talk at
  Strings 2011, Upsalla}.

\bibitem{Bhattacharyya:2008jc}
S.~Bhattacharyya, V.~E. Hubeny, S.~Minwalla and M.~Rangamani, {\it {Nonlinear
  Fluid Dynamics from Gravity}},  {\em JHEP} {\bf 0802} (2008) 045
  [\href{http://arXiv.org/abs/0712.2456}{{\tt 0712.2456}}].

\bibitem{Hubeny:2010ry}
V.~E. Hubeny and M.~Rangamani, {\it {A Holographic view on physics out of
  equilibrium}},  {\em Adv.High Energy Phys.} {\bf 2010} (2010) 297916
  [\href{http://arXiv.org/abs/1006.3675}{{\tt 1006.3675}}].

\bibitem{Rangamani:2009xk}
M.~Rangamani, {\it {Gravity and Hydrodynamics: Lectures on the fluid-gravity
  correspondence}},  {\em Class. Quant. Grav.} {\bf 26} (2009) 224003
  [\href{http://arXiv.org/abs/0905.4352}{{\tt 0905.4352}}].

\bibitem{Bhattacharyya:2007vs}
S.~Bhattacharyya, S.~Lahiri, R.~Loganayagam and S.~Minwalla, {\it {Large
  rotating AdS black holes from fluid mechanics}},  {\em JHEP} {\bf 0809}
  (2008) 054 [\href{http://arXiv.org/abs/0708.1770}{{\tt 0708.1770}}].

\bibitem{Loganayagam:2012pz}
R.~Loganayagam and P.~Surowka, {\it {Anomaly/Transport in an Ideal Weyl gas}},
  \href{http://arXiv.org/abs/1201.2812}{{\tt 1201.2812}}.

\bibitem{Jensen:2011xb}
K.~Jensen, M.~Kaminski, P.~Kovtun, R.~Meyer, A.~Ritz {\em et.~al.}, {\it
  {Parity-Violating Hydrodynamics in 2+1 Dimensions}},
  \href{http://arXiv.org/abs/1112.4498}{{\tt 1112.4498}}.

\bibitem{Gao:2012ix}
J.-H. Gao, Z.-T. Liang, S.~Pu, Q.~Wang and X.-N. Wang, {\it {Chiral anomaly and
  local polarization effect from quantum kinetic approach}},
  \href{http://arXiv.org/abs/1203.0725}{{\tt 1203.0725}}.

\bibitem{Pu:2010as}
S.~Pu, J.-h. Gao and Q.~Wang, {\it {A consistent description of kinetic
  equation with triangle anomaly}},  {\em Phys.Rev.} {\bf D83} (2011) 094017
  [\href{http://arXiv.org/abs/1008.2418}{{\tt 1008.2418}}].

\bibitem{Neiman:2010zi}
Y.~Neiman and Y.~Oz, {\it {Relativistic Hydrodynamics with General Anomalous
  Charges}},  {\em JHEP} {\bf 1103} (2011) 023
  [\href{http://arXiv.org/abs/1011.5107}{{\tt 1011.5107}}].

\bibitem{Bardeen:1984pm}
W.~A. Bardeen and B.~Zumino, {\it {Consistent and Covariant Anomalies in Gauge
  and Gravitational Theories}},  {\em Nucl.Phys.} {\bf B244} (1984) 421.
  Revised version.

\bibitem{Dubovsky:2011sj}
S.~Dubovsky, L.~Hui, A.~Nicolis and D.~T. Son, {\it {Effective field theory for
  hydrodynamics: thermodynamics, and the derivative expansion}},
  \href{http://arXiv.org/abs/1107.0731}{{\tt 1107.0731}}.

\bibitem{Lin:2011aa}
S.~Lin, {\it {An anomalous hydrodynamics for chiral superfluid}},  {\em
  Phys.Rev.} {\bf D85} (2012) 045015
  [\href{http://arXiv.org/abs/1112.3215}{{\tt 1112.3215}}]. 21 pages.

\bibitem{Bhattacharyya:2008xc}
S.~Bhattacharyya {\em et.~al.}, {\it {Local Fluid Dynamical Entropy from
  Gravity}},  {\em JHEP} {\bf 06} (2008) 055
  [\href{http://arXiv.org/abs/0803.2526}{{\tt 0803.2526}}].

\bibitem{Chapman:2012my}
S.~Chapman, Y.~Neiman and Y.~Oz, {\it {Fluid/Gravity Correspondence, Local Wald
  Entropy Current and Gravitational Anomaly}},
  \href{http://arXiv.org/abs/1202.2469}{{\tt 1202.2469}}.

\end{thebibliography}\endgroup

\end{document}